\documentclass[aps,prl,floatfix,superscriptaddress,twocolumn,footinbib]{revtex4-2}
\newcommand{\equalcontribution}{These authors contributed equally to this work.}
\usepackage{amssymb}
\usepackage{amsmath}
\usepackage{amsfonts}
\usepackage{appendix}
\usepackage{bm}
\usepackage{braket}
\usepackage{graphicx}
\usepackage{epsfig}
\usepackage{epstopdf}
\usepackage{balance}
\usepackage[dvipsnames]{xcolor}
\usepackage{calc}
\usepackage{natbib}
\usepackage[misc]{ifsym}
\usepackage[colorlinks,
            linkcolor=blue,
            anchorcolor=blue,
            citecolor=blue,
            urlcolor=blue]{hyperref}
\usepackage{lipsum}

\begin{document}
\title{Sliding-tuned  Quantum Geometry in  Moir\'e Systems: Nonlinear Hall Effect and Quantum Metric Control}

\author{Shi-Ping Ding}
\altaffiliation{\equalcontribution}
\affiliation{School of Physics and Wuhan National High Magnetic Field Center,
    Huazhong University of Science and Technology, Wuhan 430074,  China}
\author{Miao Liang}
\altaffiliation{\equalcontribution}
\affiliation{School of Physics and Wuhan National High Magnetic Field Center,
    Huazhong University of Science and Technology, Wuhan 430074,  China}  
\author{Tian-Le Wu}
\affiliation{School of Physics and Wuhan National High Magnetic Field Center,
    Huazhong University of Science and Technology, Wuhan 430074,  China}
\author{Meng-Hao Wu}
\affiliation{School of Physics and School of Chemistry,
    Huazhong University of Science and Technology, Wuhan 430074,  China}
\author{Jing-Tao L{\"u}}    
\email{jtlu@hust.edu.cn}
\affiliation{School of Physics and Wuhan National High Magnetic Field Center,
    Huazhong University of Science and Technology, Wuhan 430074,  China}
\author{Jin-Hua Gao}
\email{jinhua@hust.edu.cn}
\affiliation{School of Physics and Wuhan National High Magnetic Field Center,
    Huazhong University of Science and Technology, Wuhan 430074, China}
 \author{X. C. Xie}
 \affiliation{Interdisciplinary Center for Theoretical Physics and Information Sciences (ICTPIS), Fudan University
, Shanghai 200433, China}
\affiliation{International Center for Quantum Materials, School of Physics, Peking University, Beijing 100871, China}
 \affiliation{Hefei National Laboratory, Hefei 230088, China}

\begin{abstract}
Sliding is a ubiquitous phenomenon in moir\'e systems, but its direct influence on moir\'e bands, especially in multi-twist moir\'e systems, has been largely overlooked to date. Here, we theoretically show that sliding provides a unique pathway to engineer the quantum geometry (Berry curvature and quantum metric) of moir\'e bands, exhibiting distinct advantages over conventional strategies. Specifically, we first suggest alternating twisted trilayer $\mathrm{MoTe_2}$ (AT3L-$\mathrm{MoTe_2}$) and chirally twisted triple bilayer graphene (CT3BLG)  as two ideal paradigmatic systems for probing sliding-engineered quantum geometric phenomena. Then, two sliding-induced exotic quantum geometry  phenomena are predicted: (1) an intrinsic nonlinear Hall effect via sliding-produced non-zero Berry curvature dipole, with CT3BLG as an ideal platform; (2) significant quantum metric modulation in AT3L-$\mathrm{MoTe_2}$, enabling tests of quantum geometric criteria for fractional Chern insulating state (FCIS). Our work establishes sliding as a new degree of freedom  for manipulating quantum geometry of moir\'e bands,  which emerges as a signature phenomenon of multi-twist moir\'e systems.
\end{abstract}
\maketitle

\emph{Introduction.}---Sliding is a widespread phenomenon in moir\'e systems, as a rigid sliding of a van der Waals (vdW) layer in moir\'e heterostructures only induces a global shift of the moir\'e pattern~\cite{doi:10.1073/pnas.1108174108,Topolo_Wu_prl2019}, without altering the total energy of the system. Previous studies have already shown that sliding can lead to a variety of intriguing phenomena, e.g., sliding ferroelectricity~\cite{doi:10.1021/acsnano.7b02756,doi:10.1126/science.abd3230,doi:10.1126/science.abe8177,wu2021sliding,fei2018ferroelectric,PhysRevLett.131.096801}. However, surprisingly, the direct effects of sliding on moir\'e flat bands have rarely been explored to date. The reason may lie in the fact that the influence of sliding is  effectively encoded as a phase factor of the interlayer hopping~\cite{doi:10.1073/pnas.1108174108,Topolo_Wu_prl2019,PhysRevB.97.035306,doi:10.7566/JPSJ.84.121001,Koshino_2015,carr2020electronic,PhysRevB.97.035306}. Yet, in most current moir\'e systems with only a single moir\'e interface, this sliding-induced phase factor can be eliminated by a suitable gauge choice. 
Interestingly, this scenario changes completely in multi-twist moir\'e systems~\cite{PhysRevB.104.035139,PhysRevB.108.155106,li2024dynamical,doi:10.1021/acs.nanolett.5c01944,PhysRevB.110.115136,PhysRevB.104.045413}, since the phase factor induced by sliding can no longer be removed through gauge transformations due to the existence of multiple moir\'e interfaces. 

In this work, we theoretically reveal that sliding is a highly effective means of manipulating the quantum geometry of moir\'e flat bands in multi-twist moir\'e systems. For solid-state systems, quantum geometry refers to the geometric properties of the eigen Bloch wave functions, encapsulated by the quantum metric (geometry) tensor~\cite{provost1980riemannian,resta2011insulating,PhysRevLett.131.240001,10.1093/nsr/nwae334,torma2022superconductivity} . The imaginary part of this tensor corresponds to the celebrated Berry curvature~\cite{berry1984quantal,RevModPhys.82.1959}, while the real part defines the quantum metric~\cite{provost1980riemannian,resta2011insulating,PhysRevLett.131.240001,10.1093/nsr/nwae334,torma2022superconductivity}. Here, through calculations in two typical multi-twist moir\'e systems, we show that sliding can drastically alter the Berry curvature and quantum metric of the moir\'e flat bands, while exerting minimal influence on their bandwidth and shape. 
Thus, it is quite unlike conventional methods such as twist angle and electrostatic gating~\cite{devakul2021magic,xiao2020berry,PhysRevB.111.085412,PhysRevB.110.125142,MA202118}, thereby offering a distinct and powerful approach for exploring quantum geometry related phenomena in moir\'e systems.

\begin{figure}[b!]
    \centering
    \includegraphics[scale=0.35]{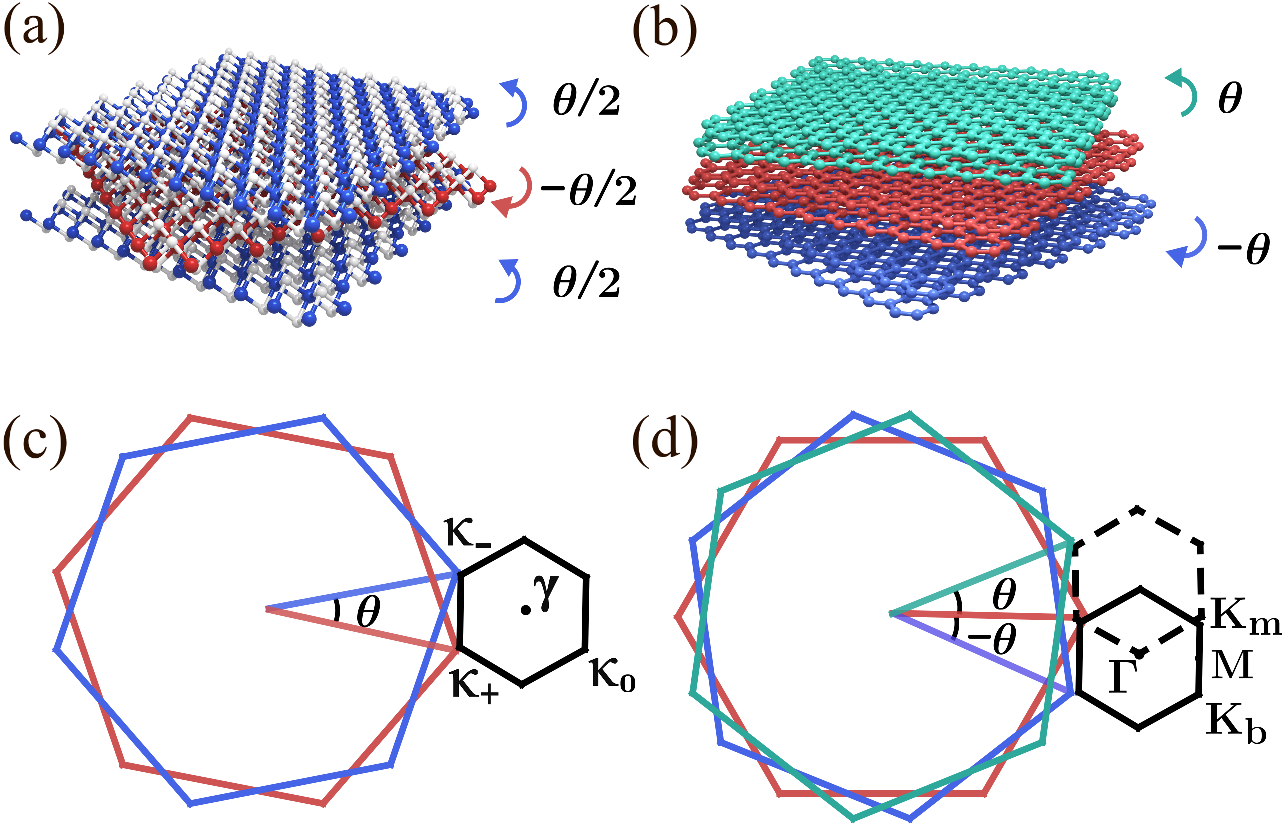}
    \caption{(a) is the schematic of AT3L-$\mathrm{MoTe_2}$ and (c) is the corresponding Brillouin Zone (BZ). 
    (b) is the schematic of CT3BLG and (d) is the corresponding BZ.}
    \label{fig:Schematic}
\end{figure}

Specifically, we first suggest that the alternating twisted trilayer $\mathrm{MoTe_2}$ (AT3L-$\mathrm{MoTe_2}$)~\cite{PhysRevB.111.085412,PhysRevB.108.155106,fedorko2025engineeringmoirekagomesuperlattices} and the chirally twisted triple bilayer graphene (CT3BLG)~\cite{PhysRevB.105.195422,PhysRevLett.132.246501,ZHOU2024100015,lin2025flatbandmanybodygap,PhysRevB.111.L161120}, as illustrated in Fig.~\ref{fig:Schematic}, are the two most promising experimental platforms for investigating the impact of sliding on quantum geometry. One main reason is that, compared to other multi-twist moir\'e systems~\cite{PhysRevB.105.195422,ma2023doubled,PhysRevB.108.195119,xie2022alternating,PhysRevB.100.085109,PhysRevB.107.245139}, these two systems possess the simplest moir\'e band structures near the Fermi level, each hosting one or two isolated moir\'e flat bands at small twist angles. Moreover, these two moir\'e systems exhibit entirely distinct quantum geometric features, which can lead to diverse sliding-induced phenomena. 

We then predict two intriguing sliding induced quantum geometric phenomena in multi-twist moir\'e systems: 
(1) \emph{Sliding-Driven Nonlinear Hall effect.} Contrary to conventional understanding, sliding can directly induce an intrinsic nonlinear Hall effect (NHE)~\cite{PhysRevLett.115.216806,ma2019observation,kang2019nonlinear} in multi-twist moir\'e systems without the need  for strain to break symmetry~\cite{sinha2022berry,hu2022nonlinear,Chakraborty_2022,PMID:37180357,  PhysRevB.106.L041111,PhysRevB.103.205403,PhysRevB.106.L161101,PhysRevLett.131.066301,PhysRevLett.123.036806,Qin_2021,PhysRevB.98.121109,ho2021hall,PhysRevB.111.L041403,PhysRevLett.123.196403,du2021engineering,zhou2024nonlinear}. This originates from sliding induced Berry curvature redistribution and the resulted Berry curvature dipole (BCD)~\cite{PhysRevB.109.075415,PhysRevB.110.235418}. In CT3BLG, the nonlinear Hall effect is especially significant, where the BCD is of the order of $10~\mathring{\mathrm{A}}$, rendering it an ideal platform for measuring this phenomenon. \newline
(2) \emph{Quantum Metric Control.} Sliding can significantly alter the quantum metric while keeping the bandwidth of the moir\'e flat bands largely unchanged, thereby providing an excellent opportunity to examine the quantum metric related phenomena in such moir\'e flat band system. One possible example is the quantum geometric criteria required for realizing fractional Chern insulators (FCI)~\cite{PhysRevB.90.165139,PhysRevLett.114.236802,PhysRevB.85.241308}. Theoretically, it is believed that only when a flat Chern band possesses the same ideal quantum geometry (mainly its quantum metric) as that of Landau levels can the fractional Chern insulator (FCI) state be realized, which is so called quantum geometric criterion. AT3L-$\mathrm{MoTe_2}$  exhibits an isolated  flat Chern band near the Fermi level at suitable small twist angles, quite like the twisted bilayer $\mathrm{MoTe_2}$. Meanwhile, sliding can significantly modulate the quantum metric of the moir\'e flat band while maintaining the correlation effects (i.e.~bandwidth of flat band) largely unchanged. So, AT3L-$\mathrm{MoTe_2}$ offers an ideal platform to realize the FCI state and test the quantum geometry criteria.

Overall, our work clearly reveals the potential of sliding as a versatile tool for quantum geometric engineering in moir\'e materials.

\begin{figure}[t]
    \centering
    \includegraphics[scale=0.36]{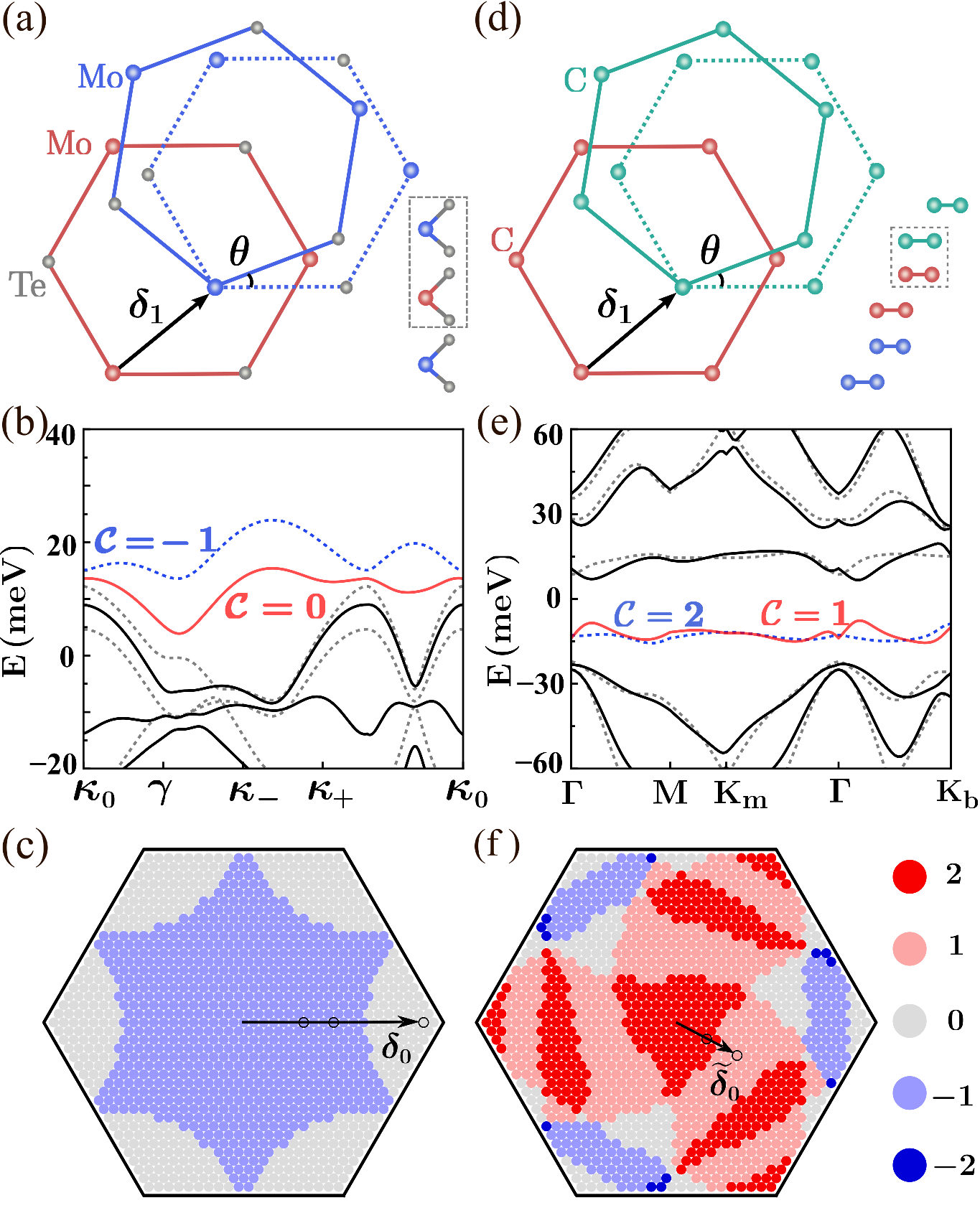}
    \caption{(a) and (d) are the schematic of sliding configuration for AT3L-$\mathrm{MoTe_2}$ and CT3BLG, respectively. In (a), blue (maroon) represents the top (middle) $\mathrm{MoTe_2}$ layer, $\theta$ is the twist angle, $\boldsymbol{\delta_1}$ is the sliding vector. In (d), green (maroon) represents the top (middle) BLG. (b) is moir\'e bands of AT3L-$\mathrm{MoTe_2}$ at $\theta=3.15^\circ$, where solid (dotted) lines are for $\boldsymbol{\delta_1}=\boldsymbol{\delta_0}$ ($\boldsymbol{\delta_1}=\boldsymbol{0}$) with $\boldsymbol{\delta_0}\equiv0.3a_0\times (\frac{3}{2},\frac{\sqrt{3}}{2})$. (c) is Chern number of topmost moir\'e flat band as a function of $\boldsymbol{\delta_1}$ for AT3L-$\mathrm{MoTe_2}$. (e) is moir\'e bands of CT3BLG, where solid (dotted) lines are for $\boldsymbol{\delta_1}=\boldsymbol{\widetilde{\delta}_0}$ ($\boldsymbol{\delta_1}=\boldsymbol{0}$) with $\boldsymbol{\widetilde{\delta}_0}=0.2\widetilde{a}_0\times (1,0)$. The voltage between adjacent vdW layers is $V_0=10$ meV. (f) is Chern number of topmost valence flat band of CT3BLG as a function of $\boldsymbol{\delta_1}$. $\boldsymbol{\delta_1}$ is defined in the WS cell of  $\mathrm{MoTe_2}$ monolayer (or BLG). Chern Numbers are labeled by different color.}
    \label{BS}
\end{figure}

\emph{Hamiltonian of AT3L-\textrm{MoTe$_2$}.} As shown in Fig.~\ref{fig:Schematic} (a), the AT3L-$\mathrm{MoTe_2}$ consists of three $\mathrm{MoTe_2}$ monolayers, stacked with two equal twist angles in an alternating manner~\cite{PhysRevB.111.085412,PhysRevB.108.155106,nakatsuji2025moirebandengineeringtwisted,fedorko2025engineeringmoirekagomesuperlattices,PhysRevB.109.085118,10.1093/nsr/nwz117,hao2024robust},  and the corresponding BZ is shown in Fig.~\ref{fig:Schematic} (c). For the $+K$ valley, the moir\'e Hamiltonian is 
\begin{equation}
	\begin{aligned}\label{H_N}
		\mathcal{H}\left(\mathbf{k}, \boldsymbol{\delta}_1, \boldsymbol{\delta}_3\right) =\left(\begin{array}{ccc}
			\mathcal{H}_1 & T_{12} & 0 \\
			T_{12}^{+} & \mathcal{H}_{2} & T_{32}^{+} \\
			0 & T_{32} & \mathcal{H}_3
		\end{array}\right),
	\end{aligned}
\end{equation}
where $\mathcal{H}_{i=1,2,3}$ describes the valence band of the three $\mathrm{MoTe_2}$ monolayers with $i$ being the layer index,
\begin{equation}
    \begin{aligned}\label{H_1}
        \mathcal{H}_{1/3}&=-\frac{\hbar^2\left(\mathbf{k}-\mathbf{\kappa}_{+} \right)^2}{2 m^*}+\Delta_{1/3}\left(\boldsymbol{\delta}_{1/3}\right),\\
\mathcal{H}_{2}&=-\frac{\hbar^2\left(\mathbf{k}-\mathbf{\kappa}_{-} \right)^2}{2 m^*}+\Delta_{2 }\left(\boldsymbol{\delta}_1\right)+\Delta_{2 }\left(\boldsymbol{\delta}_3\right).
    \end{aligned}
\end{equation}
Here, $\mathbf{\kappa}_{\pm}$ are the momentum offsets as shown in Fig.~\ref{fig:Schematic} (c), $\boldsymbol{\delta_{1,3}}$ are the sliding vectors of the top and bottom layers, and $\Delta_i\left(\boldsymbol{\delta}_{i'}\right)$ describes the layer-dependent moir\'e potential 
\begin{equation}\label{Delta_i}
	\begin{aligned}
			\Delta_i\left(\boldsymbol{\delta}_{i'}\right)=
   2 V \sum_{j=1,3,5} \cos \left[\boldsymbol{g}_j \cdot \boldsymbol{r}+\boldsymbol{G}_j \cdot \boldsymbol{\delta}_{i'} +(-1)^{i+1} \psi \right]
    \end{aligned}
\end{equation}
where $\mathbf{g}_1 = (4\pi /\sqrt{3} a_M,0)$ is one moir\'e reciprocal lattice vector with $a_M\approx a_0/\theta$ ($a_0=3.472~\mathring{\rm{A}}$), $\mathbf{G}_1 = (0,4\pi /\sqrt{3} a_0)$  is one first-shell reciprocal lattice vectors of a $\mathrm{MoTe_2}$ monolayer. $\mathbf{g}_j$ and $\mathbf{G}_j$ are $(j-1)\pi/3$ counterclockwise rotation of $\mathbf{g}_1$ and $\mathbf{G}_1$, respectively. For $\rm{MoTe_2}$, we set $a_0=3.472 \mathring{\rm{A}}$, $V=8~\rm{meV}$, $\psi=-89.6^{\circ}$, and the effective mass $m^* =0.62~m_e$~\cite{Topolo_Wu_prl2019,Three_prb085433_2013}. 
$T_{12}$ and $T_{32}$ are the interlayer tunneling described as
\begin{equation}\label{T_i2}
    \begin{aligned}
        T_{i2}(\boldsymbol{\delta}_{i})=w\left[1+e^{-i ( \mathbf{g}_2 \cdot \mathbf{r}+\boldsymbol{G}_2 \cdot \boldsymbol{\delta}_{i})}+e^{-i ( \mathbf{g}_3 \cdot \mathbf{r} + \boldsymbol{G}_3 \cdot \boldsymbol{\delta}_{i})} \right],
    \end{aligned}
\end{equation}
with $w=-8.5$ meV~\cite{Topolo_Wu_prl2019}.

\begin{figure*}[ht]
    \centering
    \includegraphics[scale=0.37]{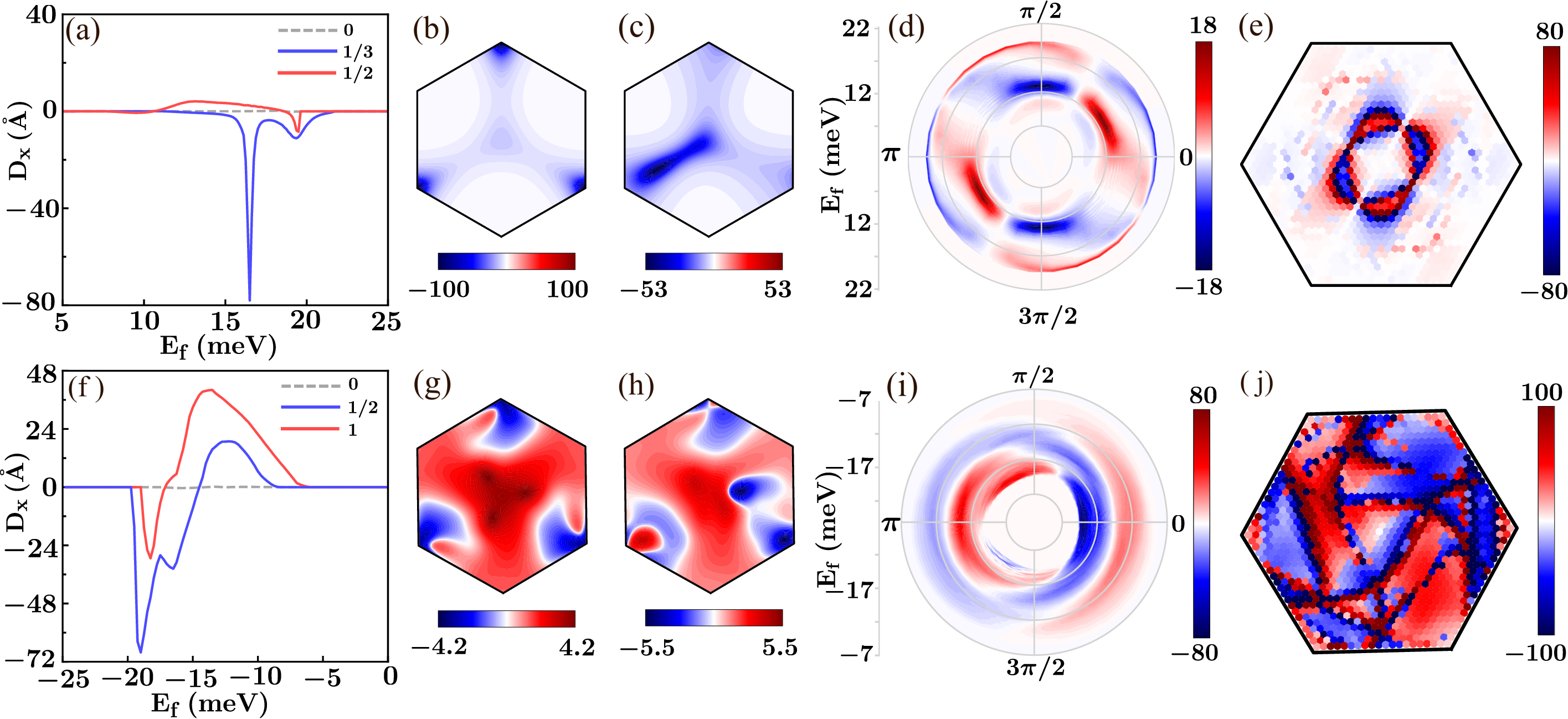}
    \caption{(a-e) are the Berry curvature dipole (BCD) of AT3L-$\mathrm{MoTe_2}$. (a) is $D_x$ for various sliding vector $\boldsymbol{\delta_1}=\alpha\boldsymbol{\delta_0}$. (b) is the Berry curvature distribution of first moir\'e band in Brillouin zone (BZ) with zero sliding, and (c) is that for $\alpha=1/2$. (d) is $D_x$ as a function of $E_f$ and $\theta_s \equiv \angle(\mathbf{\delta_1}, \mathbf{\delta_0})$, with a  fixed $\alpha=1/2$. (e) is the maximum of $D_x$ for  all possible sliding with optimal $E_f$, where $\boldsymbol{\delta_1}$ is defined in a WS cell of the $\mathrm{MoTe_2}$ monolayer. (f-j) are the BCD of CT3BLG. (f) is $D_x$ for various $\boldsymbol{\delta_1}=\alpha\boldsymbol{\widetilde{\delta}_0}$. The corresponding Berry curvature for $\alpha=0$ and $\alpha=1$ is shown in (g) and (h), respectively, on a natural logarithmic scale. $\theta_s \equiv \angle(\boldsymbol{\delta_1}, \boldsymbol{\widetilde{\delta}_0})$ and $\alpha=1/2$ in (i), (j) is the maximum of $D_x$ for  all possible sliding $\boldsymbol{\delta_1}$ that is defined in a WS cell of the top BLG. }
    \label{NLE}
\end{figure*}

As shown in Fig.~\ref{BS} (a), we take the middle layer as the reference, and the sliding vector $\boldsymbol{\delta}_1$ ($\boldsymbol{\delta}_3$) describes the sliding of the first (third) layer relative to the middle layer. In fact, the synchronous sliding with $\boldsymbol{\delta}_1 = \boldsymbol{\delta}_3$ does not affect the moir\'e bands, similar to the single twist cases. Thus, only the relative sliding between the first and third layer has an impact on the moir\'e bands. For simplicity, we set $\boldsymbol{\delta}_3=0$ and use the sliding vector $\boldsymbol{\delta}_1$ to denote the relative sliding. Furthermore, 
we observe that the effects of interlayer sliding manifest solely in the interlayer potential and the tunneling terms through phase factors like $\exp{(i\boldsymbol{G_j}\cdot \boldsymbol{\delta}_i)}$, which implies that each sliding vector  $\boldsymbol{\delta}_1$  is always equivalent to a vector in the Wigner-Seitz (WS) cell of $\rm{MoTe_2}$ monolayer in real space (see Supplementary Material~\cite{supplemental} for details), see Fig.~\ref{BS} (c). It enables a convenient description of all sliding cases.

\emph{Hamiltonian of CT3BLG}. Another interesting system is CT3BLG, which  consists of three bilayer graphene (BLG) chirally rotated with the same twist angles~\cite{PhysRevB.105.195422,PhysRevLett.132.246501}, as shown in Fig.~\ref{fig:Schematic} (b). When $\theta$ is small, we can use an approximate BZ, see Fig.~\ref{fig:Schematic} (d), to calculate its moir\'e bands. Based on the continuum model method, its Hamiltonian for the $+K$ valley is 
\begin{equation}\label{Hamiltonian_2+2+2}
    H\left(\mathbf{k}, \boldsymbol{\delta}_1, \boldsymbol{\delta}_3\right)=\left(
    \begin{array}{ccc}
        H_1&U_{12}&0\\
        U_{12}^\dag&H_2&U_{23}\\
        0&U_{23}^\dag&H_3
    \end{array}
    \right)+H_V,
\end{equation}
where $H_l$ is the $k\cdot p$ Hamiltonian of $l\textrm{th}$ BLG, $H_V=V_0\times \rm{diag}(-\frac{5}{2},-\frac{3}{2},-\frac{1}{2},\frac{1}{2},\frac{3}{2},\frac{5}{2})\otimes\sigma_0$ represents the gating voltage, and
$U_{12/23}=\frac{1}{2}(\sigma_x-i\sigma_y)\otimes U(\boldsymbol{\delta}_{1/3})$ denotes the interlayer tunneling between the 1st-2nd or 2nd-3rd BLGs with
\begin{equation}
    U(\boldsymbol{\delta}_l)=\sum_{n=0}^2 U_n
    e^{i\mathbf{\widetilde{g}}_n\cdot \mathbf{r}}e^{i \eta_l \mathbf{\widetilde{G}_n}\cdot \boldsymbol{\delta}_l }.
\end{equation}
Here, $U_n=\omega_{\mathrm{AA}}\sigma_0+\omega_{\mathrm{AB}}[\cos(2n\pi/3)\sigma_x+\sin(2n\pi/3)\sigma_y]$,
$\eta_ l=\textrm{sign}(\theta_l)$, 
$\mathbf{\widetilde{g}}_n$ ($\mathbf{\widetilde{G}_n}$) is three reciprocal lattice vector of the moir\'e pattern (BLG), and $\boldsymbol{\delta}_{l}$ denotes the sliding of $l\textrm{th}$ BLG. For simplicity, we focus on the cases that only the top BLG is shifted, so we set $\boldsymbol{\delta}_{3}=\boldsymbol{0}$ and use the sliding vector $\boldsymbol{\delta_1}$ to represent the sliding configuration, see Fig.~\ref{BS} (f). All the details of the Hamiltonian are given in the Supplementary Material~\cite{supplemental}. Note that, in order to isolate the two moir\'e flat bands here, a gating voltage is applied.

Here, the quantum geometry tensor is $Q_{\mu\nu} \equiv \braket{\partial_\mu\psi | \partial_\nu\psi} - \braket{\partial_\mu\psi | \psi}\braket{\psi | \partial_\nu\psi}$,
where $\mu,\nu\in\{k_x,k_y\}$ and $\ket{\psi}$ is the wavefunction of moir\'e band. $g_{\mu\nu}=\textrm{Re}{Q_{\mu\nu}}$ is the quantum metric  and $\Omega_{\mu\nu}=-2\textrm{Im}{Q_{\mu\nu}}$ is the Berry curvature~\cite{PhysRevLett.131.240001,10.1093/nsr/nwae334} .  

\emph{Sliding-Driven Nonlinear Hall effect.} As seen from the Hamiltonian, the effect of sliding on the system is reflected in the additional phase factors of the interlayer tunneling (potential), similar to the influence of a magnetic field on a lattice. Therefore, we can anticipate that sliding will significantly affect the quantum geometric properties of the wave functions. 

First, sliding can modify the Chern number (for each valley) of the moir\'e flat bands, demonstrating its ability to substantially modulate the Berry curvature  of the bands.

AT3L-\textrm{MoTe$_2$} exhibits an isolated moir\'e flat band near $E_f$ at small twist angles~\cite{PhysRevB.111.085412}. In Fig.~\ref{BS}(b), we  plot the moir\'e bands of AT3L-\textrm{MoTe$_2$} at $\theta=3.15^\circ$ without sliding ($\boldsymbol{\delta}_1=\bm{0}$), shown as dotted lines. Here, the Chern number of the moir\'e band near $E_f$ (blue dotted line) is $-1$. With a finite sliding $\boldsymbol{\delta}_1=\boldsymbol{\delta}_0$, $\boldsymbol{\delta}_0\equiv0.3a_0\times (\frac{3}{2},\frac{\sqrt{3}}{2})$, the corresponding moir\'e bands are plotted in Fig.~\ref{BS}(b) using solid lines, where the Chern number of the first moir\'e band becomes zero, see the red solid line. We then plot the Chern number as a function of sliding vector $\boldsymbol{\delta}_1$ in Fig.~\ref{BS}(c), which provides a topological phase diagram of the first moir\'e band. Note that, any sliding here can  be presented as a sliding vector $\boldsymbol{\delta}_1$ in the WS cell of one vdW layer, as mentioned above. Such phase diagram shows that the Chern number in this case is either $-1$ (blue) or $0$ (gray), depending on the value of $\boldsymbol{\delta}_1$. Notably, numerical calculations indicate that although sliding can significantly tune the Chern number, the corresponding change in bandwidth remains consistently very small, as seen in Fig.~\ref{BS} (b). In fact, for any sliding, the band width varies by less than 5 meV here. This feature provides a unique advantage for   quantum geometry control.

The case of CT3BLG is similar, which hosts two nearly degenerate flat bands at $E_f$~\cite{PhysRevB.105.195422}. Here, we assume a finite electric field to isolate the two moir\'e bands further. In Fig.~\ref{BS}(e), we plot the moir\'e bands of the case without sliding using dotted lines, where the Chern number of the top valence band is $2$ (blue dotted line). For a finite sliding $\boldsymbol{\delta}_1=\boldsymbol{\widetilde{\delta}_0}$, $\boldsymbol{\widetilde{\delta}_0}\equiv0.2\widetilde{a}_0\times(1,0)$ with $\widetilde{a}_0=2.46\, \mathring{\mathrm{A}}$, the shape of the moir\'e bands remains nearly unchanged, see the solid lines in Fig.~\ref{BS} (e), while the Chern number is changed from $2$ to $1$. We plot the topological phase diagram of the first valence moir\'e band in Fig.~\ref{BS} (f). Depending on $\boldsymbol{\delta}_1$, the Chern number varies within the range of $+2$ to $-2$. Note that CT3BLG has been realized in experiment recently~\cite{PhysRevLett.132.246501,lin2025flatbandmanybodygap,PhysRevB.111.L161120}. Our results here reveal that when considering the correlated Chern insulating states in CT3BLG, sliding should be a key factor that can not be ignored. 

Second, sliding is not only able to alter the Berry curvature distribution but also can break the $C_3$rotational symmetry. Therefore, it will naturally induce a non-zero Berry curvature dipole, leading to an intrinsic nonlinear Hall effect~\cite{PhysRevLett.115.216806,du2024nonlinear,du2021nonlinear,du2021quantum,qute.202100056,PhysRevLett.121.266601,layek2025quantum,PhysRevB.110.245304}.

\begin{figure}[ht!]
    \centering
    \includegraphics[scale=0.4]{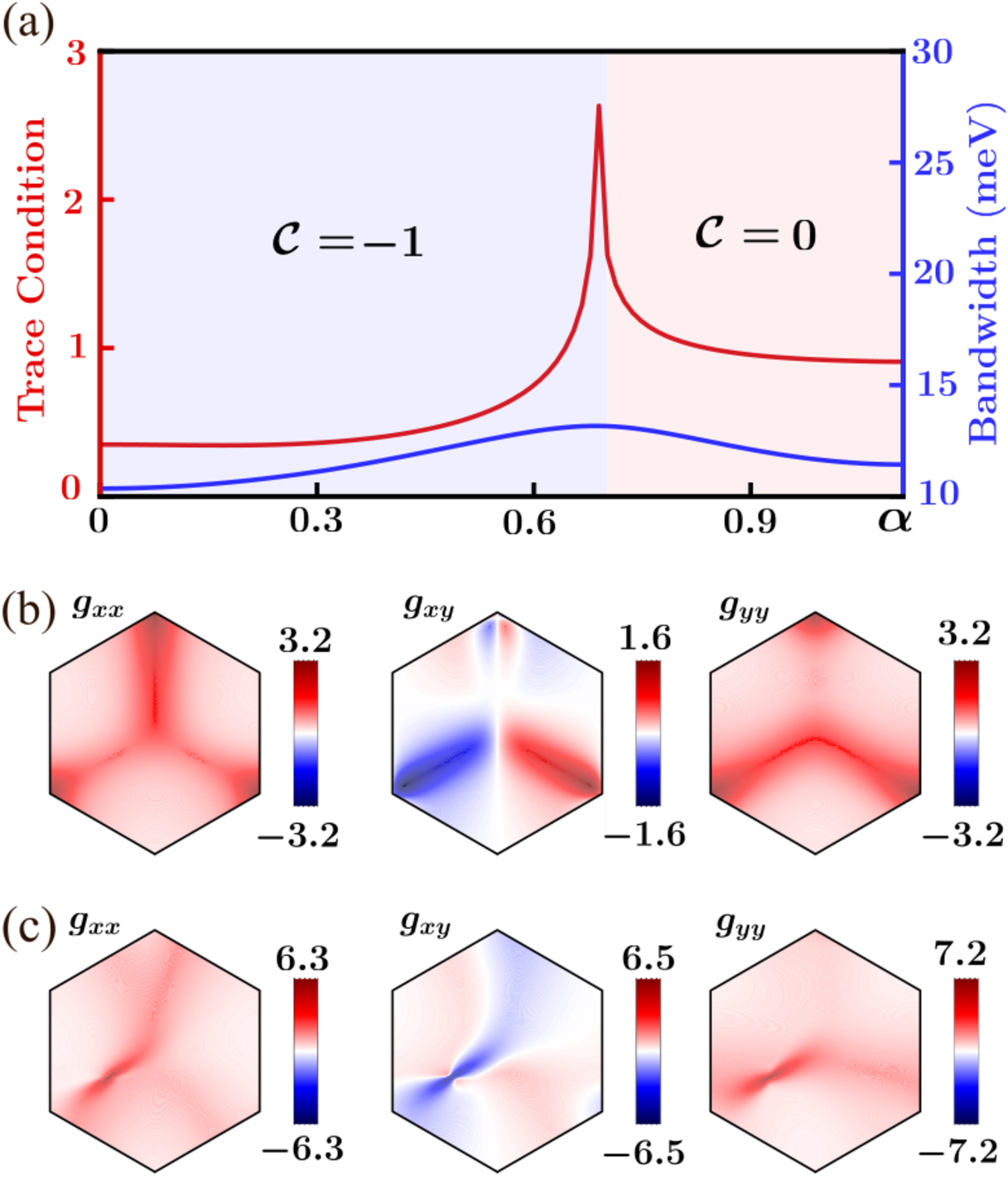}
    \caption{(a) is the trace condition $T$ (blue) and band width (red) of the first moir\'e band of AT3L-$\textrm{MoTe$_2$}$, plotting as functions of $\boldsymbol{\delta_1}=\alpha \boldsymbol{\delta_0}$. (b)  is the quantum metric $g_{xx},g_{xy},g_{yy}$ distribution in the whole BZ for $\boldsymbol{\delta_1}=\boldsymbol{0}$, and (c) is that at finite sliding $\alpha=2/3$. Data shown in (b) and (c) are natural logarithm-transformed values. }
    \label{QG}
\end{figure}

The NHE induced by quantum geometric effects has recently attracted significant attention~\cite{PhysRevLett.132.026301,Wang_2025,BANDYOPADHYAY2024100101,jiang2025revealingquantumgeometrynonlinear,gao2025quantumgeometryphenomenacondensed,PhysRevLett.127.277201,gong2025nonlineartransporttheoryorder,PhysRevB.110.L081301,du2019disorder,PhysRevApplied.13.024053,PhysRevLett.131.076601,PhysRevApplied.13.024053,saha2023nonlinear,PhysRevLett.132.266802,chichinadze2025observationgiantnonlinearhall}. Unlike the conventional Hall effect, the NHE does not require breaking time-reversal symmetry but necessitates breaking spatial-inversion symmetry. Theoretically, it is proved that, when the Berry curvature distribution is nonuniform (i.e., a BCD exists), a longitudinal ac electric field $\boldsymbol{E}(t)=\mathrm{Re}\{ \boldsymbol{\varepsilon} e^{i\omega t} \}$ with frequency $\omega$ can induce a transverse Hall current $\boldsymbol{j}(t)=\mathrm{Re}\{ \boldsymbol{j}^0 + \boldsymbol{j}^{2\omega} e^{2i\omega t}\}$, where $\boldsymbol{j}^{2\omega}= \chi_{abc} \varepsilon_b \varepsilon_c$ is the second-harmonic component with frequency $2\omega$~\cite{PhysRevLett.115.216806,PhysRevLett.123.196403,PhysRevB.106.L041111}. Here, $\chi_{abc}=-\frac{e^3 \tau}{2(1+i\omega \tau)}\varepsilon_{adc}D_{bd}$ refers to the nonlinear Hall susceptibility, where $\varepsilon_{adc}$ is the Levi-Civita tensor, $\tau$ is the scattering time and $a,b,c,d\in \{x,y,z \}$. In two dimensional  systems, the BCD is defined as $D_b \equiv D_{bz}$  $(b=x,y)$ ~\cite{PhysRevLett.115.216806}
\begin{equation}
    D_{x}=\sum_n\int_{\mathrm{mBZ}}\frac{d^2\bm{k}}{(2\pi)^2}\partial_x\Omega_z^n\times f_0(E_f),
\end{equation}
where $n$ is the band index and $\Omega^n_z$ is the Berry curvature and $f_0$ is the Fermi-Dirac distribution function. The BCD directly represents the NHE.

Due to the $C_3$ symmetry, the BCD in the moir\'e systems like twisted bilayer graphene (TBG), CT3BLG and AT3L-$\textrm{MoTe$_2$}$ vanishes in the absence of interlayer sliding~\cite{sinha2022berry,hu2022nonlinear,Chakraborty_2022,PMID:37180357,  PhysRevB.106.L041111,PhysRevB.103.205403,PhysRevB.106.L161101,PhysRevLett.131.066301,PhysRevLett.123.036806,Qin_2021,PhysRevB.98.121109,ho2021hall,PhysRevB.111.L041403,PhysRevLett.123.196403,du2021engineering,zhou2024nonlinear}. In previous studies, strain is believed to be necessary to break the $C_3$ symmetry to get nonzero BCD in TBG~\cite{PhysRevLett.131.066301}. However, in multi-twisted moir\'e systems we considered here, sliding  provides a more convenient way to break the $C_3$ symmetry, thereby inducing nonzero nonlinear Hall effect. Note that sliding can not induce a nonzero BCD in TBG, because the sliding effects in single-twisted moir\'e systems can be eliminated via an appropriate gauge transformation, as mentioned before.

Compared to strain engineering, sliding is a superior approach to modulating the NHE and Berry curvature in moir\'e systems. On the one hand, sliding directly controls the phase factor of interlayer hopping terms, effectively acting as an artificial gauge field, and thus can reshape the Berry curvature distribution in a more direct way. On the other hand, from an experimental perspective, interlayer sliding can be precisely controlled in situ and reversibly through an applied electric field, as illustrated in sliding ferroelectricity~\cite{doi:10.1126/science.abd3230,doi:10.1126/science.abe8177,wu2021sliding,fei2018ferroelectric}.

Numerical results in Fig.~\ref{NLE} show that sliding can indeed generate a nonzero BCD. Fig.~\ref{NLE} (a-e) present the results of AT3L-$\textrm{MoTe$_2$}$, while panels (f-j) correspond to the case of CT3BLG. In Fig.~\ref{NLE} (a), we plot the calculated BCD for various sliding vectors $\boldsymbol{\delta}_1=\alpha  \boldsymbol{\delta_0}$ ($\alpha=0,\frac{1}{3},\frac{1}{2}$). As we expected, with zero sliding ($\alpha=0$), the BCD is exactly zero due to the $C_3$ symmetry, while a nonzero sliding does produce a nonzero BCD, which becomes as large as 80~\AA \  (blue line with $\alpha=\frac{1}{3}$, $E_f \approx 16.4$ meV). Note that the main contribution to the giant BCD peak here originates from the second valence band (see Supplementary Material~\cite{supplemental}), as the first and second moir\'e bands, though isolated, exhibit an energetic overlap when $\alpha=\frac{1}{3}$. In fact, for AT3L-$\textrm{MoTe$_2$}$, the Berry curvature of the first moir\'e band is relatively uniform, hence resulting in a small BCD (e.g.~the red line with $\alpha=\frac{1}{2}$).

The corresponding Berry curvature distribution of the first moir\'e band with different sliding $\boldsymbol{\delta_1}$ is plotted in Fig.~\ref{NLE} (b, c). In the absence of sliding,  the Berry curvature distribution exhibits pronounced $C_3$ symmetry, as shown in Fig.~\ref{NLE} (b). In contrast, a finite sliding $\alpha=\frac{1}{2}$ clearly breaks the $C_3$ symmetry and produces an obvious modification of the Berry curvature distribution, as shown in Fig.~\ref{NLE} (c), thereby inducing a nonzero NHE.

The BCD depends on the magnitude and direction of the sliding vector, as well as the position of the Fermi level. In Fig.~\ref{NLE} (d), we plot the BCD as a function of $E_f$ and $\theta_s \equiv \angle(\mathbf{\delta_1}, \mathbf{\delta_0})$ (the angle between $\boldsymbol{\delta}_1$ and $\boldsymbol{\delta}_0$ ), where $|\boldsymbol{\delta}_1|$ is fixed with $\alpha=\frac{1}{2}$. And the maxima of BCD for all possible sliding with optimal $E_f$ are given in  Fig.~\ref{NLE} (e), suggesting that the maximum of BCD can reach about $\pm~80$ \AA. The corresponding results of the CT3BLG, presented in Fig.~\ref{NLE} (f-j), exhibit a physical mechanism closely analogous to that of the AT3L-$\textrm{MoTe$_2$}$. Note that the inhomogeneous distribution of Berry curvature in graphene systems results in a larger BCD compared to that in $\textrm{MoTe$_2$}$ systems. Since the first moir\'e flat band in CT3BLG is always separated from the lower bands in energy~\cite{PhysRevB.105.195422}, the contributions from other moir\'e bands are excluded in the calculated BCD in Fig.~\ref{NLE} (f-j).

\emph{Sliding Adjusts Quantum metric.} 
Sliding can serve as an ideal tunable knob for engineering the quantum metric of flat Chern bands, thereby in principle all the quantum metric related phenomena can be examined in the two suggested systems above. Here, we propose an intriguing example that sliding can provide a new way to examine experimentally and control the quantum metric criteria for fractional Chern insulator (FCI) realization~\cite{PhysRevB.90.165139,PhysRevLett.114.236802,PhysRevB.85.241308}, proposed to explain the recently observed fractional quantum anomalous Hall (FQAH) state in twisted bilayer \textrm{MoTe$_2$} system~\cite{park2023observation,PhysRevX.13.031037,cai2023signatures,zeng2023thermodynamic,Ji_2024,Redekop_2024,holtzmann2025opticalcontrolintegerfractional,Park_2025,Xu_2025,chang2025evidencecompetinggroundstates,Adak_2024,PhysRevLett.132.036501,PhysRevB.108.085117,PhysRevB.108.245159,PhysRevB.107.L201109,PhysRevResearch.3.L032070,PhysRevLett.133.186602,PhysRevB.110.125142,Chen_2025,doi:10.1073/pnas.2316749121,PhysRevB.111.075108,PhysRevLett.134.066601,PhysRevLett.132.096602,PhysRevB.109.045147}.

The FQAH state is a FCI that does not require an external magnetic field~\cite{Lu_2024,Xie_2025}. Theoretically, FCI can be viewed as the lattice analogy of fractional quantum Hall (FQH) states. Analogous to the role of Landau levels (LL) in the FQH states, flat Chern bands play a corresponding role in FCIs. Consequently, realizing FCIs necessitates that the quantum geometric properties of flat Chern bands closely resemble those of Landau levels. So,  a quantum geometry criterion  of flat Chern band is suggested: the quantum geometry should distribute uniformly in the whole BZ and the trace condition of the quantum metric should be fulfilled in order to mimic the lowest Landau level (LLL) physics.
Specifically, the trace condition of a flat Chern band here refers to~\cite{xie2021fractional,PhysRevB.110.125142,PhysRevB.110.165142,PhysRevLett.133.156504,doi:10.1126/sciadv.adi6063,doi:10.1073/pnas.2316749121,PhysRevLett.131.136502,PhysRevLett.133.206503,parker2021fieldtunedzerofieldfractionalchern,PhysRevResearch.5.L032022,PhysRevB.90.165139,PhysRevResearch.2.023237,PhysRevB.93.235133,PhysRevB.108.205144,PhysRevResearch.5.023166}
\begin{equation}
    T=\frac{1}{2\pi}\int \mathrm{d}^2\mathbf{k} (\textrm{Tr}\{g(\mathbf{k})\}-|\Omega(\mathbf{k})|),
\end{equation}
where $g(\mathbf{k})$ is the quantum metric, and the smaller the value of $T$, the closer the flat Chern band approaches the LLL.


The AT3L-\textrm{MoTe$_2$} offers an ideal platform to examine the quantum geometry criterion. As shown in Fig.~\ref{BS} (b), its first moir\'e flat band has a Chern number of -1, while exhibiting a uniformly distributed Berry curvature. It is quite like that of the twisted bilayer \textrm{MoTe$_2$}~\cite{devakul2021magic,PhysRevB.110.125142}, suggesting it as a promising platform of the FQAH state. More interestingly, in AT3L-\textrm{MoTe$_2$}, sliding can continuously modulate the quantum metric of the first flat Chern band (the blue dotted line in Fig.~\ref{BS} (b)), while maintaining its bandwidth nearly unchanged. For example, in Fig.~\ref{QG} (a), we plot its trace condition $T$ and bandwidth as functions of the sliding vector $\boldsymbol{\delta_1}=\alpha \boldsymbol{\delta_0}$ along the $\boldsymbol{\delta_0}$ direction. When $\alpha=0$, $T=0.35$ and the bandwidth is about 10.4~meV, which meets the requirements of FCI. Meanwhile, with a finite sliding $\alpha=\frac{2}{3}$, $T$ becomes 1.29, while the Chern number is invariant and the bandwidth is still narrow (bandwidth variation and Berry curvature uniformity for all sliding are provided in Supplementary Material~\cite{supplemental}). In this case, if the quantum metric condition is correct, we expect there exhibit a quantum metric induced phase transition from the FCI to a trivial state. Meanwhile, the quantum metric $g_{xx},g_{xy},g_{yy}$ distribution in the whole BZ for $\boldsymbol{\delta_1}=0$ is plotted in Fig.~\ref{QG} (b), while that for $\alpha=\frac{2}{3}$ are given in Fig.~\ref{QG} (c) as a comparison. Clearly, we can see that sliding does induce significant modifications to the quantum metric.

\emph{Summary.} We conclude with two critical insights: (i) Beyond the NHE and quantum metric criterion above, sliding in multi-twist moir\'e systems should in fact modulate all quantum geometry-related phenomena. (ii) Distinct from single-twist cases, sliding in turn emerges as a non-negligible factor influencing quantum geometric features in multi-twist ones. In other words, uncontrolled sliding in devices inevitably induces substantial variations in measurements of the quantum geometric properties.







\begin{acknowledgments}
    This work was supported by the National Natural Science Foundation of China (Grants No.~12141401 and No.~22273029), the National Key Research and Development Program of China (Grants No.~2022YFA1403501 and No.~2022YFA1402400), China Postdoctoral Science Foundation (Grant No. 2024M750984), and Innovation Program for Quantum Science and Technology (Grant No. 2021ZD0302400). 
\end{acknowledgments}

\bibliography{references}

\begin{thebibliography}{129}%
\makeatletter
\providecommand \@ifxundefined [1]{%
 \@ifx{#1\undefined}
}%
\providecommand \@ifnum [1]{%
 \ifnum #1\expandafter \@firstoftwo
 \else \expandafter \@secondoftwo
 \fi
}%
\providecommand \@ifx [1]{%
 \ifx #1\expandafter \@firstoftwo
 \else \expandafter \@secondoftwo
 \fi
}%
\providecommand \natexlab [1]{#1}%
\providecommand \enquote  [1]{``#1''}%
\providecommand \bibnamefont  [1]{#1}%
\providecommand \bibfnamefont [1]{#1}%
\providecommand \citenamefont [1]{#1}%
\providecommand \href@noop [0]{\@secondoftwo}%
\providecommand \href [0]{\begingroup \@sanitize@url \@href}%
\providecommand \@href[1]{\@@startlink{#1}\@@href}%
\providecommand \@@href[1]{\endgroup#1\@@endlink}%
\providecommand \@sanitize@url [0]{\catcode `\\12\catcode `\$12\catcode
  `\&12\catcode `\#12\catcode `\^12\catcode `\_12\catcode `\%12\relax}%
\providecommand \@@startlink[1]{}%
\providecommand \@@endlink[0]{}%
\providecommand \url  [0]{\begingroup\@sanitize@url \@url }%
\providecommand \@url [1]{\endgroup\@href {#1}{\urlprefix }}%
\providecommand \urlprefix  [0]{URL }%
\providecommand \Eprint [0]{\href }%
\providecommand \doibase [0]{https://doi.org/}%
\providecommand \selectlanguage [0]{\@gobble}%
\providecommand \bibinfo  [0]{\@secondoftwo}%
\providecommand \bibfield  [0]{\@secondoftwo}%
\providecommand \translation [1]{[#1]}%
\providecommand \BibitemOpen [0]{}%
\providecommand \bibitemStop [0]{}%
\providecommand \bibitemNoStop [0]{.\EOS\space}%
\providecommand \EOS [0]{\spacefactor3000\relax}%
\providecommand \BibitemShut  [1]{\csname bibitem#1\endcsname}%
\let\auto@bib@innerbib\@empty
\bibitem [{\citenamefont {Bistritzer}\ and\ \citenamefont
  {MacDonald}(2011)}]{doi:10.1073/pnas.1108174108}%
  \BibitemOpen
  \bibfield  {author} {\bibinfo {author} {\bibfnamefont {R.}~\bibnamefont
  {Bistritzer}}\ and\ \bibinfo {author} {\bibfnamefont {A.~H.}\ \bibnamefont
  {MacDonald}},\ }\bibfield  {title} {\bibinfo {title} {Moiré bands in twisted
  double-layer graphene},\ }\href {https://doi.org/10.1073/pnas.1108174108}
  {\bibfield  {journal} {\bibinfo  {journal} {Proc. Natl. Acad. Sci. U.S.A.}\
  }\textbf {\bibinfo {volume} {108}},\ \bibinfo {pages} {12233} (\bibinfo
  {year} {2011})}\BibitemShut {NoStop}%
\bibitem [{\citenamefont {Wu}\ \emph {et~al.}(2019)\citenamefont {Wu},
  \citenamefont {Lovorn}, \citenamefont {Tutuc}, \citenamefont {Martin},\ and\
  \citenamefont {MacDonald}}]{Topolo_Wu_prl2019}%
  \BibitemOpen
  \bibfield  {author} {\bibinfo {author} {\bibfnamefont {F.}~\bibnamefont
  {Wu}}, \bibinfo {author} {\bibfnamefont {T.}~\bibnamefont {Lovorn}}, \bibinfo
  {author} {\bibfnamefont {E.}~\bibnamefont {Tutuc}}, \bibinfo {author}
  {\bibfnamefont {I.}~\bibnamefont {Martin}},\ and\ \bibinfo {author}
  {\bibfnamefont {A.~H.}\ \bibnamefont {MacDonald}},\ }\bibfield  {title}
  {\bibinfo {title} {Topological insulators in twisted transition metal
  dichalcogenide homobilayers},\ }\href
  {https://doi.org/10.1103/PhysRevLett.122.086402} {\bibfield  {journal}
  {\bibinfo  {journal} {Phys. Rev. Lett.}\ }\textbf {\bibinfo {volume} {122}},\
  \bibinfo {pages} {086402} (\bibinfo {year} {2019})}\BibitemShut {NoStop}%
\bibitem [{\citenamefont {Li}\ and\ \citenamefont
  {Wu}(2017)}]{doi:10.1021/acsnano.7b02756}%
  \BibitemOpen
  \bibfield  {author} {\bibinfo {author} {\bibfnamefont {L.}~\bibnamefont
  {Li}}\ and\ \bibinfo {author} {\bibfnamefont {M.}~\bibnamefont {Wu}},\
  }\bibfield  {title} {\bibinfo {title} {Binary compound bilayer and multilayer
  with vertical polarizations: Two-dimensional ferroelectrics, multiferroics,
  and nanogenerators},\ }\href {https://doi.org/10.1021/acsnano.7b02756}
  {\bibfield  {journal} {\bibinfo  {journal} {ACS Nano}\ }\textbf {\bibinfo
  {volume} {11}},\ \bibinfo {pages} {6382} (\bibinfo {year}
  {2017})}\BibitemShut {NoStop}%
\bibitem [{\citenamefont {Yasuda}\ \emph {et~al.}(2021)\citenamefont {Yasuda},
  \citenamefont {Wang}, \citenamefont {Watanabe}, \citenamefont {Taniguchi},\
  and\ \citenamefont {Jarillo-Herrero}}]{doi:10.1126/science.abd3230}%
  \BibitemOpen
  \bibfield  {author} {\bibinfo {author} {\bibfnamefont {K.}~\bibnamefont
  {Yasuda}}, \bibinfo {author} {\bibfnamefont {X.}~\bibnamefont {Wang}},
  \bibinfo {author} {\bibfnamefont {K.}~\bibnamefont {Watanabe}}, \bibinfo
  {author} {\bibfnamefont {T.}~\bibnamefont {Taniguchi}},\ and\ \bibinfo
  {author} {\bibfnamefont {P.}~\bibnamefont {Jarillo-Herrero}},\ }\bibfield
  {title} {\bibinfo {title} {Stacking-engineered ferroelectricity in bilayer
  boron nitride},\ }\href {https://doi.org/10.1126/science.abd3230} {\bibfield
  {journal} {\bibinfo  {journal} {Science}\ }\textbf {\bibinfo {volume}
  {372}},\ \bibinfo {pages} {1458} (\bibinfo {year} {2021})}\BibitemShut
  {NoStop}%
\bibitem [{\citenamefont {Stern}\ \emph {et~al.}(2021)\citenamefont {Stern},
  \citenamefont {Waschitz}, \citenamefont {Cao}, \citenamefont {Nevo},
  \citenamefont {Watanabe}, \citenamefont {Taniguchi}, \citenamefont {Sela},
  \citenamefont {Urbakh}, \citenamefont {Hod},\ and\ \citenamefont
  {Shalom}}]{doi:10.1126/science.abe8177}%
  \BibitemOpen
  \bibfield  {author} {\bibinfo {author} {\bibfnamefont {M.~V.}\ \bibnamefont
  {Stern}}, \bibinfo {author} {\bibfnamefont {Y.}~\bibnamefont {Waschitz}},
  \bibinfo {author} {\bibfnamefont {W.}~\bibnamefont {Cao}}, \bibinfo {author}
  {\bibfnamefont {I.}~\bibnamefont {Nevo}}, \bibinfo {author} {\bibfnamefont
  {K.}~\bibnamefont {Watanabe}}, \bibinfo {author} {\bibfnamefont
  {T.}~\bibnamefont {Taniguchi}}, \bibinfo {author} {\bibfnamefont
  {E.}~\bibnamefont {Sela}}, \bibinfo {author} {\bibfnamefont {M.}~\bibnamefont
  {Urbakh}}, \bibinfo {author} {\bibfnamefont {O.}~\bibnamefont {Hod}},\ and\
  \bibinfo {author} {\bibfnamefont {M.~B.}\ \bibnamefont {Shalom}},\ }\bibfield
   {title} {\bibinfo {title} {Interfacial ferroelectricity by van der waals
  sliding},\ }\href {https://doi.org/10.1126/science.abe8177} {\bibfield
  {journal} {\bibinfo  {journal} {Science}\ }\textbf {\bibinfo {volume}
  {372}},\ \bibinfo {pages} {1462} (\bibinfo {year} {2021})}\BibitemShut
  {NoStop}%
\bibitem [{\citenamefont {Wu}\ and\ \citenamefont {Li}(2021)}]{wu2021sliding}%
  \BibitemOpen
  \bibfield  {author} {\bibinfo {author} {\bibfnamefont {M.}~\bibnamefont
  {Wu}}\ and\ \bibinfo {author} {\bibfnamefont {J.}~\bibnamefont {Li}},\
  }\bibfield  {title} {\bibinfo {title} {Sliding ferroelectricity in 2d van der
  waals materials: Related physics and future opportunities},\ }\href
  {https://doi.org/10.1073/pnas.2115703118} {\bibfield  {journal} {\bibinfo
  {journal} {Proc. Natl. Acad. Sci.}\ }\textbf {\bibinfo {volume} {118}},\
  \bibinfo {pages} {e2115703118} (\bibinfo {year} {2021})}\BibitemShut
  {NoStop}%
\bibitem [{\citenamefont {Fei}\ \emph {et~al.}(2018)\citenamefont {Fei},
  \citenamefont {Zhao}, \citenamefont {Palomaki}, \citenamefont {Sun},
  \citenamefont {Miller}, \citenamefont {Zhao}, \citenamefont {Yan},
  \citenamefont {Xu},\ and\ \citenamefont {Cobden}}]{fei2018ferroelectric}%
  \BibitemOpen
  \bibfield  {author} {\bibinfo {author} {\bibfnamefont {Z.}~\bibnamefont
  {Fei}}, \bibinfo {author} {\bibfnamefont {W.}~\bibnamefont {Zhao}}, \bibinfo
  {author} {\bibfnamefont {T.~A.}\ \bibnamefont {Palomaki}}, \bibinfo {author}
  {\bibfnamefont {B.}~\bibnamefont {Sun}}, \bibinfo {author} {\bibfnamefont
  {M.~K.}\ \bibnamefont {Miller}}, \bibinfo {author} {\bibfnamefont
  {Z.}~\bibnamefont {Zhao}}, \bibinfo {author} {\bibfnamefont {J.}~\bibnamefont
  {Yan}}, \bibinfo {author} {\bibfnamefont {X.}~\bibnamefont {Xu}},\ and\
  \bibinfo {author} {\bibfnamefont {D.~H.}\ \bibnamefont {Cobden}},\ }\bibfield
   {title} {\bibinfo {title} {Ferroelectric switching of a two-dimensional
  metal},\ }\href {https://doi.org/https://doi.org/10.1038/s41586-018-0336-3}
  {\bibfield  {journal} {\bibinfo  {journal} {Nature}\ }\textbf {\bibinfo
  {volume} {560}},\ \bibinfo {pages} {336} (\bibinfo {year}
  {2018})}\BibitemShut {NoStop}%
\bibitem [{\citenamefont {Yang}\ \emph {et~al.}(2023)\citenamefont {Yang},
  \citenamefont {Ding}, \citenamefont {Gao},\ and\ \citenamefont
  {Wu}}]{PhysRevLett.131.096801}%
  \BibitemOpen
  \bibfield  {author} {\bibinfo {author} {\bibfnamefont {L.}~\bibnamefont
  {Yang}}, \bibinfo {author} {\bibfnamefont {S.}~\bibnamefont {Ding}}, \bibinfo
  {author} {\bibfnamefont {J.}~\bibnamefont {Gao}},\ and\ \bibinfo {author}
  {\bibfnamefont {M.}~\bibnamefont {Wu}},\ }\bibfield  {title} {\bibinfo
  {title} {Atypical sliding and moir\'e ferroelectricity in pure multilayer
  graphene},\ }\href {https://doi.org/10.1103/PhysRevLett.131.096801}
  {\bibfield  {journal} {\bibinfo  {journal} {Phys. Rev. Lett.}\ }\textbf
  {\bibinfo {volume} {131}},\ \bibinfo {pages} {096801} (\bibinfo {year}
  {2023})}\BibitemShut {NoStop}%
\bibitem [{\citenamefont {Wu}\ \emph {et~al.}(2018)\citenamefont {Wu},
  \citenamefont {Lovorn},\ and\ \citenamefont
  {MacDonald}}]{PhysRevB.97.035306}%
  \BibitemOpen
  \bibfield  {author} {\bibinfo {author} {\bibfnamefont {F.}~\bibnamefont
  {Wu}}, \bibinfo {author} {\bibfnamefont {T.}~\bibnamefont {Lovorn}},\ and\
  \bibinfo {author} {\bibfnamefont {A.~H.}\ \bibnamefont {MacDonald}},\
  }\bibfield  {title} {\bibinfo {title} {Theory of optical absorption by
  interlayer excitons in transition metal dichalcogenide heterobilayers},\
  }\href {https://doi.org/10.1103/PhysRevB.97.035306} {\bibfield  {journal}
  {\bibinfo  {journal} {Phys. Rev. B}\ }\textbf {\bibinfo {volume} {97}},\
  \bibinfo {pages} {035306} (\bibinfo {year} {2018})}\BibitemShut {NoStop}%
\bibitem [{\citenamefont {Koshino}\ and\ \citenamefont
  {Moon}(2015)}]{doi:10.7566/JPSJ.84.121001}%
  \BibitemOpen
  \bibfield  {author} {\bibinfo {author} {\bibfnamefont {M.}~\bibnamefont
  {Koshino}}\ and\ \bibinfo {author} {\bibfnamefont {P.}~\bibnamefont {Moon}},\
  }\bibfield  {title} {\bibinfo {title} {Electronic properties of
  incommensurate atomic layers},\ }\href
  {https://doi.org/10.7566/JPSJ.84.121001} {\bibfield  {journal} {\bibinfo
  {journal} {J. Phys. Soc. Jpn.}\ }\textbf {\bibinfo {volume} {84}},\ \bibinfo
  {pages} {121001} (\bibinfo {year} {2015})}\BibitemShut {NoStop}%
\bibitem [{\citenamefont {Koshino}(2015)}]{Koshino_2015}%
  \BibitemOpen
  \bibfield  {author} {\bibinfo {author} {\bibfnamefont {M.}~\bibnamefont
  {Koshino}},\ }\bibfield  {title} {\bibinfo {title} {Interlayer interaction in
  general incommensurate atomic layers},\ }\href
  {https://doi.org/10.1088/1367-2630/17/1/015014} {\bibfield  {journal}
  {\bibinfo  {journal} {New J. Phys.}\ }\textbf {\bibinfo {volume} {17}},\
  \bibinfo {pages} {015014} (\bibinfo {year} {2015})}\BibitemShut {NoStop}%
\bibitem [{\citenamefont {Carr}\ \emph {et~al.}(2020)\citenamefont {Carr},
  \citenamefont {Fang},\ and\ \citenamefont {Kaxiras}}]{carr2020electronic}%
  \BibitemOpen
  \bibfield  {author} {\bibinfo {author} {\bibfnamefont {S.}~\bibnamefont
  {Carr}}, \bibinfo {author} {\bibfnamefont {S.}~\bibnamefont {Fang}},\ and\
  \bibinfo {author} {\bibfnamefont {E.}~\bibnamefont {Kaxiras}},\ }\bibfield
  {title} {\bibinfo {title} {Electronic-structure methods for twisted moir{\'e}
  layers},\ }\href {https://doi.org/10.1038/s41578-020-0214-0} {\bibfield
  {journal} {\bibinfo  {journal} {Nat. Rev. Mater.}\ }\textbf {\bibinfo
  {volume} {5}},\ \bibinfo {pages} {748} (\bibinfo {year} {2020})}\BibitemShut
  {NoStop}%
\bibitem [{\citenamefont {Lei}\ \emph {et~al.}(2021)\citenamefont {Lei},
  \citenamefont {Linhart}, \citenamefont {Qin}, \citenamefont {Libisch},\ and\
  \citenamefont {MacDonald}}]{PhysRevB.104.035139}%
  \BibitemOpen
  \bibfield  {author} {\bibinfo {author} {\bibfnamefont {C.}~\bibnamefont
  {Lei}}, \bibinfo {author} {\bibfnamefont {L.}~\bibnamefont {Linhart}},
  \bibinfo {author} {\bibfnamefont {W.}~\bibnamefont {Qin}}, \bibinfo {author}
  {\bibfnamefont {F.}~\bibnamefont {Libisch}},\ and\ \bibinfo {author}
  {\bibfnamefont {A.~H.}\ \bibnamefont {MacDonald}},\ }\bibfield  {title}
  {\bibinfo {title} {Mirror symmetry breaking and lateral stacking shifts in
  twisted trilayer graphene},\ }\href
  {https://doi.org/10.1103/PhysRevB.104.035139} {\bibfield  {journal} {\bibinfo
   {journal} {Phys. Rev. B}\ }\textbf {\bibinfo {volume} {104}},\ \bibinfo
  {pages} {035139} (\bibinfo {year} {2021})}\BibitemShut {NoStop}%
\bibitem [{\citenamefont {AlBuhairan}\ and\ \citenamefont
  {Vogl}(2023)}]{PhysRevB.108.155106}%
  \BibitemOpen
  \bibfield  {author} {\bibinfo {author} {\bibfnamefont {H.}~\bibnamefont
  {AlBuhairan}}\ and\ \bibinfo {author} {\bibfnamefont {M.}~\bibnamefont
  {Vogl}},\ }\bibfield  {title} {\bibinfo {title} {Band structure and band
  topology in twisted homotrilayer transition metal dichalcogenides},\ }\href
  {https://doi.org/10.1103/PhysRevB.108.155106} {\bibfield  {journal} {\bibinfo
   {journal} {Phys. Rev. B}\ }\textbf {\bibinfo {volume} {108}},\ \bibinfo
  {pages} {155106} (\bibinfo {year} {2023})}\BibitemShut {NoStop}%
\bibitem [{\citenamefont {Li}\ \emph {et~al.}(2024{\natexlab{a}})\citenamefont
  {Li}, \citenamefont {Zhai}, \citenamefont {Xiao},\ and\ \citenamefont
  {Yao}}]{li2024dynamical}%
  \BibitemOpen
  \bibfield  {author} {\bibinfo {author} {\bibfnamefont {J.}~\bibnamefont
  {Li}}, \bibinfo {author} {\bibfnamefont {D.}~\bibnamefont {Zhai}}, \bibinfo
  {author} {\bibfnamefont {C.}~\bibnamefont {Xiao}},\ and\ \bibinfo {author}
  {\bibfnamefont {W.}~\bibnamefont {Yao}},\ }\bibfield  {title} {\bibinfo
  {title} {Dynamical chiral nernst effect in twisted van der waals few
  layers},\ }\href {https://doi.org/10.1007/s44214-024-00059-z} {\bibfield
  {journal} {\bibinfo  {journal} {Quantum Front.}\ }\textbf {\bibinfo {volume}
  {3}},\ \bibinfo {pages} {11} (\bibinfo {year}
  {2024}{\natexlab{a}})}\BibitemShut {NoStop}%
\bibitem [{\citenamefont {Zheng}\ \emph {et~al.}(2025)\citenamefont {Zheng},
  \citenamefont {Zhai}, \citenamefont {Xiao},\ and\ \citenamefont
  {Yao}}]{doi:10.1021/acs.nanolett.5c01944}%
  \BibitemOpen
  \bibfield  {author} {\bibinfo {author} {\bibfnamefont {H.}~\bibnamefont
  {Zheng}}, \bibinfo {author} {\bibfnamefont {D.}~\bibnamefont {Zhai}},
  \bibinfo {author} {\bibfnamefont {C.}~\bibnamefont {Xiao}},\ and\ \bibinfo
  {author} {\bibfnamefont {W.}~\bibnamefont {Yao}},\ }\bibfield  {title}
  {\bibinfo {title} {Layer coherence origin of planar hall effect: From charge
  to multipole and valley},\ }\href
  {https://doi.org/10.1021/acs.nanolett.5c01944} {\bibfield  {journal}
  {\bibinfo  {journal} {Nano Letters}\ }\textbf {\bibinfo {volume} {25}},\
  \bibinfo {pages} {10096} (\bibinfo {year} {2025})}\BibitemShut {NoStop}%
\bibitem [{\citenamefont {Shin}\ \emph {et~al.}(2024)\citenamefont {Shin},
  \citenamefont {Shin}, \citenamefont {Lee}, \citenamefont {Min},\ and\
  \citenamefont {Jung}}]{PhysRevB.110.115136}%
  \BibitemOpen
  \bibfield  {author} {\bibinfo {author} {\bibfnamefont {K.}~\bibnamefont
  {Shin}}, \bibinfo {author} {\bibfnamefont {J.}~\bibnamefont {Shin}}, \bibinfo
  {author} {\bibfnamefont {Y.}~\bibnamefont {Lee}}, \bibinfo {author}
  {\bibfnamefont {H.}~\bibnamefont {Min}},\ and\ \bibinfo {author}
  {\bibfnamefont {J.}~\bibnamefont {Jung}},\ }\bibfield  {title} {\bibinfo
  {title} {Sliding-dependent electronic structures of alternating-twist
  tetralayer graphene},\ }\href {https://doi.org/10.1103/PhysRevB.110.115136}
  {\bibfield  {journal} {\bibinfo  {journal} {Phys. Rev. B}\ }\textbf {\bibinfo
  {volume} {110}},\ \bibinfo {pages} {115136} (\bibinfo {year}
  {2024})}\BibitemShut {NoStop}%
\bibitem [{\citenamefont {Shin}\ \emph {et~al.}(2021)\citenamefont {Shin},
  \citenamefont {Chittari},\ and\ \citenamefont {Jung}}]{PhysRevB.104.045413}%
  \BibitemOpen
  \bibfield  {author} {\bibinfo {author} {\bibfnamefont {J.}~\bibnamefont
  {Shin}}, \bibinfo {author} {\bibfnamefont {B.~L.}\ \bibnamefont {Chittari}},\
  and\ \bibinfo {author} {\bibfnamefont {J.}~\bibnamefont {Jung}},\ }\bibfield
  {title} {\bibinfo {title} {Stacking and gate-tunable topological flat bands,
  gaps, and anisotropic strip patterns in twisted trilayer graphene},\ }\href
  {https://doi.org/10.1103/PhysRevB.104.045413} {\bibfield  {journal} {\bibinfo
   {journal} {Phys. Rev. B}\ }\textbf {\bibinfo {volume} {104}},\ \bibinfo
  {pages} {045413} (\bibinfo {year} {2021})}\BibitemShut {NoStop}%
\bibitem [{\citenamefont {Provost}\ and\ \citenamefont
  {Vallee}(1980)}]{provost1980riemannian}%
  \BibitemOpen
  \bibfield  {author} {\bibinfo {author} {\bibfnamefont {J.}~\bibnamefont
  {Provost}}\ and\ \bibinfo {author} {\bibfnamefont {G.}~\bibnamefont
  {Vallee}},\ }\bibfield  {title} {\bibinfo {title} {Riemannian structure on
  manifolds of quantum states},\ }\href {https://doi.org/10.1007/BF02193559}
  {\bibfield  {journal} {\bibinfo  {journal} {Commun. Math. Phys.}\ }\textbf
  {\bibinfo {volume} {76}},\ \bibinfo {pages} {289} (\bibinfo {year}
  {1980})}\BibitemShut {NoStop}%
\bibitem [{\citenamefont {Resta}(2011)}]{resta2011insulating}%
  \BibitemOpen
  \bibfield  {author} {\bibinfo {author} {\bibfnamefont {R.}~\bibnamefont
  {Resta}},\ }\bibfield  {title} {\bibinfo {title} {The insulating state of
  matter: a geometrical theory},\ }\href
  {https://doi.org/10.1140/epjb/e2010-10874-4} {\bibfield  {journal} {\bibinfo
  {journal} {Eur. Phys. J. B}\ }\textbf {\bibinfo {volume} {79}},\ \bibinfo
  {pages} {121} (\bibinfo {year} {2011})}\BibitemShut {NoStop}%
\bibitem [{\citenamefont {T\"orm\"a}(2023)}]{PhysRevLett.131.240001}%
  \BibitemOpen
  \bibfield  {author} {\bibinfo {author} {\bibfnamefont {P.}~\bibnamefont
  {T\"orm\"a}},\ }\bibfield  {title} {\bibinfo {title} {Essay: Where can
  quantum geometry lead us?},\ }\href
  {https://doi.org/10.1103/PhysRevLett.131.240001} {\bibfield  {journal}
  {\bibinfo  {journal} {Phys. Rev. Lett.}\ }\textbf {\bibinfo {volume} {131}},\
  \bibinfo {pages} {240001} (\bibinfo {year} {2023})}\BibitemShut {NoStop}%
\bibitem [{\citenamefont {Liu}\ \emph {et~al.}(2024)\citenamefont {Liu},
  \citenamefont {Qiang}, \citenamefont {Lu},\ and\ \citenamefont
  {Xie}}]{10.1093/nsr/nwae334}%
  \BibitemOpen
  \bibfield  {author} {\bibinfo {author} {\bibfnamefont {T.}~\bibnamefont
  {Liu}}, \bibinfo {author} {\bibfnamefont {X.-B.}\ \bibnamefont {Qiang}},
  \bibinfo {author} {\bibfnamefont {H.-Z.}\ \bibnamefont {Lu}},\ and\ \bibinfo
  {author} {\bibfnamefont {X.~C.}\ \bibnamefont {Xie}},\ }\bibfield  {title}
  {\bibinfo {title} {Quantum geometry in condensed matter},\ }\href
  {https://doi.org/10.1093/nsr/nwae334} {\bibfield  {journal} {\bibinfo
  {journal} {Natl. Sci. Rev.}\ }\textbf {\bibinfo {volume} {12}},\ \bibinfo
  {pages} {nwae334} (\bibinfo {year} {2024})}\BibitemShut {NoStop}%
\bibitem [{\citenamefont {T{\"o}rm{\"a}}\ \emph {et~al.}(2022)\citenamefont
  {T{\"o}rm{\"a}}, \citenamefont {Peotta},\ and\ \citenamefont
  {Bernevig}}]{torma2022superconductivity}%
  \BibitemOpen
  \bibfield  {author} {\bibinfo {author} {\bibfnamefont {P.}~\bibnamefont
  {T{\"o}rm{\"a}}}, \bibinfo {author} {\bibfnamefont {S.}~\bibnamefont
  {Peotta}},\ and\ \bibinfo {author} {\bibfnamefont {B.~A.}\ \bibnamefont
  {Bernevig}},\ }\bibfield  {title} {\bibinfo {title} {Superconductivity,
  superfluidity and quantum geometry in twisted multilayer systems},\ }\href
  {https://doi.org/10.1038/s42254-022-00466-y} {\bibfield  {journal} {\bibinfo
  {journal} {Nat Rev Phys}\ }\textbf {\bibinfo {volume} {4}},\ \bibinfo {pages}
  {528} (\bibinfo {year} {2022})}\BibitemShut {NoStop}%
\bibitem [{\citenamefont {Berry}(1984)}]{berry1984quantal}%
  \BibitemOpen
  \bibfield  {author} {\bibinfo {author} {\bibfnamefont {M.~V.}\ \bibnamefont
  {Berry}},\ }\bibfield  {title} {\bibinfo {title} {Quantal phase factors
  accompanying adiabatic changes},\ }\href
  {https://doi.org/10.1098/rspa.1984.0023} {\bibfield  {journal} {\bibinfo
  {journal} {Proc. R. Soc. Lond. A}\ }\textbf {\bibinfo {volume} {392}},\
  \bibinfo {pages} {45} (\bibinfo {year} {1984})}\BibitemShut {NoStop}%
\bibitem [{\citenamefont {Xiao}\ \emph {et~al.}(2010)\citenamefont {Xiao},
  \citenamefont {Chang},\ and\ \citenamefont {Niu}}]{RevModPhys.82.1959}%
  \BibitemOpen
  \bibfield  {author} {\bibinfo {author} {\bibfnamefont {D.}~\bibnamefont
  {Xiao}}, \bibinfo {author} {\bibfnamefont {M.-C.}\ \bibnamefont {Chang}},\
  and\ \bibinfo {author} {\bibfnamefont {Q.}~\bibnamefont {Niu}},\ }\bibfield
  {title} {\bibinfo {title} {Berry phase effects on electronic properties},\
  }\href {https://doi.org/10.1103/RevModPhys.82.1959} {\bibfield  {journal}
  {\bibinfo  {journal} {Rev. Mod. Phys.}\ }\textbf {\bibinfo {volume} {82}},\
  \bibinfo {pages} {1959} (\bibinfo {year} {2010})}\BibitemShut {NoStop}%
\bibitem [{\citenamefont {Devakul}\ \emph {et~al.}(2021)\citenamefont
  {Devakul}, \citenamefont {Cr{\'e}pel}, \citenamefont {Zhang},\ and\
  \citenamefont {Fu}}]{devakul2021magic}%
  \BibitemOpen
  \bibfield  {author} {\bibinfo {author} {\bibfnamefont {T.}~\bibnamefont
  {Devakul}}, \bibinfo {author} {\bibfnamefont {V.}~\bibnamefont {Cr{\'e}pel}},
  \bibinfo {author} {\bibfnamefont {Y.}~\bibnamefont {Zhang}},\ and\ \bibinfo
  {author} {\bibfnamefont {L.}~\bibnamefont {Fu}},\ }\bibfield  {title}
  {\bibinfo {title} {Magic in twisted transition metal dichalcogenide
  bilayers},\ }\href
  {https://www.nature.com/articles/s41467-021-27042-9#citeas} {\bibfield
  {journal} {\bibinfo  {journal} {Nat Commun}\ }\textbf {\bibinfo {volume}
  {12}},\ \bibinfo {pages} {6730} (\bibinfo {year} {2021})}\BibitemShut
  {NoStop}%
\bibitem [{\citenamefont {Xiao}\ \emph {et~al.}(2020)\citenamefont {Xiao},
  \citenamefont {Wang}, \citenamefont {Wang}, \citenamefont {Pemmaraju},
  \citenamefont {Wang}, \citenamefont {Muscher}, \citenamefont {Sie},
  \citenamefont {Nyby}, \citenamefont {Devereaux}, \citenamefont {Qian} \emph
  {et~al.}}]{xiao2020berry}%
  \BibitemOpen
  \bibfield  {author} {\bibinfo {author} {\bibfnamefont {J.}~\bibnamefont
  {Xiao}}, \bibinfo {author} {\bibfnamefont {Y.}~\bibnamefont {Wang}}, \bibinfo
  {author} {\bibfnamefont {H.}~\bibnamefont {Wang}}, \bibinfo {author}
  {\bibfnamefont {C.}~\bibnamefont {Pemmaraju}}, \bibinfo {author}
  {\bibfnamefont {S.}~\bibnamefont {Wang}}, \bibinfo {author} {\bibfnamefont
  {P.}~\bibnamefont {Muscher}}, \bibinfo {author} {\bibfnamefont {E.~J.}\
  \bibnamefont {Sie}}, \bibinfo {author} {\bibfnamefont {C.~M.}\ \bibnamefont
  {Nyby}}, \bibinfo {author} {\bibfnamefont {T.~P.}\ \bibnamefont {Devereaux}},
  \bibinfo {author} {\bibfnamefont {X.}~\bibnamefont {Qian}}, \emph {et~al.},\
  }\bibfield  {title} {\bibinfo {title} {Berry curvature memory through
  electrically driven stacking transitions},\ }\href
  {https://doi.org/10.1038/s41567-020-0947-0} {\bibfield  {journal} {\bibinfo
  {journal} {Nat. Phys.}\ }\textbf {\bibinfo {volume} {16}},\ \bibinfo {pages}
  {1028} (\bibinfo {year} {2020})}\BibitemShut {NoStop}%
\bibitem [{\citenamefont {Liang}\ \emph {et~al.}(2025)\citenamefont {Liang},
  \citenamefont {Ding}, \citenamefont {Wu}, \citenamefont {Zhao},\ and\
  \citenamefont {Gao}}]{PhysRevB.111.085412}%
  \BibitemOpen
  \bibfield  {author} {\bibinfo {author} {\bibfnamefont {M.}~\bibnamefont
  {Liang}}, \bibinfo {author} {\bibfnamefont {S.-P.}\ \bibnamefont {Ding}},
  \bibinfo {author} {\bibfnamefont {M.}~\bibnamefont {Wu}}, \bibinfo {author}
  {\bibfnamefont {C.}~\bibnamefont {Zhao}},\ and\ \bibinfo {author}
  {\bibfnamefont {J.-H.}\ \bibnamefont {Gao}},\ }\bibfield  {title} {\bibinfo
  {title} {Moir\'e flat bands in alternating twisted ${\mathrm{mote}}_{2}$
  multilayers},\ }\href {https://doi.org/10.1103/PhysRevB.111.085412}
  {\bibfield  {journal} {\bibinfo  {journal} {Phys. Rev. B}\ }\textbf {\bibinfo
  {volume} {111}},\ \bibinfo {pages} {085412} (\bibinfo {year}
  {2025})}\BibitemShut {NoStop}%
\bibitem [{\citenamefont {Sharma}\ \emph {et~al.}(2024)\citenamefont {Sharma},
  \citenamefont {Peng},\ and\ \citenamefont {Sheng}}]{PhysRevB.110.125142}%
  \BibitemOpen
  \bibfield  {author} {\bibinfo {author} {\bibfnamefont {P.}~\bibnamefont
  {Sharma}}, \bibinfo {author} {\bibfnamefont {Y.}~\bibnamefont {Peng}},\ and\
  \bibinfo {author} {\bibfnamefont {D.~N.}\ \bibnamefont {Sheng}},\ }\bibfield
  {title} {\bibinfo {title} {Topological quantum phase transitions driven by a
  displacement field in twisted $\mathrm{MoTe}_{2}$ bilayers},\ }\href
  {https://doi.org/10.1103/PhysRevB.110.125142} {\bibfield  {journal} {\bibinfo
   {journal} {Phys. Rev. B}\ }\textbf {\bibinfo {volume} {110}},\ \bibinfo
  {pages} {125142} (\bibinfo {year} {2024})}\BibitemShut {NoStop}%
\bibitem [{\citenamefont {Ma}\ \emph {et~al.}(2021)\citenamefont {Ma},
  \citenamefont {Li}, \citenamefont {Zheng}, \citenamefont {Xiao},
  \citenamefont {Jiang}, \citenamefont {Gao},\ and\ \citenamefont
  {Xie}}]{MA202118}%
  \BibitemOpen
  \bibfield  {author} {\bibinfo {author} {\bibfnamefont {Z.}~\bibnamefont
  {Ma}}, \bibinfo {author} {\bibfnamefont {S.}~\bibnamefont {Li}}, \bibinfo
  {author} {\bibfnamefont {Y.-W.}\ \bibnamefont {Zheng}}, \bibinfo {author}
  {\bibfnamefont {M.-M.}\ \bibnamefont {Xiao}}, \bibinfo {author}
  {\bibfnamefont {H.}~\bibnamefont {Jiang}}, \bibinfo {author} {\bibfnamefont
  {J.-H.}\ \bibnamefont {Gao}},\ and\ \bibinfo {author} {\bibfnamefont
  {X.}~\bibnamefont {Xie}},\ }\bibfield  {title} {\bibinfo {title} {Topological
  flat bands in twisted trilayer graphene},\ }\href
  {https://doi.org/https://doi.org/10.1016/j.scib.2020.10.004} {\bibfield
  {journal} {\bibinfo  {journal} {Sci. Bull.}\ }\textbf {\bibinfo {volume}
  {66}},\ \bibinfo {pages} {18} (\bibinfo {year} {2021})}\BibitemShut {NoStop}%
\bibitem [{\citenamefont {Fedorko}\ \emph {et~al.}()\citenamefont {Fedorko},
  \citenamefont {Liu},\ and\ \citenamefont
  {Bi}}]{fedorko2025engineeringmoirekagomesuperlattices}%
  \BibitemOpen
  \bibfield  {author} {\bibinfo {author} {\bibfnamefont {A.}~\bibnamefont
  {Fedorko}}, \bibinfo {author} {\bibfnamefont {C.-X.}\ \bibnamefont {Liu}},\
  and\ \bibinfo {author} {\bibfnamefont {Z.}~\bibnamefont {Bi}},\ }\bibfield
  {title} {\bibinfo {title} {Engineering moir\'e kagome superlattices in
  twisted transition metal dichalcogenides},\ }\href
  {https://arxiv.org/abs/2503.13422} {\ }\Eprint
  {https://arxiv.org/abs/2503.13422} {arXiv:2503.13422} \BibitemShut {NoStop}%
\bibitem [{\citenamefont {Liang}\ \emph {et~al.}(2022)\citenamefont {Liang},
  \citenamefont {Xiao}, \citenamefont {Ma},\ and\ \citenamefont
  {Gao}}]{PhysRevB.105.195422}%
  \BibitemOpen
  \bibfield  {author} {\bibinfo {author} {\bibfnamefont {M.}~\bibnamefont
  {Liang}}, \bibinfo {author} {\bibfnamefont {M.-M.}\ \bibnamefont {Xiao}},
  \bibinfo {author} {\bibfnamefont {Z.}~\bibnamefont {Ma}},\ and\ \bibinfo
  {author} {\bibfnamefont {J.-H.}\ \bibnamefont {Gao}},\ }\bibfield  {title}
  {\bibinfo {title} {Moir\'e band structures of the double twisted few-layer
  graphene},\ }\href {https://doi.org/10.1103/PhysRevB.105.195422} {\bibfield
  {journal} {\bibinfo  {journal} {Phys. Rev. B}\ }\textbf {\bibinfo {volume}
  {105}},\ \bibinfo {pages} {195422} (\bibinfo {year} {2022})}\BibitemShut
  {NoStop}%
\bibitem [{\citenamefont {Wang}\ \emph
  {et~al.}(2024{\natexlab{a}})\citenamefont {Wang}, \citenamefont {Zhou},
  \citenamefont {Lin}, \citenamefont {Feng}, \citenamefont {Wang},
  \citenamefont {Liang}, \citenamefont {Zhang}, \citenamefont {Wu},
  \citenamefont {Liu}, \citenamefont {Watanabe}, \citenamefont {Taniguchi},
  \citenamefont {Yang}, \citenamefont {Zhang}, \citenamefont {Liu},
  \citenamefont {Gao}, \citenamefont {Liu}, \citenamefont {Xie}, \citenamefont
  {Song},\ and\ \citenamefont {Lu}}]{PhysRevLett.132.246501}%
  \BibitemOpen
  \bibfield  {author} {\bibinfo {author} {\bibfnamefont {W.}~\bibnamefont
  {Wang}}, \bibinfo {author} {\bibfnamefont {G.}~\bibnamefont {Zhou}}, \bibinfo
  {author} {\bibfnamefont {W.}~\bibnamefont {Lin}}, \bibinfo {author}
  {\bibfnamefont {Z.}~\bibnamefont {Feng}}, \bibinfo {author} {\bibfnamefont
  {Y.}~\bibnamefont {Wang}}, \bibinfo {author} {\bibfnamefont {M.}~\bibnamefont
  {Liang}}, \bibinfo {author} {\bibfnamefont {Z.}~\bibnamefont {Zhang}},
  \bibinfo {author} {\bibfnamefont {M.}~\bibnamefont {Wu}}, \bibinfo {author}
  {\bibfnamefont {L.}~\bibnamefont {Liu}}, \bibinfo {author} {\bibfnamefont
  {K.}~\bibnamefont {Watanabe}}, \bibinfo {author} {\bibfnamefont
  {T.}~\bibnamefont {Taniguchi}}, \bibinfo {author} {\bibfnamefont
  {W.}~\bibnamefont {Yang}}, \bibinfo {author} {\bibfnamefont {G.}~\bibnamefont
  {Zhang}}, \bibinfo {author} {\bibfnamefont {K.}~\bibnamefont {Liu}}, \bibinfo
  {author} {\bibfnamefont {J.}~\bibnamefont {Gao}}, \bibinfo {author}
  {\bibfnamefont {Y.}~\bibnamefont {Liu}}, \bibinfo {author} {\bibfnamefont
  {X.~C.}\ \bibnamefont {Xie}}, \bibinfo {author} {\bibfnamefont
  {Z.}~\bibnamefont {Song}},\ and\ \bibinfo {author} {\bibfnamefont
  {X.}~\bibnamefont {Lu}},\ }\bibfield  {title} {\bibinfo {title} {Correlated
  charge density wave insulators in chirally twisted triple bilayer graphene},\
  }\href {https://doi.org/10.1103/PhysRevLett.132.246501} {\bibfield  {journal}
  {\bibinfo  {journal} {Phys. Rev. Lett.}\ }\textbf {\bibinfo {volume} {132}},\
  \bibinfo {pages} {246501} (\bibinfo {year} {2024}{\natexlab{a}})}\BibitemShut
  {NoStop}%
\bibitem [{\citenamefont {Zhou}\ \emph {et~al.}(2024)\citenamefont {Zhou},
  \citenamefont {Wang}, \citenamefont {Wang}, \citenamefont {Lu},\ and\
  \citenamefont {Song}}]{ZHOU2024100015}%
  \BibitemOpen
  \bibfield  {author} {\bibinfo {author} {\bibfnamefont {G.-D.}\ \bibnamefont
  {Zhou}}, \bibinfo {author} {\bibfnamefont {Y.-J.}\ \bibnamefont {Wang}},
  \bibinfo {author} {\bibfnamefont {W.-X.}\ \bibnamefont {Wang}}, \bibinfo
  {author} {\bibfnamefont {X.-B.}\ \bibnamefont {Lu}},\ and\ \bibinfo {author}
  {\bibfnamefont {Z.-D.}\ \bibnamefont {Song}},\ }\bibfield  {title} {\bibinfo
  {title} {Correlated insulators and charge density wave states in chirally
  twisted triple bilayer graphene},\ }\href
  {https://doi.org/https://doi.org/10.1016/j.mtquan.2024.100015} {\bibfield
  {journal} {\bibinfo  {journal} {Mater. Today Quantum}\ }\textbf {\bibinfo
  {volume} {4}},\ \bibinfo {pages} {100015} (\bibinfo {year}
  {2024})}\BibitemShut {NoStop}%
\bibitem [{\citenamefont {Lin}\ \emph {et~al.}()\citenamefont {Lin},
  \citenamefont {Wang}, \citenamefont {Cao}, \citenamefont {Liang},
  \citenamefont {Zhao}, \citenamefont {Watanabe}, \citenamefont {Taniguchi},
  \citenamefont {Gao}, \citenamefont {Chen}, \citenamefont {Lu},\ and\
  \citenamefont {Liu}}]{lin2025flatbandmanybodygap}%
  \BibitemOpen
  \bibfield  {author} {\bibinfo {author} {\bibfnamefont {W.}~\bibnamefont
  {Lin}}, \bibinfo {author} {\bibfnamefont {W.}~\bibnamefont {Wang}}, \bibinfo
  {author} {\bibfnamefont {S.}~\bibnamefont {Cao}}, \bibinfo {author}
  {\bibfnamefont {M.}~\bibnamefont {Liang}}, \bibinfo {author} {\bibfnamefont
  {L.}~\bibnamefont {Zhao}}, \bibinfo {author} {\bibfnamefont {K.}~\bibnamefont
  {Watanabe}}, \bibinfo {author} {\bibfnamefont {T.}~\bibnamefont {Taniguchi}},
  \bibinfo {author} {\bibfnamefont {J.}~\bibnamefont {Gao}}, \bibinfo {author}
  {\bibfnamefont {J.}~\bibnamefont {Chen}}, \bibinfo {author} {\bibfnamefont
  {X.}~\bibnamefont {Lu}},\ and\ \bibinfo {author} {\bibfnamefont
  {Y.}~\bibnamefont {Liu}},\ }\href@noop {} {\bibinfo {title} {Flat band and
  many-body gap in chirally twisted triple bilayer graphene}},\ \Eprint
  {https://arxiv.org/abs/2501.06852} {arXiv:2501.06852} \BibitemShut {NoStop}%
\bibitem [{\citenamefont {Davydov}\ \emph {et~al.}(2025)\citenamefont
  {Davydov}, \citenamefont {Zhu}, \citenamefont {Friedman}, \citenamefont
  {Gramowski}, \citenamefont {Li}, \citenamefont {Tavakley}, \citenamefont
  {Watanabe}, \citenamefont {Taniguchi}, \citenamefont {Luskin}, \citenamefont
  {Kaxiras},\ and\ \citenamefont {Wang}}]{PhysRevB.111.L161120}%
  \BibitemOpen
  \bibfield  {author} {\bibinfo {author} {\bibfnamefont {K.}~\bibnamefont
  {Davydov}}, \bibinfo {author} {\bibfnamefont {Z.}~\bibnamefont {Zhu}},
  \bibinfo {author} {\bibfnamefont {N.}~\bibnamefont {Friedman}}, \bibinfo
  {author} {\bibfnamefont {E.}~\bibnamefont {Gramowski}}, \bibinfo {author}
  {\bibfnamefont {Y.}~\bibnamefont {Li}}, \bibinfo {author} {\bibfnamefont
  {J.}~\bibnamefont {Tavakley}}, \bibinfo {author} {\bibfnamefont
  {K.}~\bibnamefont {Watanabe}}, \bibinfo {author} {\bibfnamefont
  {T.}~\bibnamefont {Taniguchi}}, \bibinfo {author} {\bibfnamefont
  {M.}~\bibnamefont {Luskin}}, \bibinfo {author} {\bibfnamefont
  {E.}~\bibnamefont {Kaxiras}},\ and\ \bibinfo {author} {\bibfnamefont
  {K.}~\bibnamefont {Wang}},\ }\bibfield  {title} {\bibinfo {title} {Tunable
  atomically enhanced moir\'e berry curvatures in twisted triple bilayer
  graphene},\ }\href {https://doi.org/10.1103/PhysRevB.111.L161120} {\bibfield
  {journal} {\bibinfo  {journal} {Phys. Rev. B}\ }\textbf {\bibinfo {volume}
  {111}},\ \bibinfo {pages} {L161120} (\bibinfo {year} {2025})}\BibitemShut
  {NoStop}%
\bibitem [{\citenamefont {Ma}\ \emph {et~al.}(2023{\natexlab{a}})\citenamefont
  {Ma}, \citenamefont {Li}, \citenamefont {Lu}, \citenamefont {Xu},
  \citenamefont {Gao},\ and\ \citenamefont {Xie}}]{ma2023doubled}%
  \BibitemOpen
  \bibfield  {author} {\bibinfo {author} {\bibfnamefont {Z.}~\bibnamefont
  {Ma}}, \bibinfo {author} {\bibfnamefont {S.}~\bibnamefont {Li}}, \bibinfo
  {author} {\bibfnamefont {M.}~\bibnamefont {Lu}}, \bibinfo {author}
  {\bibfnamefont {D.-H.}\ \bibnamefont {Xu}}, \bibinfo {author} {\bibfnamefont
  {J.-H.}\ \bibnamefont {Gao}},\ and\ \bibinfo {author} {\bibfnamefont
  {X.}~\bibnamefont {Xie}},\ }\bibfield  {title} {\bibinfo {title} {Doubled
  moir{\'e} flat bands in double-twisted few-layer graphite},\ }\href
  {https://doi.org/10.1007/s11433-022-1993-7} {\bibfield  {journal} {\bibinfo
  {journal} {Sci. China Phys. Mech. Astron.}\ }\textbf {\bibinfo {volume}
  {66}},\ \bibinfo {pages} {227211} (\bibinfo {year}
  {2023}{\natexlab{a}})}\BibitemShut {NoStop}%
\bibitem [{\citenamefont {Ding}\ \emph {et~al.}(2023)\citenamefont {Ding},
  \citenamefont {Liang}, \citenamefont {Ma}, \citenamefont {L\"u},\ and\
  \citenamefont {Gao}}]{PhysRevB.108.195119}%
  \BibitemOpen
  \bibfield  {author} {\bibinfo {author} {\bibfnamefont {S.-P.}\ \bibnamefont
  {Ding}}, \bibinfo {author} {\bibfnamefont {M.}~\bibnamefont {Liang}},
  \bibinfo {author} {\bibfnamefont {Z.}~\bibnamefont {Ma}}, \bibinfo {author}
  {\bibfnamefont {J.-T.}\ \bibnamefont {L\"u}},\ and\ \bibinfo {author}
  {\bibfnamefont {J.-H.}\ \bibnamefont {Gao}},\ }\bibfield  {title} {\bibinfo
  {title} {Mirror symmetry decomposition in double-twisted multilayer graphene
  systems},\ }\href {https://doi.org/10.1103/PhysRevB.108.195119} {\bibfield
  {journal} {\bibinfo  {journal} {Phys. Rev. B}\ }\textbf {\bibinfo {volume}
  {108}},\ \bibinfo {pages} {195119} (\bibinfo {year} {2023})}\BibitemShut
  {NoStop}%
\bibitem [{\citenamefont {Xie}\ \emph {et~al.}(2022)\citenamefont {Xie},
  \citenamefont {Peng}, \citenamefont {Zhang},\ and\ \citenamefont
  {Liu}}]{xie2022alternating}%
  \BibitemOpen
  \bibfield  {author} {\bibinfo {author} {\bibfnamefont {B.}~\bibnamefont
  {Xie}}, \bibinfo {author} {\bibfnamefont {R.}~\bibnamefont {Peng}}, \bibinfo
  {author} {\bibfnamefont {S.}~\bibnamefont {Zhang}},\ and\ \bibinfo {author}
  {\bibfnamefont {J.}~\bibnamefont {Liu}},\ }\bibfield  {title} {\bibinfo
  {title} {Alternating twisted multilayer graphene: generic partition rules,
  double flat bands, and orbital magnetoelectric effect},\ }\href
  {https://doi.org/10.1038/s41524-022-00789-5} {\bibfield  {journal} {\bibinfo
  {journal} {npj Comput. Mater.}\ }\textbf {\bibinfo {volume} {8}},\ \bibinfo
  {pages} {110} (\bibinfo {year} {2022})}\BibitemShut {NoStop}%
\bibitem [{\citenamefont {Khalaf}\ \emph {et~al.}(2019)\citenamefont {Khalaf},
  \citenamefont {Kruchkov}, \citenamefont {Tarnopolsky},\ and\ \citenamefont
  {Vishwanath}}]{PhysRevB.100.085109}%
  \BibitemOpen
  \bibfield  {author} {\bibinfo {author} {\bibfnamefont {E.}~\bibnamefont
  {Khalaf}}, \bibinfo {author} {\bibfnamefont {A.~J.}\ \bibnamefont
  {Kruchkov}}, \bibinfo {author} {\bibfnamefont {G.}~\bibnamefont
  {Tarnopolsky}},\ and\ \bibinfo {author} {\bibfnamefont {A.}~\bibnamefont
  {Vishwanath}},\ }\bibfield  {title} {\bibinfo {title} {Magic angle hierarchy
  in twisted graphene multilayers},\ }\href
  {https://doi.org/10.1103/PhysRevB.100.085109} {\bibfield  {journal} {\bibinfo
   {journal} {Phys. Rev. B}\ }\textbf {\bibinfo {volume} {100}},\ \bibinfo
  {pages} {085109} (\bibinfo {year} {2019})}\BibitemShut {NoStop}%
\bibitem [{\citenamefont {Shin}\ \emph {et~al.}(2023)\citenamefont {Shin},
  \citenamefont {Jang}, \citenamefont {Shin}, \citenamefont {Jung},\ and\
  \citenamefont {Min}}]{PhysRevB.107.245139}%
  \BibitemOpen
  \bibfield  {author} {\bibinfo {author} {\bibfnamefont {K.}~\bibnamefont
  {Shin}}, \bibinfo {author} {\bibfnamefont {Y.}~\bibnamefont {Jang}}, \bibinfo
  {author} {\bibfnamefont {J.}~\bibnamefont {Shin}}, \bibinfo {author}
  {\bibfnamefont {J.}~\bibnamefont {Jung}},\ and\ \bibinfo {author}
  {\bibfnamefont {H.}~\bibnamefont {Min}},\ }\bibfield  {title} {\bibinfo
  {title} {Electronic structure of biased alternating-twist multilayer
  graphene},\ }\href {https://doi.org/10.1103/PhysRevB.107.245139} {\bibfield
  {journal} {\bibinfo  {journal} {Phys. Rev. B}\ }\textbf {\bibinfo {volume}
  {107}},\ \bibinfo {pages} {245139} (\bibinfo {year} {2023})}\BibitemShut
  {NoStop}%
\bibitem [{\citenamefont {Sodemann}\ and\ \citenamefont
  {Fu}(2015)}]{PhysRevLett.115.216806}%
  \BibitemOpen
  \bibfield  {author} {\bibinfo {author} {\bibfnamefont {I.}~\bibnamefont
  {Sodemann}}\ and\ \bibinfo {author} {\bibfnamefont {L.}~\bibnamefont {Fu}},\
  }\bibfield  {title} {\bibinfo {title} {Quantum nonlinear hall effect induced
  by berry curvature dipole in time-reversal invariant materials},\ }\href
  {https://doi.org/10.1103/PhysRevLett.115.216806} {\bibfield  {journal}
  {\bibinfo  {journal} {Phys. Rev. Lett.}\ }\textbf {\bibinfo {volume} {115}},\
  \bibinfo {pages} {216806} (\bibinfo {year} {2015})}\BibitemShut {NoStop}%
\bibitem [{\citenamefont {Ma}\ \emph {et~al.}(2019)\citenamefont {Ma},
  \citenamefont {Xu}, \citenamefont {Shen}, \citenamefont {MacNeill},
  \citenamefont {Fatemi}, \citenamefont {Chang}, \citenamefont {Mier~Valdivia},
  \citenamefont {Wu}, \citenamefont {Du}, \citenamefont {Hsu} \emph
  {et~al.}}]{ma2019observation}%
  \BibitemOpen
  \bibfield  {author} {\bibinfo {author} {\bibfnamefont {Q.}~\bibnamefont
  {Ma}}, \bibinfo {author} {\bibfnamefont {S.-Y.}\ \bibnamefont {Xu}}, \bibinfo
  {author} {\bibfnamefont {H.}~\bibnamefont {Shen}}, \bibinfo {author}
  {\bibfnamefont {D.}~\bibnamefont {MacNeill}}, \bibinfo {author}
  {\bibfnamefont {V.}~\bibnamefont {Fatemi}}, \bibinfo {author} {\bibfnamefont
  {T.-R.}\ \bibnamefont {Chang}}, \bibinfo {author} {\bibfnamefont {A.~M.}\
  \bibnamefont {Mier~Valdivia}}, \bibinfo {author} {\bibfnamefont
  {S.}~\bibnamefont {Wu}}, \bibinfo {author} {\bibfnamefont {Z.}~\bibnamefont
  {Du}}, \bibinfo {author} {\bibfnamefont {C.-H.}\ \bibnamefont {Hsu}}, \emph
  {et~al.},\ }\bibfield  {title} {\bibinfo {title} {Observation of the
  nonlinear hall effect under time-reversal-symmetric conditions},\ }\href
  {https://doi.org/10.1038/s41586-018-0807-6} {\bibfield  {journal} {\bibinfo
  {journal} {Nature}\ }\textbf {\bibinfo {volume} {565}},\ \bibinfo {pages}
  {337} (\bibinfo {year} {2019})}\BibitemShut {NoStop}%
\bibitem [{\citenamefont {Kang}\ \emph {et~al.}(2019)\citenamefont {Kang},
  \citenamefont {Li}, \citenamefont {Sohn}, \citenamefont {Shan},\ and\
  \citenamefont {Mak}}]{kang2019nonlinear}%
  \BibitemOpen
  \bibfield  {author} {\bibinfo {author} {\bibfnamefont {K.}~\bibnamefont
  {Kang}}, \bibinfo {author} {\bibfnamefont {T.}~\bibnamefont {Li}}, \bibinfo
  {author} {\bibfnamefont {E.}~\bibnamefont {Sohn}}, \bibinfo {author}
  {\bibfnamefont {J.}~\bibnamefont {Shan}},\ and\ \bibinfo {author}
  {\bibfnamefont {K.~F.}\ \bibnamefont {Mak}},\ }\bibfield  {title} {\bibinfo
  {title} {Nonlinear anomalous hall effect in few-layer $\mathrm{WTe}_2$},\
  }\href {https://doi.org/10.1038/s41563-019-0294-7} {\bibfield  {journal}
  {\bibinfo  {journal} {Nat. Mater.}\ }\textbf {\bibinfo {volume} {18}},\
  \bibinfo {pages} {324} (\bibinfo {year} {2019})}\BibitemShut {NoStop}%
\bibitem [{\citenamefont {Sinha}\ \emph {et~al.}(2022)\citenamefont {Sinha},
  \citenamefont {Adak}, \citenamefont {Chakraborty}, \citenamefont {Das},
  \citenamefont {Debnath}, \citenamefont {Sangani}, \citenamefont {Watanabe},
  \citenamefont {Taniguchi}, \citenamefont {Waghmare}, \citenamefont {Agarwal}
  \emph {et~al.}}]{sinha2022berry}%
  \BibitemOpen
  \bibfield  {author} {\bibinfo {author} {\bibfnamefont {S.}~\bibnamefont
  {Sinha}}, \bibinfo {author} {\bibfnamefont {P.~C.}\ \bibnamefont {Adak}},
  \bibinfo {author} {\bibfnamefont {A.}~\bibnamefont {Chakraborty}}, \bibinfo
  {author} {\bibfnamefont {K.}~\bibnamefont {Das}}, \bibinfo {author}
  {\bibfnamefont {K.}~\bibnamefont {Debnath}}, \bibinfo {author} {\bibfnamefont
  {L.~V.}\ \bibnamefont {Sangani}}, \bibinfo {author} {\bibfnamefont
  {K.}~\bibnamefont {Watanabe}}, \bibinfo {author} {\bibfnamefont
  {T.}~\bibnamefont {Taniguchi}}, \bibinfo {author} {\bibfnamefont {U.~V.}\
  \bibnamefont {Waghmare}}, \bibinfo {author} {\bibfnamefont {A.}~\bibnamefont
  {Agarwal}}, \emph {et~al.},\ }\bibfield  {title} {\bibinfo {title} {Berry
  curvature dipole senses topological transition in a moir{\'e} superlattice},\
  }\href {https://doi.org/https://doi.org/10.1038/s41567-022-01606-y}
  {\bibfield  {journal} {\bibinfo  {journal} {Nat. Phys.}\ }\textbf {\bibinfo
  {volume} {18}},\ \bibinfo {pages} {765} (\bibinfo {year} {2022})}\BibitemShut
  {NoStop}%
\bibitem [{\citenamefont {Hu}\ \emph {et~al.}(2022)\citenamefont {Hu},
  \citenamefont {Zhang}, \citenamefont {Xie},\ and\ \citenamefont
  {Law}}]{hu2022nonlinear}%
  \BibitemOpen
  \bibfield  {author} {\bibinfo {author} {\bibfnamefont {J.-X.}\ \bibnamefont
  {Hu}}, \bibinfo {author} {\bibfnamefont {C.-P.}\ \bibnamefont {Zhang}},
  \bibinfo {author} {\bibfnamefont {Y.-M.}\ \bibnamefont {Xie}},\ and\ \bibinfo
  {author} {\bibfnamefont {K.}~\bibnamefont {Law}},\ }\bibfield  {title}
  {\bibinfo {title} {Nonlinear hall effects in strained twisted bilayer
  $\mathrm{WSe}_2$},\ }\href {https://doi.org/10.1038/s42005-022-01034-7}
  {\bibfield  {journal} {\bibinfo  {journal} {Commun Phys}\ }\textbf {\bibinfo
  {volume} {5}},\ \bibinfo {pages} {255} (\bibinfo {year} {2022})}\BibitemShut
  {NoStop}%
\bibitem [{\citenamefont {Chakraborty}\ \emph {et~al.}(2022)\citenamefont
  {Chakraborty}, \citenamefont {Das}, \citenamefont {Sinha}, \citenamefont
  {Adak}, \citenamefont {Deshmukh},\ and\ \citenamefont
  {Agarwal}}]{Chakraborty_2022}%
  \BibitemOpen
  \bibfield  {author} {\bibinfo {author} {\bibfnamefont {A.}~\bibnamefont
  {Chakraborty}}, \bibinfo {author} {\bibfnamefont {K.}~\bibnamefont {Das}},
  \bibinfo {author} {\bibfnamefont {S.}~\bibnamefont {Sinha}}, \bibinfo
  {author} {\bibfnamefont {P.~C.}\ \bibnamefont {Adak}}, \bibinfo {author}
  {\bibfnamefont {M.~M.}\ \bibnamefont {Deshmukh}},\ and\ \bibinfo {author}
  {\bibfnamefont {A.}~\bibnamefont {Agarwal}},\ }\bibfield  {title} {\bibinfo
  {title} {Nonlinear anomalous hall effects probe topological phase-transitions
  in twisted double bilayer graphene},\ }\href
  {https://doi.org/10.1088/2053-1583/ac8b93} {\bibfield  {journal} {\bibinfo
  {journal} {2D Mater.}\ }\textbf {\bibinfo {volume} {9}},\ \bibinfo {pages}
  {045020} (\bibinfo {year} {2022})}\BibitemShut {NoStop}%
\bibitem [{\citenamefont {Huang}\ \emph
  {et~al.}(2023{\natexlab{a}})\citenamefont {Huang}, \citenamefont {Wu},
  \citenamefont {Hu}, \citenamefont {Cai}, \citenamefont {Li}, \citenamefont
  {An}, \citenamefont {Feng}, \citenamefont {Ye}, \citenamefont {Lin},
  \citenamefont {Law},\ and\ \citenamefont {Wang}}]{PMID:37180357}%
  \BibitemOpen
  \bibfield  {author} {\bibinfo {author} {\bibfnamefont {M.}~\bibnamefont
  {Huang}}, \bibinfo {author} {\bibfnamefont {Z.}~\bibnamefont {Wu}}, \bibinfo
  {author} {\bibfnamefont {J.}~\bibnamefont {Hu}}, \bibinfo {author}
  {\bibfnamefont {X.}~\bibnamefont {Cai}}, \bibinfo {author} {\bibfnamefont
  {E.}~\bibnamefont {Li}}, \bibinfo {author} {\bibfnamefont {L.}~\bibnamefont
  {An}}, \bibinfo {author} {\bibfnamefont {X.}~\bibnamefont {Feng}}, \bibinfo
  {author} {\bibfnamefont {Z.}~\bibnamefont {Ye}}, \bibinfo {author}
  {\bibfnamefont {N.}~\bibnamefont {Lin}}, \bibinfo {author} {\bibfnamefont
  {K.~T.}\ \bibnamefont {Law}},\ and\ \bibinfo {author} {\bibfnamefont
  {N.}~\bibnamefont {Wang}},\ }\bibfield  {title} {\bibinfo {title} {Giant
  nonlinear hall effect in twisted bilayer $\mathrm{WSe}_2$},\ }\href
  {https://doi.org/10.1093/nsr/nwac232} {\bibfield  {journal} {\bibinfo
  {journal} {Natl. Sci. Rev.}\ }\textbf {\bibinfo {volume} {10}},\ \bibinfo
  {pages} {nwac232} (\bibinfo {year} {2023}{\natexlab{a}})}\BibitemShut
  {NoStop}%
\bibitem [{\citenamefont {Zhang}\ \emph {et~al.}(2022)\citenamefont {Zhang},
  \citenamefont {Xiao}, \citenamefont {Zhou}, \citenamefont {Hu}, \citenamefont
  {Xie}, \citenamefont {Yan},\ and\ \citenamefont
  {Law}}]{PhysRevB.106.L041111}%
  \BibitemOpen
  \bibfield  {author} {\bibinfo {author} {\bibfnamefont {C.-P.}\ \bibnamefont
  {Zhang}}, \bibinfo {author} {\bibfnamefont {J.}~\bibnamefont {Xiao}},
  \bibinfo {author} {\bibfnamefont {B.~T.}\ \bibnamefont {Zhou}}, \bibinfo
  {author} {\bibfnamefont {J.-X.}\ \bibnamefont {Hu}}, \bibinfo {author}
  {\bibfnamefont {Y.-M.}\ \bibnamefont {Xie}}, \bibinfo {author} {\bibfnamefont
  {B.}~\bibnamefont {Yan}},\ and\ \bibinfo {author} {\bibfnamefont {K.~T.}\
  \bibnamefont {Law}},\ }\bibfield  {title} {\bibinfo {title} {Giant nonlinear
  hall effect in strained twisted bilayer graphene},\ }\href
  {https://doi.org/10.1103/PhysRevB.106.L041111} {\bibfield  {journal}
  {\bibinfo  {journal} {Phys. Rev. B}\ }\textbf {\bibinfo {volume} {106}},\
  \bibinfo {pages} {L041111} (\bibinfo {year} {2022})}\BibitemShut {NoStop}%
\bibitem [{\citenamefont {Pantale\'on}\ \emph {et~al.}(2021)\citenamefont
  {Pantale\'on}, \citenamefont {Low},\ and\ \citenamefont
  {Guinea}}]{PhysRevB.103.205403}%
  \BibitemOpen
  \bibfield  {author} {\bibinfo {author} {\bibfnamefont {P.~A.}\ \bibnamefont
  {Pantale\'on}}, \bibinfo {author} {\bibfnamefont {T.}~\bibnamefont {Low}},\
  and\ \bibinfo {author} {\bibfnamefont {F.}~\bibnamefont {Guinea}},\
  }\bibfield  {title} {\bibinfo {title} {Tunable large berry dipole in strained
  twisted bilayer graphene},\ }\href
  {https://doi.org/10.1103/PhysRevB.103.205403} {\bibfield  {journal} {\bibinfo
   {journal} {Phys. Rev. B}\ }\textbf {\bibinfo {volume} {103}},\ \bibinfo
  {pages} {205403} (\bibinfo {year} {2021})}\BibitemShut {NoStop}%
\bibitem [{\citenamefont {Pantale\'on}\ \emph {et~al.}(2022)\citenamefont
  {Pantale\'on}, \citenamefont {Phong}, \citenamefont {Naumis},\ and\
  \citenamefont {Guinea}}]{PhysRevB.106.L161101}%
  \BibitemOpen
  \bibfield  {author} {\bibinfo {author} {\bibfnamefont {P.~A.}\ \bibnamefont
  {Pantale\'on}}, \bibinfo {author} {\bibfnamefont {V.~o.~T.}\ \bibnamefont
  {Phong}}, \bibinfo {author} {\bibfnamefont {G.~G.}\ \bibnamefont {Naumis}},\
  and\ \bibinfo {author} {\bibfnamefont {F.}~\bibnamefont {Guinea}},\
  }\bibfield  {title} {\bibinfo {title} {Interaction-enhanced topological hall
  effects in strained twisted bilayer graphene},\ }\href
  {https://doi.org/10.1103/PhysRevB.106.L161101} {\bibfield  {journal}
  {\bibinfo  {journal} {Phys. Rev. B}\ }\textbf {\bibinfo {volume} {106}},\
  \bibinfo {pages} {L161101} (\bibinfo {year} {2022})}\BibitemShut {NoStop}%
\bibitem [{\citenamefont {Huang}\ \emph
  {et~al.}(2023{\natexlab{b}})\citenamefont {Huang}, \citenamefont {Wu},
  \citenamefont {Zhang}, \citenamefont {Feng}, \citenamefont {Zhou},
  \citenamefont {Wang}, \citenamefont {Chen}, \citenamefont {Cheng},
  \citenamefont {Sun}, \citenamefont {Meng},\ and\ \citenamefont
  {Wang}}]{PhysRevLett.131.066301}%
  \BibitemOpen
  \bibfield  {author} {\bibinfo {author} {\bibfnamefont {M.}~\bibnamefont
  {Huang}}, \bibinfo {author} {\bibfnamefont {Z.}~\bibnamefont {Wu}}, \bibinfo
  {author} {\bibfnamefont {X.}~\bibnamefont {Zhang}}, \bibinfo {author}
  {\bibfnamefont {X.}~\bibnamefont {Feng}}, \bibinfo {author} {\bibfnamefont
  {Z.}~\bibnamefont {Zhou}}, \bibinfo {author} {\bibfnamefont {S.}~\bibnamefont
  {Wang}}, \bibinfo {author} {\bibfnamefont {Y.}~\bibnamefont {Chen}}, \bibinfo
  {author} {\bibfnamefont {C.}~\bibnamefont {Cheng}}, \bibinfo {author}
  {\bibfnamefont {K.}~\bibnamefont {Sun}}, \bibinfo {author} {\bibfnamefont
  {Z.~Y.}\ \bibnamefont {Meng}},\ and\ \bibinfo {author} {\bibfnamefont
  {N.}~\bibnamefont {Wang}},\ }\bibfield  {title} {\bibinfo {title} {Intrinsic
  nonlinear hall effect and gate-switchable berry curvature sliding in twisted
  bilayer graphene},\ }\href {https://doi.org/10.1103/PhysRevLett.131.066301}
  {\bibfield  {journal} {\bibinfo  {journal} {Phys. Rev. Lett.}\ }\textbf
  {\bibinfo {volume} {131}},\ \bibinfo {pages} {066301} (\bibinfo {year}
  {2023}{\natexlab{b}})}\BibitemShut {NoStop}%
\bibitem [{\citenamefont {Son}\ \emph {et~al.}(2019)\citenamefont {Son},
  \citenamefont {Kim}, \citenamefont {Ahn}, \citenamefont {Lee},\ and\
  \citenamefont {Lee}}]{PhysRevLett.123.036806}%
  \BibitemOpen
  \bibfield  {author} {\bibinfo {author} {\bibfnamefont {J.}~\bibnamefont
  {Son}}, \bibinfo {author} {\bibfnamefont {K.-H.}\ \bibnamefont {Kim}},
  \bibinfo {author} {\bibfnamefont {Y.~H.}\ \bibnamefont {Ahn}}, \bibinfo
  {author} {\bibfnamefont {H.-W.}\ \bibnamefont {Lee}},\ and\ \bibinfo {author}
  {\bibfnamefont {J.}~\bibnamefont {Lee}},\ }\bibfield  {title} {\bibinfo
  {title} {Strain engineering of the berry curvature dipole and valley
  magnetization in monolayer ${\mathrm{mos}}_{2}$},\ }\href
  {https://doi.org/10.1103/PhysRevLett.123.036806} {\bibfield  {journal}
  {\bibinfo  {journal} {Phys. Rev. Lett.}\ }\textbf {\bibinfo {volume} {123}},\
  \bibinfo {pages} {036806} (\bibinfo {year} {2019})}\BibitemShut {NoStop}%
\bibitem [{\citenamefont {Qin}\ \emph {et~al.}(2021)\citenamefont {Qin},
  \citenamefont {Zhu}, \citenamefont {Ye}, \citenamefont {Xu}, \citenamefont
  {Song}, \citenamefont {Liang}, \citenamefont {Liu},\ and\ \citenamefont
  {Liao}}]{Qin_2021}%
  \BibitemOpen
  \bibfield  {author} {\bibinfo {author} {\bibfnamefont {M.-S.}\ \bibnamefont
  {Qin}}, \bibinfo {author} {\bibfnamefont {P.-F.}\ \bibnamefont {Zhu}},
  \bibinfo {author} {\bibfnamefont {X.-G.}\ \bibnamefont {Ye}}, \bibinfo
  {author} {\bibfnamefont {W.-Z.}\ \bibnamefont {Xu}}, \bibinfo {author}
  {\bibfnamefont {Z.-H.}\ \bibnamefont {Song}}, \bibinfo {author}
  {\bibfnamefont {J.}~\bibnamefont {Liang}}, \bibinfo {author} {\bibfnamefont
  {K.}~\bibnamefont {Liu}},\ and\ \bibinfo {author} {\bibfnamefont {Z.-M.}\
  \bibnamefont {Liao}},\ }\bibfield  {title} {\bibinfo {title} {Strain tunable
  berry curvature dipole, orbital magnetization and nonlinear hall effect in
  $\mathrm{WSe}_2$ monolayer},\ }\href
  {https://doi.org/10.1088/0256-307X/38/1/017301} {\bibfield  {journal}
  {\bibinfo  {journal} {Chin. Phys. Lett.}\ }\textbf {\bibinfo {volume} {38}},\
  \bibinfo {pages} {017301} (\bibinfo {year} {2021})}\BibitemShut {NoStop}%
\bibitem [{\citenamefont {You}\ \emph {et~al.}(2018)\citenamefont {You},
  \citenamefont {Fang}, \citenamefont {Xu}, \citenamefont {Kaxiras},\ and\
  \citenamefont {Low}}]{PhysRevB.98.121109}%
  \BibitemOpen
  \bibfield  {author} {\bibinfo {author} {\bibfnamefont {J.-S.}\ \bibnamefont
  {You}}, \bibinfo {author} {\bibfnamefont {S.}~\bibnamefont {Fang}}, \bibinfo
  {author} {\bibfnamefont {S.-Y.}\ \bibnamefont {Xu}}, \bibinfo {author}
  {\bibfnamefont {E.}~\bibnamefont {Kaxiras}},\ and\ \bibinfo {author}
  {\bibfnamefont {T.}~\bibnamefont {Low}},\ }\bibfield  {title} {\bibinfo
  {title} {Berry curvature dipole current in the transition metal
  dichalcogenides family},\ }\href {https://doi.org/10.1103/PhysRevB.98.121109}
  {\bibfield  {journal} {\bibinfo  {journal} {Phys. Rev. B}\ }\textbf {\bibinfo
  {volume} {98}},\ \bibinfo {pages} {121109} (\bibinfo {year}
  {2018})}\BibitemShut {NoStop}%
\bibitem [{\citenamefont {Ho}\ \emph {et~al.}(2021)\citenamefont {Ho},
  \citenamefont {Chang}, \citenamefont {Hsieh}, \citenamefont {Lo},
  \citenamefont {Huang}, \citenamefont {Vu}, \citenamefont {Ortix},\ and\
  \citenamefont {Chen}}]{ho2021hall}%
  \BibitemOpen
  \bibfield  {author} {\bibinfo {author} {\bibfnamefont {S.-C.}\ \bibnamefont
  {Ho}}, \bibinfo {author} {\bibfnamefont {C.-H.}\ \bibnamefont {Chang}},
  \bibinfo {author} {\bibfnamefont {Y.-C.}\ \bibnamefont {Hsieh}}, \bibinfo
  {author} {\bibfnamefont {S.-T.}\ \bibnamefont {Lo}}, \bibinfo {author}
  {\bibfnamefont {B.}~\bibnamefont {Huang}}, \bibinfo {author} {\bibfnamefont
  {T.-H.-Y.}\ \bibnamefont {Vu}}, \bibinfo {author} {\bibfnamefont
  {C.}~\bibnamefont {Ortix}},\ and\ \bibinfo {author} {\bibfnamefont {T.-M.}\
  \bibnamefont {Chen}},\ }\bibfield  {title} {\bibinfo {title} {Hall effects in
  artificially corrugated bilayer graphene without breaking time-reversal
  symmetry},\ }\href {https://doi.org/10.1038/s41928-021-00537-5} {\bibfield
  {journal} {\bibinfo  {journal} {Nat. Electron.}\ }\textbf {\bibinfo {volume}
  {4}},\ \bibinfo {pages} {116} (\bibinfo {year} {2021})}\BibitemShut {NoStop}%
\bibitem [{\citenamefont {Ye}\ \emph {et~al.}(2025)\citenamefont {Ye},
  \citenamefont {Zhang}, \citenamefont {Zhu}, \citenamefont {Xu}, \citenamefont
  {Wang},\ and\ \citenamefont {Liao}}]{PhysRevB.111.L041403}%
  \BibitemOpen
  \bibfield  {author} {\bibinfo {author} {\bibfnamefont {X.-G.}\ \bibnamefont
  {Ye}}, \bibinfo {author} {\bibfnamefont {Z.-T.}\ \bibnamefont {Zhang}},
  \bibinfo {author} {\bibfnamefont {P.-F.}\ \bibnamefont {Zhu}}, \bibinfo
  {author} {\bibfnamefont {W.-Z.}\ \bibnamefont {Xu}}, \bibinfo {author}
  {\bibfnamefont {A.-Q.}\ \bibnamefont {Wang}},\ and\ \bibinfo {author}
  {\bibfnamefont {Z.-M.}\ \bibnamefont {Liao}},\ }\bibfield  {title} {\bibinfo
  {title} {Engineering nonlinear hall effect in bilayer graphene/black
  phosphorus heterostructures},\ }\href
  {https://doi.org/10.1103/PhysRevB.111.L041403} {\bibfield  {journal}
  {\bibinfo  {journal} {Phys. Rev. B}\ }\textbf {\bibinfo {volume} {111}},\
  \bibinfo {pages} {L041403} (\bibinfo {year} {2025})}\BibitemShut {NoStop}%
\bibitem [{\citenamefont {Battilomo}\ \emph {et~al.}(2019)\citenamefont
  {Battilomo}, \citenamefont {Scopigno},\ and\ \citenamefont
  {Ortix}}]{PhysRevLett.123.196403}%
  \BibitemOpen
  \bibfield  {author} {\bibinfo {author} {\bibfnamefont {R.}~\bibnamefont
  {Battilomo}}, \bibinfo {author} {\bibfnamefont {N.}~\bibnamefont
  {Scopigno}},\ and\ \bibinfo {author} {\bibfnamefont {C.}~\bibnamefont
  {Ortix}},\ }\bibfield  {title} {\bibinfo {title} {Berry curvature dipole in
  strained graphene: A fermi surface warping effect},\ }\href
  {https://doi.org/10.1103/PhysRevLett.123.196403} {\bibfield  {journal}
  {\bibinfo  {journal} {Phys. Rev. Lett.}\ }\textbf {\bibinfo {volume} {123}},\
  \bibinfo {pages} {196403} (\bibinfo {year} {2019})}\BibitemShut {NoStop}%
\bibitem [{\citenamefont {Du}\ \emph {et~al.}(2021{\natexlab{a}})\citenamefont
  {Du}, \citenamefont {Hasan}, \citenamefont {Castellanos-Gomez}, \citenamefont
  {Liu}, \citenamefont {Yao}, \citenamefont {Lau},\ and\ \citenamefont
  {Sun}}]{du2021engineering}%
  \BibitemOpen
  \bibfield  {author} {\bibinfo {author} {\bibfnamefont {L.}~\bibnamefont
  {Du}}, \bibinfo {author} {\bibfnamefont {T.}~\bibnamefont {Hasan}}, \bibinfo
  {author} {\bibfnamefont {A.}~\bibnamefont {Castellanos-Gomez}}, \bibinfo
  {author} {\bibfnamefont {G.-B.}\ \bibnamefont {Liu}}, \bibinfo {author}
  {\bibfnamefont {Y.}~\bibnamefont {Yao}}, \bibinfo {author} {\bibfnamefont
  {C.~N.}\ \bibnamefont {Lau}},\ and\ \bibinfo {author} {\bibfnamefont
  {Z.}~\bibnamefont {Sun}},\ }\bibfield  {title} {\bibinfo {title} {Engineering
  symmetry breaking in 2$\mathrm{D}$ layered materials},\ }\href
  {https://doi.org/10.1038/s42254-020-00276-0} {\bibfield  {journal} {\bibinfo
  {journal} {Nat. Rev. Phys.}\ }\textbf {\bibinfo {volume} {3}},\ \bibinfo
  {pages} {193} (\bibinfo {year} {2021}{\natexlab{a}})}\BibitemShut {NoStop}%
\bibitem [{\citenamefont {Zhou}\ \emph {et~al.}()\citenamefont {Zhou},
  \citenamefont {Fang}, \citenamefont {Zhang}, \citenamefont {Wang},
  \citenamefont {Rong},\ and\ \citenamefont {Li}}]{zhou2024nonlinear}%
  \BibitemOpen
  \bibfield  {author} {\bibinfo {author} {\bibfnamefont {Z.}~\bibnamefont
  {Zhou}}, \bibinfo {author} {\bibfnamefont {R.}~\bibnamefont {Fang}}, \bibinfo
  {author} {\bibfnamefont {Z.}~\bibnamefont {Zhang}}, \bibinfo {author}
  {\bibfnamefont {X.}~\bibnamefont {Wang}}, \bibinfo {author} {\bibfnamefont
  {J.}~\bibnamefont {Rong}},\ and\ \bibinfo {author} {\bibfnamefont
  {X.}~\bibnamefont {Li}},\ }\href@noop {} {\bibinfo {title} {Nonlinear valley
  hall effect in a bilayer transition metal dichalcogenide}},\ \Eprint
  {https://arxiv.org/abs/2412.19502} {arXiv:2412.19502} \BibitemShut {NoStop}%
\bibitem [{\citenamefont {Pan}\ \emph {et~al.}(2024{\natexlab{a}})\citenamefont
  {Pan}, \citenamefont {Xie}, \citenamefont {Shi}, \citenamefont {Wang},
  \citenamefont {Zhang},\ and\ \citenamefont {Wang}}]{PhysRevB.109.075415}%
  \BibitemOpen
  \bibfield  {author} {\bibinfo {author} {\bibfnamefont {J.}~\bibnamefont
  {Pan}}, \bibinfo {author} {\bibfnamefont {H.}~\bibnamefont {Xie}}, \bibinfo
  {author} {\bibfnamefont {P.}~\bibnamefont {Shi}}, \bibinfo {author}
  {\bibfnamefont {X.}~\bibnamefont {Wang}}, \bibinfo {author} {\bibfnamefont
  {L.}~\bibnamefont {Zhang}},\ and\ \bibinfo {author} {\bibfnamefont
  {Z.}~\bibnamefont {Wang}},\ }\bibfield  {title} {\bibinfo {title} {Berry
  curvature dipole in bilayer graphene with interlayer sliding},\ }\href
  {https://doi.org/10.1103/PhysRevB.109.075415} {\bibfield  {journal} {\bibinfo
   {journal} {Phys. Rev. B}\ }\textbf {\bibinfo {volume} {109}},\ \bibinfo
  {pages} {075415} (\bibinfo {year} {2024}{\natexlab{a}})}\BibitemShut
  {NoStop}%
\bibitem [{\citenamefont {Pan}\ \emph {et~al.}(2024{\natexlab{b}})\citenamefont
  {Pan}, \citenamefont {Wang}, \citenamefont {Zou}, \citenamefont {Xie},
  \citenamefont {Ding}, \citenamefont {Zhang}, \citenamefont {Fang},\ and\
  \citenamefont {Wang}}]{PhysRevB.110.235418}%
  \BibitemOpen
  \bibfield  {author} {\bibinfo {author} {\bibfnamefont {J.}~\bibnamefont
  {Pan}}, \bibinfo {author} {\bibfnamefont {H.}~\bibnamefont {Wang}}, \bibinfo
  {author} {\bibfnamefont {L.}~\bibnamefont {Zou}}, \bibinfo {author}
  {\bibfnamefont {H.}~\bibnamefont {Xie}}, \bibinfo {author} {\bibfnamefont
  {Y.}~\bibnamefont {Ding}}, \bibinfo {author} {\bibfnamefont {Y.}~\bibnamefont
  {Zhang}}, \bibinfo {author} {\bibfnamefont {A.}~\bibnamefont {Fang}},\ and\
  \bibinfo {author} {\bibfnamefont {Z.}~\bibnamefont {Wang}},\ }\bibfield
  {title} {\bibinfo {title} {Inducing berry curvature dipole in multilayer
  graphene through inhomogeneous interlayer sliding},\ }\href
  {https://doi.org/10.1103/PhysRevB.110.235418} {\bibfield  {journal} {\bibinfo
   {journal} {Phys. Rev. B}\ }\textbf {\bibinfo {volume} {110}},\ \bibinfo
  {pages} {235418} (\bibinfo {year} {2024}{\natexlab{b}})}\BibitemShut
  {NoStop}%
\bibitem [{\citenamefont {Roy}(2014)}]{PhysRevB.90.165139}%
  \BibitemOpen
  \bibfield  {author} {\bibinfo {author} {\bibfnamefont {R.}~\bibnamefont
  {Roy}},\ }\bibfield  {title} {\bibinfo {title} {Band geometry of fractional
  topological insulators},\ }\href {https://doi.org/10.1103/PhysRevB.90.165139}
  {\bibfield  {journal} {\bibinfo  {journal} {Phys. Rev. B}\ }\textbf {\bibinfo
  {volume} {90}},\ \bibinfo {pages} {165139} (\bibinfo {year}
  {2014})}\BibitemShut {NoStop}%
\bibitem [{\citenamefont {Claassen}\ \emph {et~al.}(2015)\citenamefont
  {Claassen}, \citenamefont {Lee}, \citenamefont {Thomale}, \citenamefont
  {Qi},\ and\ \citenamefont {Devereaux}}]{PhysRevLett.114.236802}%
  \BibitemOpen
  \bibfield  {author} {\bibinfo {author} {\bibfnamefont {M.}~\bibnamefont
  {Claassen}}, \bibinfo {author} {\bibfnamefont {C.~H.}\ \bibnamefont {Lee}},
  \bibinfo {author} {\bibfnamefont {R.}~\bibnamefont {Thomale}}, \bibinfo
  {author} {\bibfnamefont {X.-L.}\ \bibnamefont {Qi}},\ and\ \bibinfo {author}
  {\bibfnamefont {T.~P.}\ \bibnamefont {Devereaux}},\ }\bibfield  {title}
  {\bibinfo {title} {Position-momentum duality and fractional quantum hall
  effect in chern insulators},\ }\href
  {https://doi.org/10.1103/PhysRevLett.114.236802} {\bibfield  {journal}
  {\bibinfo  {journal} {Phys. Rev. Lett.}\ }\textbf {\bibinfo {volume} {114}},\
  \bibinfo {pages} {236802} (\bibinfo {year} {2015})}\BibitemShut {NoStop}%
\bibitem [{\citenamefont {Parameswaran}\ \emph {et~al.}(2012)\citenamefont
  {Parameswaran}, \citenamefont {Roy},\ and\ \citenamefont
  {Sondhi}}]{PhysRevB.85.241308}%
  \BibitemOpen
  \bibfield  {author} {\bibinfo {author} {\bibfnamefont {S.~A.}\ \bibnamefont
  {Parameswaran}}, \bibinfo {author} {\bibfnamefont {R.}~\bibnamefont {Roy}},\
  and\ \bibinfo {author} {\bibfnamefont {S.~L.}\ \bibnamefont {Sondhi}},\
  }\bibfield  {title} {\bibinfo {title} {Fractional chern insulators and the
  ${W}_{\ensuremath{\infty}}$ algebra},\ }\href
  {https://doi.org/10.1103/PhysRevB.85.241308} {\bibfield  {journal} {\bibinfo
  {journal} {Phys. Rev. B}\ }\textbf {\bibinfo {volume} {85}},\ \bibinfo
  {pages} {241308} (\bibinfo {year} {2012})}\BibitemShut {NoStop}%
\bibitem [{\citenamefont {Nakatsuji}\ \emph {et~al.}()\citenamefont
  {Nakatsuji}, \citenamefont {Kawakami}, \citenamefont {Tateishi},
  \citenamefont {Kato},\ and\ \citenamefont
  {Koshino}}]{nakatsuji2025moirebandengineeringtwisted}%
  \BibitemOpen
  \bibfield  {author} {\bibinfo {author} {\bibfnamefont {N.}~\bibnamefont
  {Nakatsuji}}, \bibinfo {author} {\bibfnamefont {T.}~\bibnamefont {Kawakami}},
  \bibinfo {author} {\bibfnamefont {H.}~\bibnamefont {Tateishi}}, \bibinfo
  {author} {\bibfnamefont {K.}~\bibnamefont {Kato}},\ and\ \bibinfo {author}
  {\bibfnamefont {M.}~\bibnamefont {Koshino}},\ }\href@noop {} {\bibinfo
  {title} {Moir\'e band engineering in twisted trilayer $\mathrm{WSe}_2$}},\
  \Eprint {https://arxiv.org/abs/2504.20449} {arXiv:2504.20449} \BibitemShut
  {NoStop}%
\bibitem [{\citenamefont {Li}\ \emph {et~al.}(2024{\natexlab{b}})\citenamefont
  {Li}, \citenamefont {Zhan},\ and\ \citenamefont
  {Yuan}}]{PhysRevB.109.085118}%
  \BibitemOpen
  \bibfield  {author} {\bibinfo {author} {\bibfnamefont {Y.}~\bibnamefont
  {Li}}, \bibinfo {author} {\bibfnamefont {Z.}~\bibnamefont {Zhan}},\ and\
  \bibinfo {author} {\bibfnamefont {S.}~\bibnamefont {Yuan}},\ }\bibfield
  {title} {\bibinfo {title} {Tuning flat bands by interlayer interaction,
  spin-orbital coupling, and external fields in twisted homotrilayer
  $\mathrm{MoS}_{2}$},\ }\href {https://doi.org/10.1103/PhysRevB.109.085118}
  {\bibfield  {journal} {\bibinfo  {journal} {Phys. Rev. B}\ }\textbf {\bibinfo
  {volume} {109}},\ \bibinfo {pages} {085118} (\bibinfo {year}
  {2024}{\natexlab{b}})}\BibitemShut {NoStop}%
\bibitem [{\citenamefont {Yu}\ \emph {et~al.}(2019)\citenamefont {Yu},
  \citenamefont {Chen},\ and\ \citenamefont {Yao}}]{10.1093/nsr/nwz117}%
  \BibitemOpen
  \bibfield  {author} {\bibinfo {author} {\bibfnamefont {H.}~\bibnamefont
  {Yu}}, \bibinfo {author} {\bibfnamefont {M.}~\bibnamefont {Chen}},\ and\
  \bibinfo {author} {\bibfnamefont {W.}~\bibnamefont {Yao}},\ }\bibfield
  {title} {\bibinfo {title} {Giant magnetic field from moiré induced berry
  phase in homobilayer semiconductors},\ }\href
  {https://doi.org/10.1093/nsr/nwz117} {\bibfield  {journal} {\bibinfo
  {journal} {Natl. Sci. Rev.}\ }\textbf {\bibinfo {volume} {7}},\ \bibinfo
  {pages} {12} (\bibinfo {year} {2019})}\BibitemShut {NoStop}%
\bibitem [{\citenamefont {Hao}\ \emph {et~al.}(2024)\citenamefont {Hao},
  \citenamefont {Zhan}, \citenamefont {Pantale{\'o}n}, \citenamefont {He},
  \citenamefont {Zhao}, \citenamefont {Watanabe}, \citenamefont {Taniguchi},
  \citenamefont {Guinea},\ and\ \citenamefont {He}}]{hao2024robust}%
  \BibitemOpen
  \bibfield  {author} {\bibinfo {author} {\bibfnamefont {C.-Y.}\ \bibnamefont
  {Hao}}, \bibinfo {author} {\bibfnamefont {Z.}~\bibnamefont {Zhan}}, \bibinfo
  {author} {\bibfnamefont {P.~A.}\ \bibnamefont {Pantale{\'o}n}}, \bibinfo
  {author} {\bibfnamefont {J.-Q.}\ \bibnamefont {He}}, \bibinfo {author}
  {\bibfnamefont {Y.-X.}\ \bibnamefont {Zhao}}, \bibinfo {author}
  {\bibfnamefont {K.}~\bibnamefont {Watanabe}}, \bibinfo {author}
  {\bibfnamefont {T.}~\bibnamefont {Taniguchi}}, \bibinfo {author}
  {\bibfnamefont {F.}~\bibnamefont {Guinea}},\ and\ \bibinfo {author}
  {\bibfnamefont {L.}~\bibnamefont {He}},\ }\bibfield  {title} {\bibinfo
  {title} {Robust flat bands in twisted trilayer graphene moir{\'e}
  quasicrystals},\ }\href {https://doi.org/10.1038/s41467-024-52784-7}
  {\bibfield  {journal} {\bibinfo  {journal} {Nat. Commun.}\ }\textbf {\bibinfo
  {volume} {15}},\ \bibinfo {pages} {8437} (\bibinfo {year}
  {2024})}\BibitemShut {NoStop}%
\bibitem [{\citenamefont {Liu}\ \emph {et~al.}(2013)\citenamefont {Liu},
  \citenamefont {Shan}, \citenamefont {Yao}, \citenamefont {Yao},\ and\
  \citenamefont {Xiao}}]{Three_prb085433_2013}%
  \BibitemOpen
  \bibfield  {author} {\bibinfo {author} {\bibfnamefont {G.-B.}\ \bibnamefont
  {Liu}}, \bibinfo {author} {\bibfnamefont {W.-Y.}\ \bibnamefont {Shan}},
  \bibinfo {author} {\bibfnamefont {Y.}~\bibnamefont {Yao}}, \bibinfo {author}
  {\bibfnamefont {W.}~\bibnamefont {Yao}},\ and\ \bibinfo {author}
  {\bibfnamefont {D.}~\bibnamefont {Xiao}},\ }\bibfield  {title} {\bibinfo
  {title} {Three-band tight-binding model for monolayers of group-vib
  transition metal dichalcogenides},\ }\href
  {https://doi.org/10.1103/PhysRevB.88.085433} {\bibfield  {journal} {\bibinfo
  {journal} {Phys. Rev. B}\ }\textbf {\bibinfo {volume} {88}},\ \bibinfo
  {pages} {085433} (\bibinfo {year} {2013})}\BibitemShut {NoStop}%
\bibitem [{sup()}]{supplemental}%
  \BibitemOpen
  \href@noop {} {\bibinfo  {journal} {See supplemental material for more
  details about the Hamiltonian of CT3BLG, the equivalent relation between
  sliding vectors, the bandwidth variation and berry curvature standard
  deviation of AT3L-\textrm{MoTe$_2$}, and the BCD of the first and second
  valence band for AT3L-$\mathrm{MoTe_2}$}\ }\BibitemShut {NoStop}%
\bibitem [{\citenamefont {Du}\ \emph {et~al.}(2024)\citenamefont {Du},
  \citenamefont {Huang}, \citenamefont {Zhang}, \citenamefont {Ye},
  \citenamefont {Dai}, \citenamefont {Deng}, \citenamefont {Zhang},\ and\
  \citenamefont {Sun}}]{du2024nonlinear}%
  \BibitemOpen
\bibfield  {journal} {  }\bibfield  {author} {\bibinfo {author} {\bibfnamefont
  {L.}~\bibnamefont {Du}}, \bibinfo {author} {\bibfnamefont {Z.}~\bibnamefont
  {Huang}}, \bibinfo {author} {\bibfnamefont {J.}~\bibnamefont {Zhang}},
  \bibinfo {author} {\bibfnamefont {F.}~\bibnamefont {Ye}}, \bibinfo {author}
  {\bibfnamefont {Q.}~\bibnamefont {Dai}}, \bibinfo {author} {\bibfnamefont
  {H.}~\bibnamefont {Deng}}, \bibinfo {author} {\bibfnamefont {G.}~\bibnamefont
  {Zhang}},\ and\ \bibinfo {author} {\bibfnamefont {Z.}~\bibnamefont {Sun}},\
  }\bibfield  {title} {\bibinfo {title} {Nonlinear physics of moir{\'e}
  superlattices},\ }\href {https://doi.org/10.1038/s41563-024-01951-8}
  {\bibfield  {journal} {\bibinfo  {journal} {Nat. Mater.}\ }\textbf {\bibinfo
  {volume} {23}},\ \bibinfo {pages} {1179} (\bibinfo {year}
  {2024})}\BibitemShut {NoStop}%
\bibitem [{\citenamefont {Du}\ \emph {et~al.}(2021{\natexlab{b}})\citenamefont
  {Du}, \citenamefont {Lu},\ and\ \citenamefont {Xie}}]{du2021nonlinear}%
  \BibitemOpen
  \bibfield  {author} {\bibinfo {author} {\bibfnamefont {Z.}~\bibnamefont
  {Du}}, \bibinfo {author} {\bibfnamefont {H.-Z.}\ \bibnamefont {Lu}},\ and\
  \bibinfo {author} {\bibfnamefont {X.}~\bibnamefont {Xie}},\ }\bibfield
  {title} {\bibinfo {title} {Nonlinear hall effects},\ }\href
  {https://doi.org/10.1038/s42254-021-00359- 6} {\bibfield  {journal} {\bibinfo
   {journal} {Nat. Rev. Phys.}\ }\textbf {\bibinfo {volume} {3}},\ \bibinfo
  {pages} {744} (\bibinfo {year} {2021}{\natexlab{b}})}\BibitemShut {NoStop}%
\bibitem [{\citenamefont {Du}\ \emph {et~al.}(2021{\natexlab{c}})\citenamefont
  {Du}, \citenamefont {Wang}, \citenamefont {Sun}, \citenamefont {Lu},\ and\
  \citenamefont {Xie}}]{du2021quantum}%
  \BibitemOpen
  \bibfield  {author} {\bibinfo {author} {\bibfnamefont {Z.}~\bibnamefont
  {Du}}, \bibinfo {author} {\bibfnamefont {C.}~\bibnamefont {Wang}}, \bibinfo
  {author} {\bibfnamefont {H.-P.}\ \bibnamefont {Sun}}, \bibinfo {author}
  {\bibfnamefont {H.-Z.}\ \bibnamefont {Lu}},\ and\ \bibinfo {author}
  {\bibfnamefont {X.}~\bibnamefont {Xie}},\ }\bibfield  {title} {\bibinfo
  {title} {Quantum theory of the nonlinear hall effect},\ }\href
  {https://doi.org/10.1038/s41467-021-25273-4} {\bibfield  {journal} {\bibinfo
  {journal} {Nat. Commun.}\ }\textbf {\bibinfo {volume} {12}},\ \bibinfo
  {pages} {5038} (\bibinfo {year} {2021}{\natexlab{c}})}\BibitemShut {NoStop}%
\bibitem [{\citenamefont {Ortix}(2021)}]{qute.202100056}%
  \BibitemOpen
  \bibfield  {author} {\bibinfo {author} {\bibfnamefont {C.}~\bibnamefont
  {Ortix}},\ }\bibfield  {title} {\bibinfo {title} {Nonlinear hall effect with
  time-reversal symmetry: Theory and material realizations},\ }\href
  {https://doi.org/https://doi.org/10.1002/qute.202100056} {\bibfield
  {journal} {\bibinfo  {journal} {Adv. Quantum Technol.}\ }\textbf {\bibinfo
  {volume} {4}},\ \bibinfo {pages} {2100056} (\bibinfo {year}
  {2021})}\BibitemShut {NoStop}%
\bibitem [{\citenamefont {Du}\ \emph {et~al.}(2018)\citenamefont {Du},
  \citenamefont {Wang}, \citenamefont {Lu},\ and\ \citenamefont
  {Xie}}]{PhysRevLett.121.266601}%
  \BibitemOpen
  \bibfield  {author} {\bibinfo {author} {\bibfnamefont {Z.~Z.}\ \bibnamefont
  {Du}}, \bibinfo {author} {\bibfnamefont {C.~M.}\ \bibnamefont {Wang}},
  \bibinfo {author} {\bibfnamefont {H.-Z.}\ \bibnamefont {Lu}},\ and\ \bibinfo
  {author} {\bibfnamefont {X.~C.}\ \bibnamefont {Xie}},\ }\bibfield  {title}
  {\bibinfo {title} {Band signatures for strong nonlinear hall effect in
  bilayer $\mathrm{WTe}_{2}$},\ }\href
  {https://doi.org/10.1103/PhysRevLett.121.266601} {\bibfield  {journal}
  {\bibinfo  {journal} {Phys. Rev. Lett.}\ }\textbf {\bibinfo {volume} {121}},\
  \bibinfo {pages} {266601} (\bibinfo {year} {2018})}\BibitemShut {NoStop}%
\bibitem [{\citenamefont {Layek}\ \emph {et~al.}(2025)\citenamefont {Layek},
  \citenamefont {Sinha}, \citenamefont {Chakraborty}, \citenamefont
  {Mukherjee}, \citenamefont {Agarwal}, \citenamefont {Watanabe}, \citenamefont
  {Taniguchi}, \citenamefont {Agarwal},\ and\ \citenamefont
  {Deshmukh}}]{layek2025quantum}%
  \BibitemOpen
  \bibfield  {author} {\bibinfo {author} {\bibfnamefont {S.}~\bibnamefont
  {Layek}}, \bibinfo {author} {\bibfnamefont {S.}~\bibnamefont {Sinha}},
  \bibinfo {author} {\bibfnamefont {A.}~\bibnamefont {Chakraborty}}, \bibinfo
  {author} {\bibfnamefont {A.}~\bibnamefont {Mukherjee}}, \bibinfo {author}
  {\bibfnamefont {H.}~\bibnamefont {Agarwal}}, \bibinfo {author} {\bibfnamefont
  {K.}~\bibnamefont {Watanabe}}, \bibinfo {author} {\bibfnamefont
  {T.}~\bibnamefont {Taniguchi}}, \bibinfo {author} {\bibfnamefont
  {A.}~\bibnamefont {Agarwal}},\ and\ \bibinfo {author} {\bibfnamefont {M.~M.}\
  \bibnamefont {Deshmukh}},\ }\bibfield  {title} {\bibinfo {title} {Quantum
  geometric moment encodes stacking order of moir{\'e} matter},\ }\href
  {https://doi.org/10.1002/adma.202417682} {\bibfield  {journal} {\bibinfo
  {journal} {Adv. Mater.}\ ,\ \bibinfo {pages} {2417682}} (\bibinfo {year}
  {2025})}\BibitemShut {NoStop}%
\bibitem [{\citenamefont {Zhu}\ \emph {et~al.}(2024)\citenamefont {Zhu},
  \citenamefont {Chen},\ and\ \citenamefont {Zhou}}]{PhysRevB.110.245304}%
  \BibitemOpen
  \bibfield  {author} {\bibinfo {author} {\bibfnamefont {J.-Y.}\ \bibnamefont
  {Zhu}}, \bibinfo {author} {\bibfnamefont {R.}~\bibnamefont {Chen}},\ and\
  \bibinfo {author} {\bibfnamefont {B.}~\bibnamefont {Zhou}},\ }\bibfield
  {title} {\bibinfo {title} {Nonlinear hall effect in kagome and lieb lattices
  with staggered hopping},\ }\href
  {https://doi.org/10.1103/PhysRevB.110.245304} {\bibfield  {journal} {\bibinfo
   {journal} {Phys. Rev. B}\ }\textbf {\bibinfo {volume} {110}},\ \bibinfo
  {pages} {245304} (\bibinfo {year} {2024})}\BibitemShut {NoStop}%
\bibitem [{\citenamefont {Kaplan}\ \emph {et~al.}(2024)\citenamefont {Kaplan},
  \citenamefont {Holder},\ and\ \citenamefont {Yan}}]{PhysRevLett.132.026301}%
  \BibitemOpen
  \bibfield  {author} {\bibinfo {author} {\bibfnamefont {D.}~\bibnamefont
  {Kaplan}}, \bibinfo {author} {\bibfnamefont {T.}~\bibnamefont {Holder}},\
  and\ \bibinfo {author} {\bibfnamefont {B.}~\bibnamefont {Yan}},\ }\bibfield
  {title} {\bibinfo {title} {Unification of nonlinear anomalous hall effect and
  nonreciprocal magnetoresistance in metals by the quantum geometry},\ }\href
  {https://doi.org/10.1103/PhysRevLett.132.026301} {\bibfield  {journal}
  {\bibinfo  {journal} {Phys. Rev. Lett.}\ }\textbf {\bibinfo {volume} {132}},\
  \bibinfo {pages} {026301} (\bibinfo {year} {2024})}\BibitemShut {NoStop}%
\bibitem [{\citenamefont {Wang}\ \emph {et~al.}(2025)\citenamefont {Wang},
  \citenamefont {Niu},\ and\ \citenamefont {Fang}}]{Wang_2025}%
  \BibitemOpen
  \bibfield  {author} {\bibinfo {author} {\bibfnamefont {S.}~\bibnamefont
  {Wang}}, \bibinfo {author} {\bibfnamefont {W.}~\bibnamefont {Niu}},\ and\
  \bibinfo {author} {\bibfnamefont {Y.-W.}\ \bibnamefont {Fang}},\ }\bibfield
  {title} {\bibinfo {title} {Nonlinear hall effect in two-dimensional
  materials},\ }\href {http://dx.doi.org/10.20517/microstructures.2024.129}
  {\bibfield  {journal} {\bibinfo  {journal} {Microstructures}\ }\textbf
  {\bibinfo {volume} {5}} (\bibinfo {year} {2025})}\BibitemShut {NoStop}%
\bibitem [{\citenamefont {Bandyopadhyay}\ \emph {et~al.}(2024)\citenamefont
  {Bandyopadhyay}, \citenamefont {Joseph},\ and\ \citenamefont
  {Narayan}}]{BANDYOPADHYAY2024100101}%
  \BibitemOpen
  \bibfield  {author} {\bibinfo {author} {\bibfnamefont {A.}~\bibnamefont
  {Bandyopadhyay}}, \bibinfo {author} {\bibfnamefont {N.~B.}\ \bibnamefont
  {Joseph}},\ and\ \bibinfo {author} {\bibfnamefont {A.}~\bibnamefont
  {Narayan}},\ }\bibfield  {title} {\bibinfo {title} {Non-linear hall effects:
  Mechanisms and materials},\ }\href
  {https://doi.org/https://doi.org/10.1016/j.mtelec.2024.100101} {\bibfield
  {journal} {\bibinfo  {journal} {Mater. Today Electro.}\ }\textbf {\bibinfo
  {volume} {8}},\ \bibinfo {pages} {100101} (\bibinfo {year}
  {2024})}\BibitemShut {NoStop}%
\bibitem [{\citenamefont {Jiang}\ \emph {et~al.}()\citenamefont {Jiang},
  \citenamefont {Holder},\ and\ \citenamefont
  {Yan}}]{jiang2025revealingquantumgeometrynonlinear}%
  \BibitemOpen
  \bibfield  {author} {\bibinfo {author} {\bibfnamefont {Y.}~\bibnamefont
  {Jiang}}, \bibinfo {author} {\bibfnamefont {T.}~\bibnamefont {Holder}},\ and\
  \bibinfo {author} {\bibfnamefont {B.}~\bibnamefont {Yan}},\ }\href@noop {}
  {\bibinfo {title} {Revealing quantum geometry in nonlinear quantum
  materials}},\ \Eprint {https://arxiv.org/abs/2503.04943} {arXiv:2503.04943}
  \BibitemShut {NoStop}%
\bibitem [{\citenamefont {Gao}\ \emph {et~al.}()\citenamefont {Gao},
  \citenamefont {Nagaosa}, \citenamefont {Ni},\ and\ \citenamefont
  {Xu}}]{gao2025quantumgeometryphenomenacondensed}%
  \BibitemOpen
  \bibfield  {author} {\bibinfo {author} {\bibfnamefont {A.}~\bibnamefont
  {Gao}}, \bibinfo {author} {\bibfnamefont {N.}~\bibnamefont {Nagaosa}},
  \bibinfo {author} {\bibfnamefont {N.}~\bibnamefont {Ni}},\ and\ \bibinfo
  {author} {\bibfnamefont {S.-Y.}\ \bibnamefont {Xu}},\ }\href@noop {}
  {\bibinfo {title} {Quantum geometry phenomena in condensed matter systems}},\
  \Eprint {https://arxiv.org/abs/2508.00469} {arXiv:2508.00469} \BibitemShut
  {NoStop}%
\bibitem [{\citenamefont {Wang}\ \emph {et~al.}(2021)\citenamefont {Wang},
  \citenamefont {Gao},\ and\ \citenamefont {Xiao}}]{PhysRevLett.127.277201}%
  \BibitemOpen
  \bibfield  {author} {\bibinfo {author} {\bibfnamefont {C.}~\bibnamefont
  {Wang}}, \bibinfo {author} {\bibfnamefont {Y.}~\bibnamefont {Gao}},\ and\
  \bibinfo {author} {\bibfnamefont {D.}~\bibnamefont {Xiao}},\ }\bibfield
  {title} {\bibinfo {title} {Intrinsic nonlinear hall effect in
  antiferromagnetic tetragonal cumnas},\ }\href
  {https://doi.org/10.1103/PhysRevLett.127.277201} {\bibfield  {journal}
  {\bibinfo  {journal} {Phys. Rev. Lett.}\ }\textbf {\bibinfo {volume} {127}},\
  \bibinfo {pages} {277201} (\bibinfo {year} {2021})}\BibitemShut {NoStop}%
\bibitem [{\citenamefont {Gong}\ \emph {et~al.}()\citenamefont {Gong},
  \citenamefont {Du}, \citenamefont {Sun}, \citenamefont {Lu},\ and\
  \citenamefont {Xie}}]{gong2025nonlineartransporttheoryorder}%
  \BibitemOpen
  \bibfield  {author} {\bibinfo {author} {\bibfnamefont {Z.-H.}\ \bibnamefont
  {Gong}}, \bibinfo {author} {\bibfnamefont {Z.~Z.}\ \bibnamefont {Du}},
  \bibinfo {author} {\bibfnamefont {H.-P.}\ \bibnamefont {Sun}}, \bibinfo
  {author} {\bibfnamefont {H.-Z.}\ \bibnamefont {Lu}},\ and\ \bibinfo {author}
  {\bibfnamefont {X.~C.}\ \bibnamefont {Xie}},\ }\href@noop {} {\bibinfo
  {title} {Nonlinear transport theory at the order of quantum metric}},\
  \Eprint {https://arxiv.org/abs/2410.04995} {arXiv:2410.04995} \BibitemShut
  {NoStop}%
\bibitem [{\citenamefont {Chen}\ \emph {et~al.}(2024)\citenamefont {Chen},
  \citenamefont {Du}, \citenamefont {Sun}, \citenamefont {Lu},\ and\
  \citenamefont {Xie}}]{PhysRevB.110.L081301}%
  \BibitemOpen
  \bibfield  {author} {\bibinfo {author} {\bibfnamefont {R.}~\bibnamefont
  {Chen}}, \bibinfo {author} {\bibfnamefont {Z.~Z.}\ \bibnamefont {Du}},
  \bibinfo {author} {\bibfnamefont {H.-P.}\ \bibnamefont {Sun}}, \bibinfo
  {author} {\bibfnamefont {H.-Z.}\ \bibnamefont {Lu}},\ and\ \bibinfo {author}
  {\bibfnamefont {X.~C.}\ \bibnamefont {Xie}},\ }\bibfield  {title} {\bibinfo
  {title} {Nonlinear hall effect on a disordered lattice},\ }\href
  {https://doi.org/10.1103/PhysRevB.110.L081301} {\bibfield  {journal}
  {\bibinfo  {journal} {Phys. Rev. B}\ }\textbf {\bibinfo {volume} {110}},\
  \bibinfo {pages} {L081301} (\bibinfo {year} {2024})}\BibitemShut {NoStop}%
\bibitem [{\citenamefont {Du}\ \emph {et~al.}(2019)\citenamefont {Du},
  \citenamefont {Wang}, \citenamefont {Li}, \citenamefont {Lu},\ and\
  \citenamefont {Xie}}]{du2019disorder}%
  \BibitemOpen
  \bibfield  {author} {\bibinfo {author} {\bibfnamefont {Z.}~\bibnamefont
  {Du}}, \bibinfo {author} {\bibfnamefont {C.}~\bibnamefont {Wang}}, \bibinfo
  {author} {\bibfnamefont {S.}~\bibnamefont {Li}}, \bibinfo {author}
  {\bibfnamefont {H.-Z.}\ \bibnamefont {Lu}},\ and\ \bibinfo {author}
  {\bibfnamefont {X.}~\bibnamefont {Xie}},\ }\bibfield  {title} {\bibinfo
  {title} {Disorder-induced nonlinear hall effect with time-reversal
  symmetry},\ }\href {https://doi.org/10.1038/s41467-019-10941-3} {\bibfield
  {journal} {\bibinfo  {journal} {Nat. Commun.}\ }\textbf {\bibinfo {volume}
  {10}},\ \bibinfo {pages} {3047} (\bibinfo {year} {2019})}\BibitemShut
  {NoStop}%
\bibitem [{\citenamefont {Zhou}\ \emph {et~al.}(2020)\citenamefont {Zhou},
  \citenamefont {Zhang},\ and\ \citenamefont {Law}}]{PhysRevApplied.13.024053}%
  \BibitemOpen
  \bibfield  {author} {\bibinfo {author} {\bibfnamefont {B.~T.}\ \bibnamefont
  {Zhou}}, \bibinfo {author} {\bibfnamefont {C.-P.}\ \bibnamefont {Zhang}},\
  and\ \bibinfo {author} {\bibfnamefont {K.}~\bibnamefont {Law}},\ }\bibfield
  {title} {\bibinfo {title} {Highly tunable nonlinear hall effects induced by
  spin-orbit couplings in strained polar transition-metal dichalcogenides},\
  }\href {https://doi.org/10.1103/PhysRevApplied.13.024053} {\bibfield
  {journal} {\bibinfo  {journal} {Phys. Rev. Appl.}\ }\textbf {\bibinfo
  {volume} {13}},\ \bibinfo {pages} {024053} (\bibinfo {year}
  {2020})}\BibitemShut {NoStop}%
\bibitem [{\citenamefont {Ma}\ \emph {et~al.}(2023{\natexlab{b}})\citenamefont
  {Ma}, \citenamefont {Arora}, \citenamefont {Vignale},\ and\ \citenamefont
  {Song}}]{PhysRevLett.131.076601}%
  \BibitemOpen
  \bibfield  {author} {\bibinfo {author} {\bibfnamefont {D.}~\bibnamefont
  {Ma}}, \bibinfo {author} {\bibfnamefont {A.}~\bibnamefont {Arora}}, \bibinfo
  {author} {\bibfnamefont {G.}~\bibnamefont {Vignale}},\ and\ \bibinfo {author}
  {\bibfnamefont {J.~C.~W.}\ \bibnamefont {Song}},\ }\bibfield  {title}
  {\bibinfo {title} {Anomalous skew-scattering nonlinear hall effect and chiral
  photocurrents in $\mathcal{PT}$-symmetric antiferromagnets},\ }\href
  {https://doi.org/10.1103/PhysRevLett.131.076601} {\bibfield  {journal}
  {\bibinfo  {journal} {Phys. Rev. Lett.}\ }\textbf {\bibinfo {volume} {131}},\
  \bibinfo {pages} {076601} (\bibinfo {year} {2023}{\natexlab{b}})}\BibitemShut
  {NoStop}%
\bibitem [{\citenamefont {Saha}\ and\ \citenamefont
  {Narayan}(2023)}]{saha2023nonlinear}%
  \BibitemOpen
  \bibfield  {author} {\bibinfo {author} {\bibfnamefont {S.}~\bibnamefont
  {Saha}}\ and\ \bibinfo {author} {\bibfnamefont {A.}~\bibnamefont {Narayan}},\
  }\bibfield  {title} {\bibinfo {title} {Nonlinear hall effect in rashba
  systems with hexagonal warping},\ }\href
  {https://iopscience.iop.org/article/10.1088/1361-648X/acf1eb} {\bibfield
  {journal} {\bibinfo  {journal} {J. Phys.-Condes. Matter}\ }\textbf {\bibinfo
  {volume} {35}},\ \bibinfo {pages} {485301} (\bibinfo {year}
  {2023})}\BibitemShut {NoStop}%
\bibitem [{\citenamefont {Wang}\ \emph
  {et~al.}(2024{\natexlab{b}})\citenamefont {Wang}, \citenamefont {Zeng},
  \citenamefont {Duan},\ and\ \citenamefont {Huang}}]{PhysRevLett.132.266802}%
  \BibitemOpen
  \bibfield  {author} {\bibinfo {author} {\bibfnamefont {E.}~\bibnamefont
  {Wang}}, \bibinfo {author} {\bibfnamefont {H.}~\bibnamefont {Zeng}}, \bibinfo
  {author} {\bibfnamefont {W.}~\bibnamefont {Duan}},\ and\ \bibinfo {author}
  {\bibfnamefont {H.}~\bibnamefont {Huang}},\ }\bibfield  {title} {\bibinfo
  {title} {Spontaneous inversion symmetry breaking and emergence of berry
  curvature and orbital magnetization in topological $\mathrm{ZrTe}_{5}$
  films},\ }\href {https://doi.org/10.1103/PhysRevLett.132.266802} {\bibfield
  {journal} {\bibinfo  {journal} {Phys. Rev. Lett.}\ }\textbf {\bibinfo
  {volume} {132}},\ \bibinfo {pages} {266802} (\bibinfo {year}
  {2024}{\natexlab{b}})}\BibitemShut {NoStop}%
\bibitem [{\citenamefont {Chichinadze}\ \emph {et~al.}()\citenamefont
  {Chichinadze}, \citenamefont {Zhang}, \citenamefont {Lin}, \citenamefont
  {Morissette}, \citenamefont {Wang}, \citenamefont {Watanabe}, \citenamefont
  {Taniguchi}, \citenamefont {Vafek},\ and\ \citenamefont
  {Li}}]{chichinadze2025observationgiantnonlinearhall}%
  \BibitemOpen
  \bibfield  {author} {\bibinfo {author} {\bibfnamefont {D.~V.}\ \bibnamefont
  {Chichinadze}}, \bibinfo {author} {\bibfnamefont {N.~J.}\ \bibnamefont
  {Zhang}}, \bibinfo {author} {\bibfnamefont {J.-X.}\ \bibnamefont {Lin}},
  \bibinfo {author} {\bibfnamefont {E.}~\bibnamefont {Morissette}}, \bibinfo
  {author} {\bibfnamefont {X.}~\bibnamefont {Wang}}, \bibinfo {author}
  {\bibfnamefont {K.}~\bibnamefont {Watanabe}}, \bibinfo {author}
  {\bibfnamefont {T.}~\bibnamefont {Taniguchi}}, \bibinfo {author}
  {\bibfnamefont {O.}~\bibnamefont {Vafek}},\ and\ \bibinfo {author}
  {\bibfnamefont {J.~I.~A.}\ \bibnamefont {Li}},\ }\href@noop {} {\bibinfo
  {title} {Observation of giant nonlinear hall conductivity in bernal bilayer
  graphene}},\ \Eprint {https://arxiv.org/abs/2411.11156} {arXiv:2411.11156}
  \BibitemShut {NoStop}%
\bibitem [{\citenamefont {Park}\ \emph {et~al.}(2023)\citenamefont {Park},
  \citenamefont {Cai}, \citenamefont {Anderson}, \citenamefont {Zhang},
  \citenamefont {Zhu}, \citenamefont {Liu}, \citenamefont {Wang}, \citenamefont
  {Holtzmann}, \citenamefont {Hu}, \citenamefont {Liu} \emph
  {et~al.}}]{park2023observation}%
  \BibitemOpen
  \bibfield  {author} {\bibinfo {author} {\bibfnamefont {H.}~\bibnamefont
  {Park}}, \bibinfo {author} {\bibfnamefont {J.}~\bibnamefont {Cai}}, \bibinfo
  {author} {\bibfnamefont {E.}~\bibnamefont {Anderson}}, \bibinfo {author}
  {\bibfnamefont {Y.}~\bibnamefont {Zhang}}, \bibinfo {author} {\bibfnamefont
  {J.}~\bibnamefont {Zhu}}, \bibinfo {author} {\bibfnamefont {X.}~\bibnamefont
  {Liu}}, \bibinfo {author} {\bibfnamefont {C.}~\bibnamefont {Wang}}, \bibinfo
  {author} {\bibfnamefont {W.}~\bibnamefont {Holtzmann}}, \bibinfo {author}
  {\bibfnamefont {C.}~\bibnamefont {Hu}}, \bibinfo {author} {\bibfnamefont
  {Z.}~\bibnamefont {Liu}}, \emph {et~al.},\ }\bibfield  {title} {\bibinfo
  {title} {Observation of fractionally quantized anomalous hall effect},\
  }\href {https://doi.org/10.1038/s41586-023-06536-0} {\bibfield  {journal}
  {\bibinfo  {journal} {Nature}\ }\textbf {\bibinfo {volume} {622}},\ \bibinfo
  {pages} {74} (\bibinfo {year} {2023})}\BibitemShut {NoStop}%
\bibitem [{\citenamefont {Xu}\ \emph {et~al.}(2023)\citenamefont {Xu},
  \citenamefont {Sun}, \citenamefont {Jia}, \citenamefont {Liu}, \citenamefont
  {Xu}, \citenamefont {Li}, \citenamefont {Gu}, \citenamefont {Watanabe},
  \citenamefont {Taniguchi}, \citenamefont {Tong}, \citenamefont {Jia},
  \citenamefont {Shi}, \citenamefont {Jiang}, \citenamefont {Zhang},
  \citenamefont {Liu},\ and\ \citenamefont {Li}}]{PhysRevX.13.031037}%
  \BibitemOpen
  \bibfield  {author} {\bibinfo {author} {\bibfnamefont {F.}~\bibnamefont
  {Xu}}, \bibinfo {author} {\bibfnamefont {Z.}~\bibnamefont {Sun}}, \bibinfo
  {author} {\bibfnamefont {T.}~\bibnamefont {Jia}}, \bibinfo {author}
  {\bibfnamefont {C.}~\bibnamefont {Liu}}, \bibinfo {author} {\bibfnamefont
  {C.}~\bibnamefont {Xu}}, \bibinfo {author} {\bibfnamefont {C.}~\bibnamefont
  {Li}}, \bibinfo {author} {\bibfnamefont {Y.}~\bibnamefont {Gu}}, \bibinfo
  {author} {\bibfnamefont {K.}~\bibnamefont {Watanabe}}, \bibinfo {author}
  {\bibfnamefont {T.}~\bibnamefont {Taniguchi}}, \bibinfo {author}
  {\bibfnamefont {B.}~\bibnamefont {Tong}}, \bibinfo {author} {\bibfnamefont
  {J.}~\bibnamefont {Jia}}, \bibinfo {author} {\bibfnamefont {Z.}~\bibnamefont
  {Shi}}, \bibinfo {author} {\bibfnamefont {S.}~\bibnamefont {Jiang}}, \bibinfo
  {author} {\bibfnamefont {Y.}~\bibnamefont {Zhang}}, \bibinfo {author}
  {\bibfnamefont {X.}~\bibnamefont {Liu}},\ and\ \bibinfo {author}
  {\bibfnamefont {T.}~\bibnamefont {Li}},\ }\bibfield  {title} {\bibinfo
  {title} {Observation of integer and fractional quantum anomalous hall effects
  in twisted bilayer $\mathrm{MoTe}_{2}$},\ }\href
  {https://doi.org/10.1103/PhysRevX.13.031037} {\bibfield  {journal} {\bibinfo
  {journal} {Phys. Rev. X}\ }\textbf {\bibinfo {volume} {13}},\ \bibinfo
  {pages} {031037} (\bibinfo {year} {2023})}\BibitemShut {NoStop}%
\bibitem [{\citenamefont {Cai}\ \emph {et~al.}(2023)\citenamefont {Cai},
  \citenamefont {Anderson}, \citenamefont {Wang}, \citenamefont {Zhang},
  \citenamefont {Liu}, \citenamefont {Holtzmann}, \citenamefont {Zhang},
  \citenamefont {Fan}, \citenamefont {Taniguchi}, \citenamefont {Watanabe}
  \emph {et~al.}}]{cai2023signatures}%
  \BibitemOpen
  \bibfield  {author} {\bibinfo {author} {\bibfnamefont {J.}~\bibnamefont
  {Cai}}, \bibinfo {author} {\bibfnamefont {E.}~\bibnamefont {Anderson}},
  \bibinfo {author} {\bibfnamefont {C.}~\bibnamefont {Wang}}, \bibinfo {author}
  {\bibfnamefont {X.}~\bibnamefont {Zhang}}, \bibinfo {author} {\bibfnamefont
  {X.}~\bibnamefont {Liu}}, \bibinfo {author} {\bibfnamefont {W.}~\bibnamefont
  {Holtzmann}}, \bibinfo {author} {\bibfnamefont {Y.}~\bibnamefont {Zhang}},
  \bibinfo {author} {\bibfnamefont {F.}~\bibnamefont {Fan}}, \bibinfo {author}
  {\bibfnamefont {T.}~\bibnamefont {Taniguchi}}, \bibinfo {author}
  {\bibfnamefont {K.}~\bibnamefont {Watanabe}}, \emph {et~al.},\ }\bibfield
  {title} {\bibinfo {title} {Signatures of fractional quantum anomalous hall
  states in twisted $\mathrm{MoTe}_{2}$},\ }\href
  {https://doi.org/10.1038/s41586-023-06289-w} {\bibfield  {journal} {\bibinfo
  {journal} {Nature}\ }\textbf {\bibinfo {volume} {622}},\ \bibinfo {pages}
  {63} (\bibinfo {year} {2023})}\BibitemShut {NoStop}%
\bibitem [{\citenamefont {Zeng}\ \emph {et~al.}(2023)\citenamefont {Zeng},
  \citenamefont {Xia}, \citenamefont {Kang}, \citenamefont {Zhu}, \citenamefont
  {Kn{\"u}ppel}, \citenamefont {Vaswani}, \citenamefont {Watanabe},
  \citenamefont {Taniguchi}, \citenamefont {Mak},\ and\ \citenamefont
  {Shan}}]{zeng2023thermodynamic}%
  \BibitemOpen
  \bibfield  {author} {\bibinfo {author} {\bibfnamefont {Y.}~\bibnamefont
  {Zeng}}, \bibinfo {author} {\bibfnamefont {Z.}~\bibnamefont {Xia}}, \bibinfo
  {author} {\bibfnamefont {K.}~\bibnamefont {Kang}}, \bibinfo {author}
  {\bibfnamefont {J.}~\bibnamefont {Zhu}}, \bibinfo {author} {\bibfnamefont
  {P.}~\bibnamefont {Kn{\"u}ppel}}, \bibinfo {author} {\bibfnamefont
  {C.}~\bibnamefont {Vaswani}}, \bibinfo {author} {\bibfnamefont
  {K.}~\bibnamefont {Watanabe}}, \bibinfo {author} {\bibfnamefont
  {T.}~\bibnamefont {Taniguchi}}, \bibinfo {author} {\bibfnamefont {K.~F.}\
  \bibnamefont {Mak}},\ and\ \bibinfo {author} {\bibfnamefont {J.}~\bibnamefont
  {Shan}},\ }\bibfield  {title} {\bibinfo {title} {Thermodynamic evidence of
  fractional chern insulator in moir{\'e} $\mathrm{MoTe}_{2}$},\ }\href
  {https://doi.org/10.1038/s41586-023-06452-3} {\bibfield  {journal} {\bibinfo
  {journal} {Nature}\ }\textbf {\bibinfo {volume} {622}},\ \bibinfo {pages}
  {69} (\bibinfo {year} {2023})}\BibitemShut {NoStop}%
\bibitem [{\citenamefont {Ji}\ \emph {et~al.}(2024)\citenamefont {Ji},
  \citenamefont {Park}, \citenamefont {Barber}, \citenamefont {Hu},
  \citenamefont {Watanabe}, \citenamefont {Taniguchi}, \citenamefont {Chu},
  \citenamefont {Xu},\ and\ \citenamefont {Shen}}]{Ji_2024}%
  \BibitemOpen
  \bibfield  {author} {\bibinfo {author} {\bibfnamefont {Z.}~\bibnamefont
  {Ji}}, \bibinfo {author} {\bibfnamefont {H.}~\bibnamefont {Park}}, \bibinfo
  {author} {\bibfnamefont {M.~E.}\ \bibnamefont {Barber}}, \bibinfo {author}
  {\bibfnamefont {C.}~\bibnamefont {Hu}}, \bibinfo {author} {\bibfnamefont
  {K.}~\bibnamefont {Watanabe}}, \bibinfo {author} {\bibfnamefont
  {T.}~\bibnamefont {Taniguchi}}, \bibinfo {author} {\bibfnamefont {J.-H.}\
  \bibnamefont {Chu}}, \bibinfo {author} {\bibfnamefont {X.}~\bibnamefont
  {Xu}},\ and\ \bibinfo {author} {\bibfnamefont {Z.-X.}\ \bibnamefont {Shen}},\
  }\bibfield  {title} {\bibinfo {title} {Local probe of bulk and edge states in
  a fractional chern insulator},\ }\href
  {https://doi.org/10.1038/s41586-024-08092-7} {\bibfield  {journal} {\bibinfo
  {journal} {Nature}\ }\textbf {\bibinfo {volume} {635}},\ \bibinfo {pages}
  {578} (\bibinfo {year} {2024})}\BibitemShut {NoStop}%
\bibitem [{\citenamefont {Redekop}\ \emph {et~al.}(2024)\citenamefont
  {Redekop}, \citenamefont {Zhang}, \citenamefont {Park}, \citenamefont {Cai},
  \citenamefont {Anderson}, \citenamefont {Sheekey}, \citenamefont {Arp},
  \citenamefont {Babikyan}, \citenamefont {Salters}, \citenamefont {Watanabe},
  \citenamefont {Taniguchi}, \citenamefont {Huber}, \citenamefont {Xu},\ and\
  \citenamefont {Young}}]{Redekop_2024}%
  \BibitemOpen
  \bibfield  {author} {\bibinfo {author} {\bibfnamefont {E.}~\bibnamefont
  {Redekop}}, \bibinfo {author} {\bibfnamefont {C.}~\bibnamefont {Zhang}},
  \bibinfo {author} {\bibfnamefont {H.}~\bibnamefont {Park}}, \bibinfo {author}
  {\bibfnamefont {J.}~\bibnamefont {Cai}}, \bibinfo {author} {\bibfnamefont
  {E.}~\bibnamefont {Anderson}}, \bibinfo {author} {\bibfnamefont
  {O.}~\bibnamefont {Sheekey}}, \bibinfo {author} {\bibfnamefont
  {T.}~\bibnamefont {Arp}}, \bibinfo {author} {\bibfnamefont {G.}~\bibnamefont
  {Babikyan}}, \bibinfo {author} {\bibfnamefont {S.}~\bibnamefont {Salters}},
  \bibinfo {author} {\bibfnamefont {K.}~\bibnamefont {Watanabe}}, \bibinfo
  {author} {\bibfnamefont {T.}~\bibnamefont {Taniguchi}}, \bibinfo {author}
  {\bibfnamefont {M.~E.}\ \bibnamefont {Huber}}, \bibinfo {author}
  {\bibfnamefont {X.}~\bibnamefont {Xu}},\ and\ \bibinfo {author}
  {\bibfnamefont {A.~F.}\ \bibnamefont {Young}},\ }\bibfield  {title} {\bibinfo
  {title} {Direct magnetic imaging of fractional chern insulators in twisted
  $\mathrm{MoTe}_{2}$},\ }\href {https://doi.org/10.1038/s41586-024-08153-x}
  {\bibfield  {journal} {\bibinfo  {journal} {Nature}\ }\textbf {\bibinfo
  {volume} {635}},\ \bibinfo {pages} {584} (\bibinfo {year}
  {2024})}\BibitemShut {NoStop}%
\bibitem [{\citenamefont {Holtzmann}\ \emph {et~al.}()\citenamefont
  {Holtzmann}, \citenamefont {Li}, \citenamefont {Anderson}, \citenamefont
  {Cai}, \citenamefont {Park}, \citenamefont {Hu}, \citenamefont {Taniguchi},
  \citenamefont {Watanabe}, \citenamefont {Chu}, \citenamefont {Xiao},
  \citenamefont {Cao},\ and\ \citenamefont
  {Xu}}]{holtzmann2025opticalcontrolintegerfractional}%
  \BibitemOpen
  \bibfield  {author} {\bibinfo {author} {\bibfnamefont {W.}~\bibnamefont
  {Holtzmann}}, \bibinfo {author} {\bibfnamefont {W.}~\bibnamefont {Li}},
  \bibinfo {author} {\bibfnamefont {E.}~\bibnamefont {Anderson}}, \bibinfo
  {author} {\bibfnamefont {J.}~\bibnamefont {Cai}}, \bibinfo {author}
  {\bibfnamefont {H.}~\bibnamefont {Park}}, \bibinfo {author} {\bibfnamefont
  {C.}~\bibnamefont {Hu}}, \bibinfo {author} {\bibfnamefont {T.}~\bibnamefont
  {Taniguchi}}, \bibinfo {author} {\bibfnamefont {K.}~\bibnamefont {Watanabe}},
  \bibinfo {author} {\bibfnamefont {J.-H.}\ \bibnamefont {Chu}}, \bibinfo
  {author} {\bibfnamefont {D.}~\bibnamefont {Xiao}}, \bibinfo {author}
  {\bibfnamefont {T.}~\bibnamefont {Cao}},\ and\ \bibinfo {author}
  {\bibfnamefont {X.}~\bibnamefont {Xu}},\ }\href@noop {} {\bibinfo {title}
  {Optical control of integer and fractional chern insulators}},\ \Eprint
  {https://arxiv.org/abs/2508.18639} {arXiv:2508.18639} \BibitemShut {NoStop}%
\bibitem [{\citenamefont {Park}\ \emph {et~al.}(2025)\citenamefont {Park},
  \citenamefont {Cai}, \citenamefont {Anderson}, \citenamefont {Zhang},
  \citenamefont {Liu}, \citenamefont {Holtzmann}, \citenamefont {Li},
  \citenamefont {Wang}, \citenamefont {Hu}, \citenamefont {Zhao}, \citenamefont
  {Taniguchi}, \citenamefont {Watanabe}, \citenamefont {Yang}, \citenamefont
  {Cobden}, \citenamefont {Chu}, \citenamefont {Regnault}, \citenamefont
  {Bernevig}, \citenamefont {Fu}, \citenamefont {Cao}, \citenamefont {Xiao},\
  and\ \citenamefont {Xu}}]{Park_2025}%
  \BibitemOpen
  \bibfield  {author} {\bibinfo {author} {\bibfnamefont {H.}~\bibnamefont
  {Park}}, \bibinfo {author} {\bibfnamefont {J.}~\bibnamefont {Cai}}, \bibinfo
  {author} {\bibfnamefont {E.}~\bibnamefont {Anderson}}, \bibinfo {author}
  {\bibfnamefont {X.-W.}\ \bibnamefont {Zhang}}, \bibinfo {author}
  {\bibfnamefont {X.}~\bibnamefont {Liu}}, \bibinfo {author} {\bibfnamefont
  {W.}~\bibnamefont {Holtzmann}}, \bibinfo {author} {\bibfnamefont
  {W.}~\bibnamefont {Li}}, \bibinfo {author} {\bibfnamefont {C.}~\bibnamefont
  {Wang}}, \bibinfo {author} {\bibfnamefont {C.}~\bibnamefont {Hu}}, \bibinfo
  {author} {\bibfnamefont {Y.}~\bibnamefont {Zhao}}, \bibinfo {author}
  {\bibfnamefont {T.}~\bibnamefont {Taniguchi}}, \bibinfo {author}
  {\bibfnamefont {K.}~\bibnamefont {Watanabe}}, \bibinfo {author}
  {\bibfnamefont {J.}~\bibnamefont {Yang}}, \bibinfo {author} {\bibfnamefont
  {D.}~\bibnamefont {Cobden}}, \bibinfo {author} {\bibfnamefont {J.-h.}\
  \bibnamefont {Chu}}, \bibinfo {author} {\bibfnamefont {N.}~\bibnamefont
  {Regnault}}, \bibinfo {author} {\bibfnamefont {B.~A.}\ \bibnamefont
  {Bernevig}}, \bibinfo {author} {\bibfnamefont {L.}~\bibnamefont {Fu}},
  \bibinfo {author} {\bibfnamefont {T.}~\bibnamefont {Cao}}, \bibinfo {author}
  {\bibfnamefont {D.}~\bibnamefont {Xiao}},\ and\ \bibinfo {author}
  {\bibfnamefont {X.}~\bibnamefont {Xu}},\ }\bibfield  {title} {\bibinfo
  {title} {Ferromagnetism and topology of the higher flat band in a fractional
  chern insulator},\ }\href {https://doi.org/10.1038/s41567-025-02804-0}
  {\bibfield  {journal} {\bibinfo  {journal} {Nat. Phys.}\ }\textbf {\bibinfo
  {volume} {21}},\ \bibinfo {pages} {549} (\bibinfo {year} {2025})}\BibitemShut
  {NoStop}%
\bibitem [{\citenamefont {Xu}\ \emph {et~al.}(2025{\natexlab{a}})\citenamefont
  {Xu}, \citenamefont {Chang}, \citenamefont {Xiao}, \citenamefont {Zhang},
  \citenamefont {Liu}, \citenamefont {Sun}, \citenamefont {Mao}, \citenamefont
  {Peshcherenko}, \citenamefont {Li}, \citenamefont {Watanabe}, \citenamefont
  {Taniguchi}, \citenamefont {Tong}, \citenamefont {Lu}, \citenamefont {Jia},
  \citenamefont {Qian}, \citenamefont {Shi}, \citenamefont {Zhang},
  \citenamefont {Liu}, \citenamefont {Jiang},\ and\ \citenamefont
  {Li}}]{Xu_2025}%
  \BibitemOpen
  \bibfield  {author} {\bibinfo {author} {\bibfnamefont {F.}~\bibnamefont
  {Xu}}, \bibinfo {author} {\bibfnamefont {X.}~\bibnamefont {Chang}}, \bibinfo
  {author} {\bibfnamefont {J.}~\bibnamefont {Xiao}}, \bibinfo {author}
  {\bibfnamefont {Y.}~\bibnamefont {Zhang}}, \bibinfo {author} {\bibfnamefont
  {F.}~\bibnamefont {Liu}}, \bibinfo {author} {\bibfnamefont {Z.}~\bibnamefont
  {Sun}}, \bibinfo {author} {\bibfnamefont {N.}~\bibnamefont {Mao}}, \bibinfo
  {author} {\bibfnamefont {N.}~\bibnamefont {Peshcherenko}}, \bibinfo {author}
  {\bibfnamefont {J.}~\bibnamefont {Li}}, \bibinfo {author} {\bibfnamefont
  {K.}~\bibnamefont {Watanabe}}, \bibinfo {author} {\bibfnamefont
  {T.}~\bibnamefont {Taniguchi}}, \bibinfo {author} {\bibfnamefont
  {B.}~\bibnamefont {Tong}}, \bibinfo {author} {\bibfnamefont {L.}~\bibnamefont
  {Lu}}, \bibinfo {author} {\bibfnamefont {J.}~\bibnamefont {Jia}}, \bibinfo
  {author} {\bibfnamefont {D.}~\bibnamefont {Qian}}, \bibinfo {author}
  {\bibfnamefont {Z.}~\bibnamefont {Shi}}, \bibinfo {author} {\bibfnamefont
  {Y.}~\bibnamefont {Zhang}}, \bibinfo {author} {\bibfnamefont
  {X.}~\bibnamefont {Liu}}, \bibinfo {author} {\bibfnamefont {S.}~\bibnamefont
  {Jiang}},\ and\ \bibinfo {author} {\bibfnamefont {T.}~\bibnamefont {Li}},\
  }\bibfield  {title} {\bibinfo {title} {Interplay between topology and
  correlations in the second moiré band of twisted bilayer
  $\mathrm{MoTe}_{2}$},\ }\href {https://doi.org/10.1038/s41567-025-02803-1}
  {\bibfield  {journal} {\bibinfo  {journal} {Nat. Phys.}\ }\textbf {\bibinfo
  {volume} {21}},\ \bibinfo {pages} {542} (\bibinfo {year}
  {2025}{\natexlab{a}})}\BibitemShut {NoStop}%
\bibitem [{\citenamefont {Chang}\ \emph {et~al.}()\citenamefont {Chang},
  \citenamefont {Liu}, \citenamefont {Xu}, \citenamefont {Xu}, \citenamefont
  {Xiao}, \citenamefont {Sun}, \citenamefont {Jiao}, \citenamefont {Zhang},
  \citenamefont {Wang}, \citenamefont {Shen}, \citenamefont {He}, \citenamefont
  {Watanabe}, \citenamefont {Taniguchi}, \citenamefont {Zhong}, \citenamefont
  {Jia}, \citenamefont {Shi}, \citenamefont {Liu}, \citenamefont {Zhang},
  \citenamefont {Qian}, \citenamefont {Li},\ and\ \citenamefont
  {Jiang}}]{chang2025evidencecompetinggroundstates}%
  \BibitemOpen
  \bibfield  {author} {\bibinfo {author} {\bibfnamefont {X.}~\bibnamefont
  {Chang}}, \bibinfo {author} {\bibfnamefont {F.}~\bibnamefont {Liu}}, \bibinfo
  {author} {\bibfnamefont {F.}~\bibnamefont {Xu}}, \bibinfo {author}
  {\bibfnamefont {C.}~\bibnamefont {Xu}}, \bibinfo {author} {\bibfnamefont
  {J.}~\bibnamefont {Xiao}}, \bibinfo {author} {\bibfnamefont {Z.}~\bibnamefont
  {Sun}}, \bibinfo {author} {\bibfnamefont {P.}~\bibnamefont {Jiao}}, \bibinfo
  {author} {\bibfnamefont {Y.}~\bibnamefont {Zhang}}, \bibinfo {author}
  {\bibfnamefont {S.}~\bibnamefont {Wang}}, \bibinfo {author} {\bibfnamefont
  {B.}~\bibnamefont {Shen}}, \bibinfo {author} {\bibfnamefont {R.}~\bibnamefont
  {He}}, \bibinfo {author} {\bibfnamefont {K.}~\bibnamefont {Watanabe}},
  \bibinfo {author} {\bibfnamefont {T.}~\bibnamefont {Taniguchi}}, \bibinfo
  {author} {\bibfnamefont {R.}~\bibnamefont {Zhong}}, \bibinfo {author}
  {\bibfnamefont {J.}~\bibnamefont {Jia}}, \bibinfo {author} {\bibfnamefont
  {Z.}~\bibnamefont {Shi}}, \bibinfo {author} {\bibfnamefont {X.}~\bibnamefont
  {Liu}}, \bibinfo {author} {\bibfnamefont {Y.}~\bibnamefont {Zhang}}, \bibinfo
  {author} {\bibfnamefont {D.}~\bibnamefont {Qian}}, \bibinfo {author}
  {\bibfnamefont {T.}~\bibnamefont {Li}},\ and\ \bibinfo {author}
  {\bibfnamefont {S.}~\bibnamefont {Jiang}},\ }\href@noop {} {\bibinfo {title}
  {Evidence of competing ground states between fractional chern insulator and
  antiferromagnetism in moir\'e $\mathrm{MoTe}_{2}$}},\ \Eprint
  {https://arxiv.org/abs/2503.13213} {arXiv:2503.13213} \BibitemShut {NoStop}%
\bibitem [{\citenamefont {Adak}\ \emph {et~al.}(2024)\citenamefont {Adak},
  \citenamefont {Sinha}, \citenamefont {Agarwal},\ and\ \citenamefont
  {Deshmukh}}]{Adak_2024}%
  \BibitemOpen
  \bibfield  {author} {\bibinfo {author} {\bibfnamefont {P.~C.}\ \bibnamefont
  {Adak}}, \bibinfo {author} {\bibfnamefont {S.}~\bibnamefont {Sinha}},
  \bibinfo {author} {\bibfnamefont {A.}~\bibnamefont {Agarwal}},\ and\ \bibinfo
  {author} {\bibfnamefont {M.~M.}\ \bibnamefont {Deshmukh}},\ }\bibfield
  {title} {\bibinfo {title} {Tunable moiré materials for probing berry physics
  and topology},\ }\href {https://doi.org/10.1038/s41578-024-00671-4}
  {\bibfield  {journal} {\bibinfo  {journal} {Nat. Rev. Mater.}\ }\textbf
  {\bibinfo {volume} {9}},\ \bibinfo {pages} {481} (\bibinfo {year}
  {2024})}\BibitemShut {NoStop}%
\bibitem [{\citenamefont {Wang}\ \emph
  {et~al.}(2024{\natexlab{c}})\citenamefont {Wang}, \citenamefont {Zhang},
  \citenamefont {Liu}, \citenamefont {He}, \citenamefont {Xu}, \citenamefont
  {Ran}, \citenamefont {Cao},\ and\ \citenamefont
  {Xiao}}]{PhysRevLett.132.036501}%
  \BibitemOpen
  \bibfield  {author} {\bibinfo {author} {\bibfnamefont {C.}~\bibnamefont
  {Wang}}, \bibinfo {author} {\bibfnamefont {X.-W.}\ \bibnamefont {Zhang}},
  \bibinfo {author} {\bibfnamefont {X.}~\bibnamefont {Liu}}, \bibinfo {author}
  {\bibfnamefont {Y.}~\bibnamefont {He}}, \bibinfo {author} {\bibfnamefont
  {X.}~\bibnamefont {Xu}}, \bibinfo {author} {\bibfnamefont {Y.}~\bibnamefont
  {Ran}}, \bibinfo {author} {\bibfnamefont {T.}~\bibnamefont {Cao}},\ and\
  \bibinfo {author} {\bibfnamefont {D.}~\bibnamefont {Xiao}},\ }\bibfield
  {title} {\bibinfo {title} {Fractional chern insulator in twisted bilayer
  $\mathrm{MoTe}_{2}$},\ }\href
  {https://doi.org/10.1103/PhysRevLett.132.036501} {\bibfield  {journal}
  {\bibinfo  {journal} {Phys. Rev. Lett.}\ }\textbf {\bibinfo {volume} {132}},\
  \bibinfo {pages} {036501} (\bibinfo {year} {2024}{\natexlab{c}})}\BibitemShut
  {NoStop}%
\bibitem [{\citenamefont {Reddy}\ \emph {et~al.}(2023)\citenamefont {Reddy},
  \citenamefont {Alsallom}, \citenamefont {Zhang}, \citenamefont {Devakul},\
  and\ \citenamefont {Fu}}]{PhysRevB.108.085117}%
  \BibitemOpen
  \bibfield  {author} {\bibinfo {author} {\bibfnamefont {A.~P.}\ \bibnamefont
  {Reddy}}, \bibinfo {author} {\bibfnamefont {F.}~\bibnamefont {Alsallom}},
  \bibinfo {author} {\bibfnamefont {Y.}~\bibnamefont {Zhang}}, \bibinfo
  {author} {\bibfnamefont {T.}~\bibnamefont {Devakul}},\ and\ \bibinfo {author}
  {\bibfnamefont {L.}~\bibnamefont {Fu}},\ }\bibfield  {title} {\bibinfo
  {title} {Fractional quantum anomalous hall states in twisted bilayer
  $\mathrm{MoTe}_{2}$ and $\mathrm{WSe}_{2}$},\ }\href
  {https://doi.org/10.1103/PhysRevB.108.085117} {\bibfield  {journal} {\bibinfo
   {journal} {Phys. Rev. B}\ }\textbf {\bibinfo {volume} {108}},\ \bibinfo
  {pages} {085117} (\bibinfo {year} {2023})}\BibitemShut {NoStop}%
\bibitem [{\citenamefont {Reddy}\ and\ \citenamefont
  {Fu}(2023)}]{PhysRevB.108.245159}%
  \BibitemOpen
  \bibfield  {author} {\bibinfo {author} {\bibfnamefont {A.~P.}\ \bibnamefont
  {Reddy}}\ and\ \bibinfo {author} {\bibfnamefont {L.}~\bibnamefont {Fu}},\
  }\bibfield  {title} {\bibinfo {title} {Toward a global phase diagram of the
  fractional quantum anomalous hall effect},\ }\href
  {https://doi.org/10.1103/PhysRevB.108.245159} {\bibfield  {journal} {\bibinfo
   {journal} {Phys. Rev. B}\ }\textbf {\bibinfo {volume} {108}},\ \bibinfo
  {pages} {245159} (\bibinfo {year} {2023})}\BibitemShut {NoStop}%
\bibitem [{\citenamefont {Cr\'epel}\ and\ \citenamefont
  {Fu}(2023)}]{PhysRevB.107.L201109}%
  \BibitemOpen
  \bibfield  {author} {\bibinfo {author} {\bibfnamefont {V.}~\bibnamefont
  {Cr\'epel}}\ and\ \bibinfo {author} {\bibfnamefont {L.}~\bibnamefont {Fu}},\
  }\bibfield  {title} {\bibinfo {title} {Anomalous hall metal and fractional
  chern insulator in twisted transition metal dichalcogenides},\ }\href
  {https://doi.org/10.1103/PhysRevB.107.L201109} {\bibfield  {journal}
  {\bibinfo  {journal} {Phys. Rev. B}\ }\textbf {\bibinfo {volume} {107}},\
  \bibinfo {pages} {L201109} (\bibinfo {year} {2023})}\BibitemShut {NoStop}%
\bibitem [{\citenamefont {Li}\ \emph {et~al.}(2021)\citenamefont {Li},
  \citenamefont {Kumar}, \citenamefont {Sun},\ and\ \citenamefont
  {Lin}}]{PhysRevResearch.3.L032070}%
  \BibitemOpen
  \bibfield  {author} {\bibinfo {author} {\bibfnamefont {H.}~\bibnamefont
  {Li}}, \bibinfo {author} {\bibfnamefont {U.}~\bibnamefont {Kumar}}, \bibinfo
  {author} {\bibfnamefont {K.}~\bibnamefont {Sun}},\ and\ \bibinfo {author}
  {\bibfnamefont {S.-Z.}\ \bibnamefont {Lin}},\ }\bibfield  {title} {\bibinfo
  {title} {Spontaneous fractional chern insulators in transition metal
  dichalcogenide moir\'e superlattices},\ }\href
  {https://doi.org/10.1103/PhysRevResearch.3.L032070} {\bibfield  {journal}
  {\bibinfo  {journal} {Phys. Rev. Res.}\ }\textbf {\bibinfo {volume} {3}},\
  \bibinfo {pages} {L032070} (\bibinfo {year} {2021})}\BibitemShut {NoStop}%
\bibitem [{\citenamefont {Lu}\ and\ \citenamefont
  {Santos}(2024)}]{PhysRevLett.133.186602}%
  \BibitemOpen
  \bibfield  {author} {\bibinfo {author} {\bibfnamefont {T.}~\bibnamefont
  {Lu}}\ and\ \bibinfo {author} {\bibfnamefont {L.~H.}\ \bibnamefont
  {Santos}},\ }\bibfield  {title} {\bibinfo {title} {Fractional chern
  insulators in twisted bilayer $\mathrm{MoTe}_{2}$: A composite fermion
  perspective},\ }\href {https://doi.org/10.1103/PhysRevLett.133.186602}
  {\bibfield  {journal} {\bibinfo  {journal} {Phys. Rev. Lett.}\ }\textbf
  {\bibinfo {volume} {133}},\ \bibinfo {pages} {186602} (\bibinfo {year}
  {2024})}\BibitemShut {NoStop}%
\bibitem [{\citenamefont {Chen}\ \emph {et~al.}(2025)\citenamefont {Chen},
  \citenamefont {Luo}, \citenamefont {Zhu},\ and\ \citenamefont
  {Sheng}}]{Chen_2025}%
  \BibitemOpen
  \bibfield  {author} {\bibinfo {author} {\bibfnamefont {F.}~\bibnamefont
  {Chen}}, \bibinfo {author} {\bibfnamefont {W.-W.}\ \bibnamefont {Luo}},
  \bibinfo {author} {\bibfnamefont {W.}~\bibnamefont {Zhu}},\ and\ \bibinfo
  {author} {\bibfnamefont {D.~N.}\ \bibnamefont {Sheng}},\ }\bibfield  {title}
  {\bibinfo {title} {Robust non-abelian even-denominator fractional chern
  insulator in twisted bilayer $\mathrm{MoTe}_{2}$},\ }\href
  {http://dx.doi.org/10.1038/s41467-025-57326-3} {\bibfield  {journal}
  {\bibinfo  {journal} {Nat. Commun.}\ }\textbf {\bibinfo {volume} {16}}
  (\bibinfo {year} {2025})}\BibitemShut {NoStop}%
\bibitem [{\citenamefont {Xu}\ \emph {et~al.}(2024)\citenamefont {Xu},
  \citenamefont {Li}, \citenamefont {Xu}, \citenamefont {Bi},\ and\
  \citenamefont {Zhang}}]{doi:10.1073/pnas.2316749121}%
  \BibitemOpen
  \bibfield  {author} {\bibinfo {author} {\bibfnamefont {C.}~\bibnamefont
  {Xu}}, \bibinfo {author} {\bibfnamefont {J.}~\bibnamefont {Li}}, \bibinfo
  {author} {\bibfnamefont {Y.}~\bibnamefont {Xu}}, \bibinfo {author}
  {\bibfnamefont {Z.}~\bibnamefont {Bi}},\ and\ \bibinfo {author}
  {\bibfnamefont {Y.}~\bibnamefont {Zhang}},\ }\bibfield  {title} {\bibinfo
  {title} {Maximally localized wannier functions, interaction models, and
  fractional quantum anomalous hall effect in twisted bilayer
  $\mathrm{MoTe}_{2}$},\ }\href {https://doi.org/10.1073/pnas.2316749121}
  {\bibfield  {journal} {\bibinfo  {journal} {Proc. Natl. Acad. Sci. U.S.A.}\
  }\textbf {\bibinfo {volume} {121}},\ \bibinfo {pages} {e2316749121} (\bibinfo
  {year} {2024})}\BibitemShut {NoStop}%
\bibitem [{\citenamefont {Pichler}\ \emph {et~al.}(2025)\citenamefont
  {Pichler}, \citenamefont {Kadow}, \citenamefont {Kuhlenkamp},\ and\
  \citenamefont {Knap}}]{PhysRevB.111.075108}%
  \BibitemOpen
  \bibfield  {author} {\bibinfo {author} {\bibfnamefont {F.}~\bibnamefont
  {Pichler}}, \bibinfo {author} {\bibfnamefont {W.}~\bibnamefont {Kadow}},
  \bibinfo {author} {\bibfnamefont {C.}~\bibnamefont {Kuhlenkamp}},\ and\
  \bibinfo {author} {\bibfnamefont {M.}~\bibnamefont {Knap}},\ }\bibfield
  {title} {\bibinfo {title} {Single-particle spectral function of fractional
  quantum anomalous hall states},\ }\href
  {https://doi.org/10.1103/PhysRevB.111.075108} {\bibfield  {journal} {\bibinfo
   {journal} {Phys. Rev. B}\ }\textbf {\bibinfo {volume} {111}},\ \bibinfo
  {pages} {075108} (\bibinfo {year} {2025})}\BibitemShut {NoStop}%
\bibitem [{\citenamefont {Xu}\ \emph {et~al.}(2025{\natexlab{b}})\citenamefont
  {Xu}, \citenamefont {Mao}, \citenamefont {Zeng},\ and\ \citenamefont
  {Zhang}}]{PhysRevLett.134.066601}%
  \BibitemOpen
  \bibfield  {author} {\bibinfo {author} {\bibfnamefont {C.}~\bibnamefont
  {Xu}}, \bibinfo {author} {\bibfnamefont {N.}~\bibnamefont {Mao}}, \bibinfo
  {author} {\bibfnamefont {T.}~\bibnamefont {Zeng}},\ and\ \bibinfo {author}
  {\bibfnamefont {Y.}~\bibnamefont {Zhang}},\ }\bibfield  {title} {\bibinfo
  {title} {Multiple chern bands in twisted $\mathrm{MoTe}_{2}$ and possible
  non-abelian states},\ }\href {https://doi.org/10.1103/PhysRevLett.134.066601}
  {\bibfield  {journal} {\bibinfo  {journal} {Phys. Rev. Lett.}\ }\textbf
  {\bibinfo {volume} {134}},\ \bibinfo {pages} {066601} (\bibinfo {year}
  {2025}{\natexlab{b}})}\BibitemShut {NoStop}%
\bibitem [{\citenamefont {Morales-Dur\'an}\ \emph {et~al.}(2024)\citenamefont
  {Morales-Dur\'an}, \citenamefont {Wei}, \citenamefont {Shi},\ and\
  \citenamefont {MacDonald}}]{PhysRevLett.132.096602}%
  \BibitemOpen
  \bibfield  {author} {\bibinfo {author} {\bibfnamefont {N.}~\bibnamefont
  {Morales-Dur\'an}}, \bibinfo {author} {\bibfnamefont {N.}~\bibnamefont
  {Wei}}, \bibinfo {author} {\bibfnamefont {J.}~\bibnamefont {Shi}},\ and\
  \bibinfo {author} {\bibfnamefont {A.~H.}\ \bibnamefont {MacDonald}},\
  }\bibfield  {title} {\bibinfo {title} {Magic angles and fractional chern
  insulators in twisted homobilayer transition metal dichalcogenides},\ }\href
  {https://doi.org/10.1103/PhysRevLett.132.096602} {\bibfield  {journal}
  {\bibinfo  {journal} {Phys. Rev. Lett.}\ }\textbf {\bibinfo {volume} {132}},\
  \bibinfo {pages} {096602} (\bibinfo {year} {2024})}\BibitemShut {NoStop}%
\bibitem [{\citenamefont {Yu}\ \emph {et~al.}(2024)\citenamefont {Yu},
  \citenamefont {Herzog-Arbeitman}, \citenamefont {Wang}, \citenamefont
  {Vafek}, \citenamefont {Bernevig},\ and\ \citenamefont
  {Regnault}}]{PhysRevB.109.045147}%
  \BibitemOpen
  \bibfield  {author} {\bibinfo {author} {\bibfnamefont {J.}~\bibnamefont
  {Yu}}, \bibinfo {author} {\bibfnamefont {J.}~\bibnamefont
  {Herzog-Arbeitman}}, \bibinfo {author} {\bibfnamefont {M.}~\bibnamefont
  {Wang}}, \bibinfo {author} {\bibfnamefont {O.}~\bibnamefont {Vafek}},
  \bibinfo {author} {\bibfnamefont {B.~A.}\ \bibnamefont {Bernevig}},\ and\
  \bibinfo {author} {\bibfnamefont {N.}~\bibnamefont {Regnault}},\ }\bibfield
  {title} {\bibinfo {title} {Fractional chern insulators versus nonmagnetic
  states in twisted bilayer $\mathrm{MoTe}_{2}$},\ }\href
  {https://doi.org/10.1103/PhysRevB.109.045147} {\bibfield  {journal} {\bibinfo
   {journal} {Phys. Rev. B}\ }\textbf {\bibinfo {volume} {109}},\ \bibinfo
  {pages} {045147} (\bibinfo {year} {2024})}\BibitemShut {NoStop}%
\bibitem [{\citenamefont {Lu}\ \emph {et~al.}(2024)\citenamefont {Lu},
  \citenamefont {Han}, \citenamefont {Yao}, \citenamefont {Reddy},
  \citenamefont {Yang}, \citenamefont {Seo}, \citenamefont {Watanabe},
  \citenamefont {Taniguchi}, \citenamefont {Fu},\ and\ \citenamefont
  {Ju}}]{Lu_2024}%
  \BibitemOpen
  \bibfield  {author} {\bibinfo {author} {\bibfnamefont {Z.}~\bibnamefont
  {Lu}}, \bibinfo {author} {\bibfnamefont {T.}~\bibnamefont {Han}}, \bibinfo
  {author} {\bibfnamefont {Y.}~\bibnamefont {Yao}}, \bibinfo {author}
  {\bibfnamefont {A.~P.}\ \bibnamefont {Reddy}}, \bibinfo {author}
  {\bibfnamefont {J.}~\bibnamefont {Yang}}, \bibinfo {author} {\bibfnamefont
  {J.}~\bibnamefont {Seo}}, \bibinfo {author} {\bibfnamefont {K.}~\bibnamefont
  {Watanabe}}, \bibinfo {author} {\bibfnamefont {T.}~\bibnamefont {Taniguchi}},
  \bibinfo {author} {\bibfnamefont {L.}~\bibnamefont {Fu}},\ and\ \bibinfo
  {author} {\bibfnamefont {L.}~\bibnamefont {Ju}},\ }\bibfield  {title}
  {\bibinfo {title} {Fractional quantum anomalous hall effect in multilayer
  graphene},\ }\href {https://doi.org/10.1038/s41586-023-07010-7} {\bibfield
  {journal} {\bibinfo  {journal} {Nature}\ }\textbf {\bibinfo {volume} {626}},\
  \bibinfo {pages} {759} (\bibinfo {year} {2024})}\BibitemShut {NoStop}%
\bibitem [{\citenamefont {Xie}\ \emph {et~al.}(2025)\citenamefont {Xie},
  \citenamefont {Huo}, \citenamefont {Lu}, \citenamefont {Feng}, \citenamefont
  {Zhang}, \citenamefont {Wang}, \citenamefont {Yang}, \citenamefont
  {Watanabe}, \citenamefont {Taniguchi}, \citenamefont {Liu}, \citenamefont
  {Song}, \citenamefont {Xie}, \citenamefont {Liu},\ and\ \citenamefont
  {Lu}}]{Xie_2025}%
  \BibitemOpen
  \bibfield  {author} {\bibinfo {author} {\bibfnamefont {J.}~\bibnamefont
  {Xie}}, \bibinfo {author} {\bibfnamefont {Z.}~\bibnamefont {Huo}}, \bibinfo
  {author} {\bibfnamefont {X.}~\bibnamefont {Lu}}, \bibinfo {author}
  {\bibfnamefont {Z.}~\bibnamefont {Feng}}, \bibinfo {author} {\bibfnamefont
  {Z.}~\bibnamefont {Zhang}}, \bibinfo {author} {\bibfnamefont
  {W.}~\bibnamefont {Wang}}, \bibinfo {author} {\bibfnamefont {Q.}~\bibnamefont
  {Yang}}, \bibinfo {author} {\bibfnamefont {K.}~\bibnamefont {Watanabe}},
  \bibinfo {author} {\bibfnamefont {T.}~\bibnamefont {Taniguchi}}, \bibinfo
  {author} {\bibfnamefont {K.}~\bibnamefont {Liu}}, \bibinfo {author}
  {\bibfnamefont {Z.}~\bibnamefont {Song}}, \bibinfo {author} {\bibfnamefont
  {X.~C.}\ \bibnamefont {Xie}}, \bibinfo {author} {\bibfnamefont
  {J.}~\bibnamefont {Liu}},\ and\ \bibinfo {author} {\bibfnamefont
  {X.}~\bibnamefont {Lu}},\ }\bibfield  {title} {\bibinfo {title} {Tunable
  fractional chern insulators in rhombohedral graphene superlattices},\ }\href
  {https://doi.org/10.1038/s41563-025-02225-7} {\bibfield  {journal} {\bibinfo
  {journal} {Nat. Mater.}\ }\textbf {\bibinfo {volume} {24}},\ \bibinfo {pages}
  {1042} (\bibinfo {year} {2025})}\BibitemShut {NoStop}%
\bibitem [{\citenamefont {Xie}\ \emph {et~al.}(2021)\citenamefont {Xie},
  \citenamefont {Pierce}, \citenamefont {Park}, \citenamefont {Parker},
  \citenamefont {Khalaf}, \citenamefont {Ledwith}, \citenamefont {Cao},
  \citenamefont {Lee}, \citenamefont {Chen}, \citenamefont {Forrester} \emph
  {et~al.}}]{xie2021fractional}%
  \BibitemOpen
  \bibfield  {author} {\bibinfo {author} {\bibfnamefont {Y.}~\bibnamefont
  {Xie}}, \bibinfo {author} {\bibfnamefont {A.~T.}\ \bibnamefont {Pierce}},
  \bibinfo {author} {\bibfnamefont {J.~M.}\ \bibnamefont {Park}}, \bibinfo
  {author} {\bibfnamefont {D.~E.}\ \bibnamefont {Parker}}, \bibinfo {author}
  {\bibfnamefont {E.}~\bibnamefont {Khalaf}}, \bibinfo {author} {\bibfnamefont
  {P.}~\bibnamefont {Ledwith}}, \bibinfo {author} {\bibfnamefont
  {Y.}~\bibnamefont {Cao}}, \bibinfo {author} {\bibfnamefont {S.~H.}\
  \bibnamefont {Lee}}, \bibinfo {author} {\bibfnamefont {S.}~\bibnamefont
  {Chen}}, \bibinfo {author} {\bibfnamefont {P.~R.}\ \bibnamefont {Forrester}},
  \emph {et~al.},\ }\bibfield  {title} {\bibinfo {title} {Fractional chern
  insulators in magic-angle twisted bilayer graphene},\ }\href
  {https://doi.org/10.1038/s41586-021-04002-3} {\bibfield  {journal} {\bibinfo
  {journal} {Nature}\ }\textbf {\bibinfo {volume} {600}},\ \bibinfo {pages}
  {439} (\bibinfo {year} {2021})}\BibitemShut {NoStop}%
\bibitem [{\citenamefont {Ma}\ \emph {et~al.}(2024)\citenamefont {Ma},
  \citenamefont {Huang}, \citenamefont {Zhu}, \citenamefont {Chen},\ and\
  \citenamefont {Yao}}]{PhysRevB.110.165142}%
  \BibitemOpen
  \bibfield  {author} {\bibinfo {author} {\bibfnamefont {J.-Z.}\ \bibnamefont
  {Ma}}, \bibinfo {author} {\bibfnamefont {R.-Z.}\ \bibnamefont {Huang}},
  \bibinfo {author} {\bibfnamefont {G.-Y.}\ \bibnamefont {Zhu}}, \bibinfo
  {author} {\bibfnamefont {J.-Y.}\ \bibnamefont {Chen}},\ and\ \bibinfo
  {author} {\bibfnamefont {D.-X.}\ \bibnamefont {Yao}},\ }\bibfield  {title}
  {\bibinfo {title} {Fractional chern insulator candidate on a twisted bilayer
  checkerboard lattice},\ }\href {https://doi.org/10.1103/PhysRevB.110.165142}
  {\bibfield  {journal} {\bibinfo  {journal} {Phys. Rev. B}\ }\textbf {\bibinfo
  {volume} {110}},\ \bibinfo {pages} {165142} (\bibinfo {year}
  {2024})}\BibitemShut {NoStop}%
\bibitem [{\citenamefont {Shavit}\ and\ \citenamefont
  {Oreg}(2024)}]{PhysRevLett.133.156504}%
  \BibitemOpen
  \bibfield  {author} {\bibinfo {author} {\bibfnamefont {G.}~\bibnamefont
  {Shavit}}\ and\ \bibinfo {author} {\bibfnamefont {Y.}~\bibnamefont {Oreg}},\
  }\bibfield  {title} {\bibinfo {title} {Quantum geometry and stabilization of
  fractional chern insulators far from the ideal limit},\ }\href
  {https://doi.org/10.1103/PhysRevLett.133.156504} {\bibfield  {journal}
  {\bibinfo  {journal} {Phys. Rev. Lett.}\ }\textbf {\bibinfo {volume} {133}},\
  \bibinfo {pages} {156504} (\bibinfo {year} {2024})}\BibitemShut {NoStop}%
\bibitem [{\citenamefont {Devakul}\ \emph {et~al.}(2023)\citenamefont
  {Devakul}, \citenamefont {Ledwith}, \citenamefont {Xia}, \citenamefont {Uri},
  \citenamefont {de~la Barrera}, \citenamefont {Jarillo-Herrero},\ and\
  \citenamefont {Fu}}]{doi:10.1126/sciadv.adi6063}%
  \BibitemOpen
  \bibfield  {author} {\bibinfo {author} {\bibfnamefont {T.}~\bibnamefont
  {Devakul}}, \bibinfo {author} {\bibfnamefont {P.~J.}\ \bibnamefont
  {Ledwith}}, \bibinfo {author} {\bibfnamefont {L.-Q.}\ \bibnamefont {Xia}},
  \bibinfo {author} {\bibfnamefont {A.}~\bibnamefont {Uri}}, \bibinfo {author}
  {\bibfnamefont {S.~C.}\ \bibnamefont {de~la Barrera}}, \bibinfo {author}
  {\bibfnamefont {P.}~\bibnamefont {Jarillo-Herrero}},\ and\ \bibinfo {author}
  {\bibfnamefont {L.}~\bibnamefont {Fu}},\ }\bibfield  {title} {\bibinfo
  {title} {Magic-angle helical trilayer graphene},\ }\href
  {https://doi.org/10.1126/sciadv.adi6063} {\bibfield  {journal} {\bibinfo
  {journal} {Sci. Adv.}\ }\textbf {\bibinfo {volume} {9}},\ \bibinfo {pages}
  {eadi6063} (\bibinfo {year} {2023})}\BibitemShut {NoStop}%
\bibitem [{\citenamefont {Dong}\ \emph
  {et~al.}(2023{\natexlab{a}})\citenamefont {Dong}, \citenamefont {Wang},
  \citenamefont {Ledwith}, \citenamefont {Vishwanath},\ and\ \citenamefont
  {Parker}}]{PhysRevLett.131.136502}%
  \BibitemOpen
  \bibfield  {author} {\bibinfo {author} {\bibfnamefont {J.}~\bibnamefont
  {Dong}}, \bibinfo {author} {\bibfnamefont {J.}~\bibnamefont {Wang}}, \bibinfo
  {author} {\bibfnamefont {P.~J.}\ \bibnamefont {Ledwith}}, \bibinfo {author}
  {\bibfnamefont {A.}~\bibnamefont {Vishwanath}},\ and\ \bibinfo {author}
  {\bibfnamefont {D.~E.}\ \bibnamefont {Parker}},\ }\bibfield  {title}
  {\bibinfo {title} {Composite fermi liquid at zero magnetic field in twisted
  $\mathrm{MoTe}_{2}$},\ }\href
  {https://doi.org/10.1103/PhysRevLett.131.136502} {\bibfield  {journal}
  {\bibinfo  {journal} {Phys. Rev. Lett.}\ }\textbf {\bibinfo {volume} {131}},\
  \bibinfo {pages} {136502} (\bibinfo {year} {2023}{\natexlab{a}})}\BibitemShut
  {NoStop}%
\bibitem [{\citenamefont {Dong}\ \emph {et~al.}(2024)\citenamefont {Dong},
  \citenamefont {Wang}, \citenamefont {Wang}, \citenamefont {Soejima},
  \citenamefont {Zaletel}, \citenamefont {Vishwanath},\ and\ \citenamefont
  {Parker}}]{PhysRevLett.133.206503}%
  \BibitemOpen
  \bibfield  {author} {\bibinfo {author} {\bibfnamefont {J.}~\bibnamefont
  {Dong}}, \bibinfo {author} {\bibfnamefont {T.}~\bibnamefont {Wang}}, \bibinfo
  {author} {\bibfnamefont {T.}~\bibnamefont {Wang}}, \bibinfo {author}
  {\bibfnamefont {T.}~\bibnamefont {Soejima}}, \bibinfo {author} {\bibfnamefont
  {M.~P.}\ \bibnamefont {Zaletel}}, \bibinfo {author} {\bibfnamefont
  {A.}~\bibnamefont {Vishwanath}},\ and\ \bibinfo {author} {\bibfnamefont
  {D.~E.}\ \bibnamefont {Parker}},\ }\bibfield  {title} {\bibinfo {title}
  {Anomalous hall crystals in rhombohedral multilayer graphene. $\mathrm{I}$.
  interaction-driven chern bands and fractional quantum hall states at zero
  magnetic field},\ }\href {https://doi.org/10.1103/PhysRevLett.133.206503}
  {\bibfield  {journal} {\bibinfo  {journal} {Phys. Rev. Lett.}\ }\textbf
  {\bibinfo {volume} {133}},\ \bibinfo {pages} {206503} (\bibinfo {year}
  {2024})}\BibitemShut {NoStop}%
\bibitem [{\citenamefont {Parker}\ \emph {et~al.}()\citenamefont {Parker},
  \citenamefont {Ledwith}, \citenamefont {Khalaf}, \citenamefont {Soejima},
  \citenamefont {Hauschild}, \citenamefont {Xie}, \citenamefont {Pierce},
  \citenamefont {Zaletel}, \citenamefont {Yacoby},\ and\ \citenamefont
  {Vishwanath}}]{parker2021fieldtunedzerofieldfractionalchern}%
  \BibitemOpen
  \bibfield  {author} {\bibinfo {author} {\bibfnamefont {D.}~\bibnamefont
  {Parker}}, \bibinfo {author} {\bibfnamefont {P.}~\bibnamefont {Ledwith}},
  \bibinfo {author} {\bibfnamefont {E.}~\bibnamefont {Khalaf}}, \bibinfo
  {author} {\bibfnamefont {T.}~\bibnamefont {Soejima}}, \bibinfo {author}
  {\bibfnamefont {J.}~\bibnamefont {Hauschild}}, \bibinfo {author}
  {\bibfnamefont {Y.}~\bibnamefont {Xie}}, \bibinfo {author} {\bibfnamefont
  {A.}~\bibnamefont {Pierce}}, \bibinfo {author} {\bibfnamefont {M.~P.}\
  \bibnamefont {Zaletel}}, \bibinfo {author} {\bibfnamefont {A.}~\bibnamefont
  {Yacoby}},\ and\ \bibinfo {author} {\bibfnamefont {A.}~\bibnamefont
  {Vishwanath}},\ }\href@noop {} {\bibinfo {title} {Field-tuned and zero-field
  fractional chern insulators in magic angle graphene}},\ \Eprint
  {https://arxiv.org/abs/2112.13837} {arXiv:2112.13837} \BibitemShut {NoStop}%
\bibitem [{\citenamefont {Morales-Dur\'an}\ \emph {et~al.}(2023)\citenamefont
  {Morales-Dur\'an}, \citenamefont {Wang}, \citenamefont {Schleder},
  \citenamefont {Angeli}, \citenamefont {Zhu}, \citenamefont {Kaxiras},
  \citenamefont {Repellin},\ and\ \citenamefont
  {Cano}}]{PhysRevResearch.5.L032022}%
  \BibitemOpen
  \bibfield  {author} {\bibinfo {author} {\bibfnamefont {N.}~\bibnamefont
  {Morales-Dur\'an}}, \bibinfo {author} {\bibfnamefont {J.}~\bibnamefont
  {Wang}}, \bibinfo {author} {\bibfnamefont {G.~R.}\ \bibnamefont {Schleder}},
  \bibinfo {author} {\bibfnamefont {M.}~\bibnamefont {Angeli}}, \bibinfo
  {author} {\bibfnamefont {Z.}~\bibnamefont {Zhu}}, \bibinfo {author}
  {\bibfnamefont {E.}~\bibnamefont {Kaxiras}}, \bibinfo {author} {\bibfnamefont
  {C.}~\bibnamefont {Repellin}},\ and\ \bibinfo {author} {\bibfnamefont
  {J.}~\bibnamefont {Cano}},\ }\bibfield  {title} {\bibinfo {title}
  {Pressure-enhanced fractional chern insulators along a magic line in moir\'e
  transition metal dichalcogenides},\ }\href
  {https://doi.org/10.1103/PhysRevResearch.5.L032022} {\bibfield  {journal}
  {\bibinfo  {journal} {Phys. Rev. Res.}\ }\textbf {\bibinfo {volume} {5}},\
  \bibinfo {pages} {L032022} (\bibinfo {year} {2023})}\BibitemShut {NoStop}%
\bibitem [{\citenamefont {Ledwith}\ \emph {et~al.}(2020)\citenamefont
  {Ledwith}, \citenamefont {Tarnopolsky}, \citenamefont {Khalaf},\ and\
  \citenamefont {Vishwanath}}]{PhysRevResearch.2.023237}%
  \BibitemOpen
  \bibfield  {author} {\bibinfo {author} {\bibfnamefont {P.~J.}\ \bibnamefont
  {Ledwith}}, \bibinfo {author} {\bibfnamefont {G.}~\bibnamefont
  {Tarnopolsky}}, \bibinfo {author} {\bibfnamefont {E.}~\bibnamefont
  {Khalaf}},\ and\ \bibinfo {author} {\bibfnamefont {A.}~\bibnamefont
  {Vishwanath}},\ }\bibfield  {title} {\bibinfo {title} {Fractional chern
  insulator states in twisted bilayer graphene: An analytical approach},\
  }\href {https://doi.org/10.1103/PhysRevResearch.2.023237} {\bibfield
  {journal} {\bibinfo  {journal} {Phys. Rev. Res.}\ }\textbf {\bibinfo {volume}
  {2}},\ \bibinfo {pages} {023237} (\bibinfo {year} {2020})}\BibitemShut
  {NoStop}%
\bibitem [{\citenamefont {Bauer}\ \emph {et~al.}(2016)\citenamefont {Bauer},
  \citenamefont {Jackson},\ and\ \citenamefont {Roy}}]{PhysRevB.93.235133}%
  \BibitemOpen
  \bibfield  {author} {\bibinfo {author} {\bibfnamefont {D.}~\bibnamefont
  {Bauer}}, \bibinfo {author} {\bibfnamefont {T.~S.}\ \bibnamefont {Jackson}},\
  and\ \bibinfo {author} {\bibfnamefont {R.}~\bibnamefont {Roy}},\ }\bibfield
  {title} {\bibinfo {title} {Quantum geometry and stability of the fractional
  quantum hall effect in the hofstadter model},\ }\href
  {https://doi.org/10.1103/PhysRevB.93.235133} {\bibfield  {journal} {\bibinfo
  {journal} {Phys. Rev. B}\ }\textbf {\bibinfo {volume} {93}},\ \bibinfo
  {pages} {235133} (\bibinfo {year} {2016})}\BibitemShut {NoStop}%
\bibitem [{\citenamefont {Ledwith}\ \emph {et~al.}(2023)\citenamefont
  {Ledwith}, \citenamefont {Vishwanath},\ and\ \citenamefont
  {Parker}}]{PhysRevB.108.205144}%
  \BibitemOpen
  \bibfield  {author} {\bibinfo {author} {\bibfnamefont {P.~J.}\ \bibnamefont
  {Ledwith}}, \bibinfo {author} {\bibfnamefont {A.}~\bibnamefont
  {Vishwanath}},\ and\ \bibinfo {author} {\bibfnamefont {D.~E.}\ \bibnamefont
  {Parker}},\ }\bibfield  {title} {\bibinfo {title} {Vortexability: A unifying
  criterion for ideal fractional chern insulators},\ }\href
  {https://doi.org/10.1103/PhysRevB.108.205144} {\bibfield  {journal} {\bibinfo
   {journal} {Phys. Rev. B}\ }\textbf {\bibinfo {volume} {108}},\ \bibinfo
  {pages} {205144} (\bibinfo {year} {2023})}\BibitemShut {NoStop}%
\bibitem [{\citenamefont {Dong}\ \emph
  {et~al.}(2023{\natexlab{b}})\citenamefont {Dong}, \citenamefont {Ledwith},
  \citenamefont {Khalaf}, \citenamefont {Lee},\ and\ \citenamefont
  {Vishwanath}}]{PhysRevResearch.5.023166}%
  \BibitemOpen
  \bibfield  {author} {\bibinfo {author} {\bibfnamefont {J.}~\bibnamefont
  {Dong}}, \bibinfo {author} {\bibfnamefont {P.~J.}\ \bibnamefont {Ledwith}},
  \bibinfo {author} {\bibfnamefont {E.}~\bibnamefont {Khalaf}}, \bibinfo
  {author} {\bibfnamefont {J.~Y.}\ \bibnamefont {Lee}},\ and\ \bibinfo {author}
  {\bibfnamefont {A.}~\bibnamefont {Vishwanath}},\ }\bibfield  {title}
  {\bibinfo {title} {Many-body ground states from decomposition of ideal higher
  chern bands: Applications to chirally twisted graphene multilayers},\ }\href
  {https://doi.org/10.1103/PhysRevResearch.5.023166} {\bibfield  {journal}
  {\bibinfo  {journal} {Phys. Rev. Res.}\ }\textbf {\bibinfo {volume} {5}},\
  \bibinfo {pages} {023166} (\bibinfo {year} {2023}{\natexlab{b}})}\BibitemShut
  {NoStop}%
\end{thebibliography}%

\clearpage
\onecolumngrid

\newcommand{\bk}{\bm{k}}
\newcommand{\bq}{\bm{q}}
\newcommand{\btk}{\widetilde{\bm{k}}}
\newcommand{\btq}{\widetilde{\bm{q}}}
\newcommand{\br}{\bm{r}}
\newcommand{\cop}{\hat{c}}
\newcommand{\dop}{\hat{d}}
\newcommand{\xmark}{\ding{55}}
\def\Red#1{\textcolor{red}{#1}}
\def\Blue#1{\textcolor{blue}{#1}}

\begin{center}
\textbf{\large Supplementary Materials for: Sliding-tuned  Quantum Geometry in  moir\'e Systems: Nonlinear Hall Effect and Quantum Metric Control}
\end{center}

\setcounter{equation}{0}
\setcounter{figure}{0}
\setcounter{table}{0}
\setcounter{page}{1}
\makeatletter
\renewcommand{\theequation}{S\arabic{equation}}
\renewcommand{\thefigure}{S\arabic{figure}}
\renewcommand{\bibnumfmt}[1]{[S#1]}

\section{I. Hamiltonian of CT3BLG}

In the main text, the Hamiltonian of CT3BLG has been given in Eq.~(\ref{Hamiltonian_2+2+2}).
In the following, we provide a detailed description of its components.
Firstly, $H_l$ in Eq.~(\ref{Hamiltonian_2+2+2}) represents the $k\cdot p$ Hamiltonian of the $l\textrm{th}$ BLG,

\begin{equation}
    H_l(\mathbf{k}_l)=\left(
    \begin{array}{cc}
    H_0(\mathbf{k}_l)&T\\
    T&H_0(\mathbf{k}_l)
    \end{array}
    \right),
\end{equation}
where $H_0=-v_F\mathbf{k}_l\cdot(\xi\mathbf{\sigma}_x,\sigma_y)$ is the Hamiltonian of monolayer graphene and $\mathbf{k}_l=R(\theta_l)(\hat{\mathbf{k}}-\mathbf{K}_l^\xi)$.
Here, $v_F=\sqrt{3}a\gamma_0/2\hbar$ is the Fermi velocity, $\mathbf{\sigma}_x(\mathbf{\sigma}_y)$ is the Pauli matrix, and $R(\theta_l)$ represents the rotation matrix. The rotation angle $\theta_l$ is defined as
$\theta_l=\{\theta,0,-\theta\}$ for the layers $l=\{1,2,3\}$. 
Additionally, $T$ is the interlayer hopping between two AB-stacked graphene layers, 
\begin{equation}
   T=\left(
    \begin{array}{cc}
    0&t_\perp\\
    0&0
    \end{array}
    \right).
\end{equation}
We set $\gamma_0=2.464~eV, t_\perp=0.4~eV$. 

Then $U_{12/23}$ in Eq.~(\ref{Hamiltonian_2+2+2}) represents moir\'e interlayer tunneling between the 1st and 2nd (2nd and 3rd) BLGs,
\begin{equation}
    U_{12/23}(\boldsymbol{\delta_l})=\left(
\begin{array}{cc}
    0 & 0 \\
    1 & 0
\end{array}
\right)
\otimes U(\boldsymbol{\delta_l})
\end{equation}
where
\begin{equation}\label{U_l}
    U(\boldsymbol{\delta}_l)=\sum_{n=0}^2 U_n
    e^{i\mathbf{\widetilde{g}}_n\cdot \mathbf{r}}e^{i \eta_l \mathbf{\widetilde{G}_n}\cdot \boldsymbol{\delta}_l },
\end{equation}
Here, the reciprocal lattice vectors of the moir\'e pattern are given by
\begin{equation}
\widetilde{\bm{g}}_n=(\delta_{n,1}+\delta_{n,2})\widetilde{\mathbf{G}}_M^1+\delta_{n,2}\widetilde{\mathbf{G}}_M^2,
\end{equation}
the component $\widetilde{\mathbf{G}}_M^i$ take the forms as
\begin{equation}\label{G_M}
\widetilde{\mathbf{G}}_M^i=R(-\theta)\widetilde{\mathbf{G}}_C^i-\widetilde{\mathbf{G}}_C^i(i=1,2)
\end{equation}
And $\widetilde{\mathbf{G}}_n$ in Eq.~(\ref{U_l}) is the reciprocal lattice vector of BLG, defined by
\begin{equation}\label{G_n}
\widetilde{\mathbf{G}}_n=(\delta_{n,1}+\delta_{n,2})\widetilde{\mathbf{G}}_C^1+\delta_{n,2}\widetilde{\mathbf{G}}_C^2
\end{equation}
where, $\widetilde{\mathbf{G}}_C^i$ in Eq.~(\ref{G_M}) and Eq.~(\ref{G_n}) represent the reciprocal lattice vectors of the top BLG,
\begin{equation}
\begin{aligned}
\widetilde{\mathbf{G}}_C^1&=\frac{2\pi}{\widetilde{a}_0}\times R(\theta)(1,-1/\sqrt{3})\\
\widetilde{\mathbf{G}}_C^2&=\frac{2\pi}{\widetilde{a}_0}\times R(\theta)(0,2/\sqrt{3})
\end{aligned}
\end{equation}
The $U_n$ matrix in Eq.~(\ref{U_l}) has the form
\begin{equation}
    U_n=\left(
    \begin{array}{cc}
       \omega_{\mathrm{AA}} & \omega_{\mathrm{AB}}e^{-i\xi n\frac{2\pi}{3}}\\
       \omega_{\mathrm{AB}}e^{i\xi n\frac{2\pi}{3}} & \omega_{\mathrm{AA}} 
    \end{array}
    \right),
\end{equation}
with interlayer tunneling parameters $\omega_{AA}=0.0797~$eV, $\omega_{AB}=0.0975~$eV. 
In this paper, we only focus on the case that the sliding of top BLG,
\begin{equation}
    U(\boldsymbol{\delta}_1)=\sum_{n=0}^2 \left(
    \begin{array}{cc}
       \omega_{\mathrm{AA}} & \omega_{\mathrm{AB}}e^{-i\xi n\frac{2\pi}{3}}\\
       \omega_{\mathrm{AB}}e^{i\xi n\frac{2\pi}{3}} & \omega_{\mathrm{AA}} 
    \end{array}
    \right)
    e^{i\mathbf{\widetilde{g}}_n\cdot \mathbf{r}}e^{i \mathbf{\widetilde{G}_n}\cdot \boldsymbol{\delta}_1 }
\end{equation}
and the sliding vectors of the third layer relative to the middle layer is zero, i.e., $\boldsymbol{\delta_3}\equiv\boldsymbol{0}$ for $U_{23}(\boldsymbol{\delta_3})$.

\section{II. Equivalent relation between sliding vectors}


For the sliding lattice configurations discussed in our study, the sliding vector can theoretically take arbitrary values within the two-dimensional space. However, as previously noted in the main text, the effects of interlayer sliding manifest solely in terms associated with the phase factor $\exp{(i\boldsymbol{G_j}\cdot \boldsymbol{\delta}_i)}$. This implies that each sliding vector $\boldsymbol{\delta}_1$ is always equivalent to a vector in the Wigner-Seitz (WS) cell of monolayer in real space. Below, we present the detailed derivation.

In our paper, we only focus on the sliding of the top layer relative to the middle layer ($\boldsymbol{\delta_1}\neq \boldsymbol{0}$, $\boldsymbol{\delta_3}=\boldsymbol{0}$). Firstly,
given the lattice periodicity, any sliding vector $\boldsymbol{\delta_1}$ can be expressed as the sum of $\boldsymbol{r}_1$ within the WS cell and integer lattice vector $\boldsymbol{R}_{1}$,
\begin{equation}\label{delta_1}
   \boldsymbol{\delta_1}=\boldsymbol{R}_{1}+\boldsymbol{r}_1.
\end{equation}
In the $\mathrm{MoTe_2}$ system, we substitute this formula into the terms of
the Hamiltonian (Eq.~(\ref{H_N})) that depend on the sliding vector $\boldsymbol{\delta_1}$. Specifically, we insert Eq.~(\ref{delta_1}) into Eq.~(\ref{Delta_i}) and Eq.~(\ref{T_i2}) as
\begin{align}
    \Delta_{1,2}\left(\boldsymbol{R}_1+\boldsymbol{r}_1\right)
   &=2 V \sum_{j=1,3,5} \cos \left[\boldsymbol{g}_j \cdot \boldsymbol{r}+\boldsymbol{G}_j \cdot \boldsymbol{R}_1 +\boldsymbol{G}_j \cdot \boldsymbol{r}_1 \pm \psi \right]\notag\\
   T_{12}(\boldsymbol{R}_1+\boldsymbol{r}_1)
    &=w\left[1+e^{-i ( \mathbf{g}_2 \cdot \mathbf{r}+\boldsymbol{G}_2 \cdot \boldsymbol{r}_{1})}e^{-i \boldsymbol{G}_2 \cdot \boldsymbol{R}_{1}}+e^{-i ( \mathbf{g}_3 \cdot \mathbf{r} + \boldsymbol{G}_3 \cdot \boldsymbol{r}_{1})}e^{-i \boldsymbol{G}_3 \cdot \boldsymbol{R}_{1}} \right],
\end{align}
where $\boldsymbol{G}_j$ is the reciprocal lattice vector and $\boldsymbol{R}_1$ is a lattice vector ($\boldsymbol{G}_j \cdot \boldsymbol{R}_1=2\pi N (N\in \mathbb{Z})$). Due to the periodic nature of the cosine and exponential function, 
\begin{equation}
\begin{aligned}
    \Delta_{1,2}\left(\boldsymbol{R}_1+\boldsymbol{r}_1\right)&= \Delta_{1,2}\left(\boldsymbol{r}_1\right)\\
    T_{12}(\boldsymbol{R}_1+\boldsymbol{r}_1)&=T_{12}(\boldsymbol{r}_1),
\end{aligned}
\end{equation}
This means $\mathcal{H}\left(\mathbf{k}, \boldsymbol{\delta}_1, \boldsymbol{\delta}_3=\boldsymbol{0}\right)=\mathcal{H}\left(\mathbf{k}, \boldsymbol{r}_1, \boldsymbol{\delta}_3=\boldsymbol{0}\right)$, i.e., any sliding vector $\boldsymbol{\delta}_1$ is always equivalent to a vector $\boldsymbol{r}_1$ in WS cell of $\mathrm{MoTe_2}$ monolayer.

For CT3BLG, the only term associated with sliding is $U_{12}(\boldsymbol{\delta_l})$ in Eq.~(\ref{Hamiltonian_2+2+2}), defined by
\begin{equation}
U_{12}(\boldsymbol{\delta_l})=\left(
\begin{array}{cc}
    0 & 0 \\
    1 & 0
\end{array}
\right)
\otimes U(\boldsymbol{\delta_1}).
\end{equation}
Similarly, if we substitute Eq.~(\ref{delta_1}) into $U(\boldsymbol{\delta_1})$, we can obtain
\begin{align}
    U(\boldsymbol{R}_1+\boldsymbol{r}_1)&=\sum_{n=0}^2 U_n
    e^{i\mathbf{\widetilde{g}}_n\cdot \mathbf{r}}e^{i\mathbf{\widetilde{G}_n}\cdot \boldsymbol{R}_1 }e^{i\mathbf{\widetilde{G}_n}\cdot \boldsymbol{r}_1 }\\&=\sum_{n=0}^2 U_n
    e^{i\mathbf{\widetilde{g}}_n\cdot \mathbf{r}}e^{i2\pi N}e^{i\mathbf{\widetilde{G}_n}\cdot \boldsymbol{r}_1 }\\&=U(\boldsymbol{r}_1).
\end{align}
Thus, any arbitrary sliding displacement $\boldsymbol{\delta}_1$ in CT3BLG can be uniquely mapped to its corresponding position vector $\boldsymbol{r}_1$ within the WS cell.\par 

Experimentally, the presence of nonzero interlayer sliding in the sample can be determined by measuring the nonlinear Hall effect response. Specifically, the second-harmonic voltage signal is measured, and its magnitude quantitatively correlates with the sliding displacement.

\begin{figure}[h]
    \centering
    \includegraphics[scale=0.45]{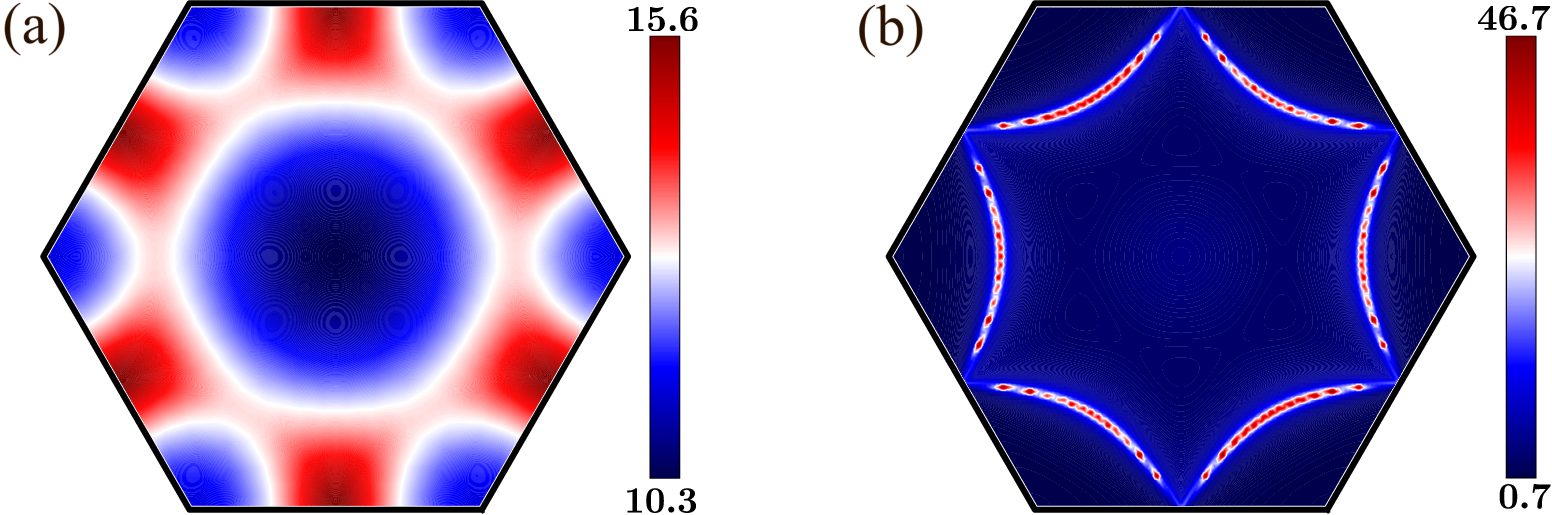}
    \caption{(a) The bandwidth variation and (b) the berry curvature standard deviation for all sliding in Wigner-Seitz cell. Both correspond to the first valence band for AT3L-\textrm{MoTe$_2$} at $\theta=3.15^\circ$.}
    \label{SM_1}
\end{figure}

\section{III. Bandwidth Variation and Berry Curvature standard deviation of AT3L-\textrm{MoTe$_2$}}

In addition to the quantum metric criterion, realizing fractional Chern insulators (FCIs) requires two additional conditions: a nearly flat band structure and a uniform Berry curvature distribution. 

Fig.~\ref{SM_1} (a) shows the bandwidth of AT3L-\textrm{MoTe$_2$}'s first valence band for all sliding values. It remains almost unchanged with a variation of only about $5$~meV, showing a nearly flat band structure. Fig.~\ref{SM_1} (b) is the corresponding Berry curvature standard deviation $\sigma=[\frac{1}{2\pi}\int d^2\bm{k} (\Omega-\bar{\Omega})^2]^{1/2}$, where $\bar{\Omega}=\frac{2\pi C}{S}$,  with C being the Chern number and S the Brillouin zone area. $\sigma$ remains small in the entire sliding region except at the critical points where Chern number transitions occur (see Fig.~\ref{BS}(c)). This directly demonstrates that the Berry curvature maintains a nearly uniform distribution for most sliding configurations.

\section{IV. BCD of the First and Second valence band for AT3L-\textrm{MoTe$_2$}}
\begin{figure}[h]
    \centering
    \includegraphics[scale=0.4]{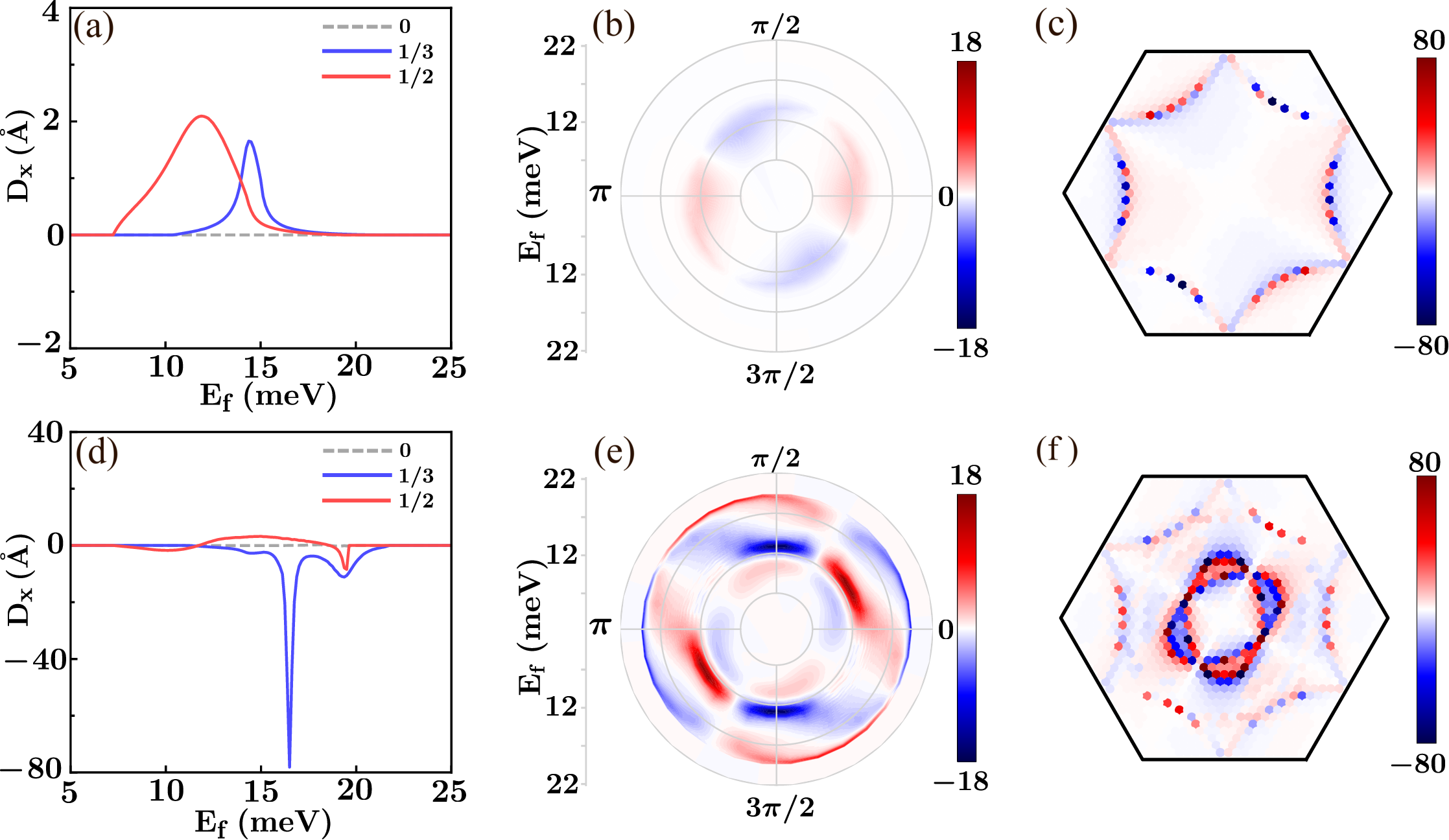}
    \caption{The Berry curvature dipole (BCD) of (a-c) the first valence band and (d-f) the second valence band in AT3L-\textrm{MoTe$_2$}. (a) and (d) show BCD as a function of Fermi level $E_f$ for different sliding vectors
    $\boldsymbol{\delta}_1=\alpha  \boldsymbol{\delta_0}$ ($\alpha=0,\frac{1}{3},\frac{1}{2}$). 
    (b) and (e) are $D_x$ plotted as a function of $E_f$ and sliding angle $\theta_s \equiv \angle(\mathbf{\delta_1}, \mathbf{\delta_0})$ between $\mathbf{\delta_1}$ and $\mathbf{\delta_0}$, with a fixed sliding vector length $\frac{1}{2}|\boldsymbol{\delta_0}|$. (c), (f) show the maximum of $D_x$ for all possible sliding in Wigner-Seitz cell with optimal $E_f$.}
    \label{SM_2}
\end{figure}

In AT3L-\textrm{MoTe$_2$}, the giant Berry curvature dipole (BCD) originates from the top two moir\'e valence bands, with their individual contributions quantitatively shown in Fig.~\ref{SM_2}.

Fig.~\ref{SM_2} (a-c) are BCD data for the first valence band and (d-f) are for the second. Fig.~\ref{SM_2} (a) and (d) are BCD as a function of $E_f$ for various sliding vectors $\boldsymbol{\delta}_1=\alpha \boldsymbol{\delta_0}$ ($\alpha=0,\frac{1}{3},\frac{1}{2}$). At zero sliding ($\alpha=0$), the BCD vanishes in both bands due to the $C_3$ symmetry. When $\alpha=\frac{1}{2}$, the absolute BCD value is very large (about $80\,\mathring{\mathrm{A}}$) for the second valence band in Fig.~\ref{SM_2} (d). This clearly shows that the total giant BCD originates mainly from the second valence band, and the contribution from the first valence band is an order of magnitude smaller. Fig.~\ref{SM_2} (b) and (e) are $D_x$ as a function of $E_f$ and sliding angle $\theta_s \equiv \angle(\mathbf{\delta_1}, \mathbf{\delta_0})$ between $\mathbf{\delta_1}$ and $\mathbf{\delta_0}$, with a fixed sliding vector length $\frac{1}{2}|\boldsymbol{\delta_0}|$. Fig.~\ref{SM_2} (c) and (f) are the maximum of $D_x$ for all possible sliding in Wigner-Seitz cell with optimal $E_f$. These maxima are noticeably larger at the critical points where Chern number transitions occur (see Fig.~\ref{BS}(c)) and sometimes have opposite signs, resulting in a smaller sum, as seen in Fig.~\ref{NLE}(e).

\end{document}